\documentclass[apj,twocolumn,twocolappendix,numberedappendix]{openjournal}
\usepackage{amsmath}
\usepackage{booktabs}
\usepackage{multirow}
\usepackage{color}
\usepackage{soul}
\usepackage{threeparttable}
\usepackage{float}
\usepackage{graphicx}
\usepackage{CJK}
\usepackage{xspace}
\usepackage{afterpage}
\usepackage{placeins}
\usepackage{amssymb}
\usepackage[breaklinks,colorlinks,citecolor=blue,urlcolor=blue,linkcolor=blue,filecolor=blue]{hyperref}
\usepackage[normalem]{ulem}

\makeatletter % Workaround to only color the year for cite commands
  \patchcmd{\NAT@citex}
    {\@citea\NAT@hyper@{%
      \NAT@nmfmt{\NAT@nm}%
      \hyper@natlinkbreak{\NAT@aysep\NAT@spacechar}{\@citeb\@extra@b@citeb}%
      \NAT@date}}
    {\@citea\NAT@nmfmt{\NAT@nm}%
    \NAT@aysep\NAT@spacechar\NAT@hyper@{\NAT@date}}{}{}

  \patchcmd{\NAT@citex}
    {\@citea\NAT@hyper@{%
      \NAT@nmfmt{\NAT@nm}%
      \hyper@natlinkbreak{\NAT@spacechar\NAT@@open\if*#1*\else#1\NAT@spacechar\fi}%
        {\@citeb\@extra@b@citeb}%
      \NAT@date}}
    {\@citea\NAT@nmfmt{\NAT@nm}%
    \NAT@spacechar\NAT@@open\if*#1*\else#1\NAT@spacechar\fi\NAT@hyper@{\NAT@date}}
    {}{}
\makeatother

\shorttitle{Little Red Dot $-$ Host Galaxy = Black Hole Star}
\shortauthors{Sun, Naidu et al.}

\usepackage[export]{adjustbox}

\newcommand{\orcidauthor}[3]{\author{\href{http://orcid.org/#1}{#2$^{#3}$}}}
\newcommand{\nion}[2]{#1\,\textsc{#2}}

\begin{document}
% \begin{CJK*}{UTF8}{gbsn}

\title{\vspace{-1cm} Little Red Dot $-$ Host Galaxy $=$ Black Hole Star:\\ A Gas-Enshrouded Heart at the Center of Every Little Red Dot \vspace{-1.75cm}}

\orcidauthor{0009-0007-3791-7890}{Wendy Q. Sun}{1, *}
\orcidauthor{0000-0003-3997-5705}{Rohan P. Naidu}{1, *,\dagger}
\orcidauthor{0000-0003-2871-127X}{Jorryt Matthee}{2}
\orcidauthor{0000-0002-2380-9801}{Anna de Graaff}{3}
\orcidauthor{0000-0002-0302-2577}{John Chisholm}{4,5}
\orcidauthor{0000-0002-5612-3427}{Jenny E. Greene}{6}
\orcidauthor{0000-0001-5851-6649}{Pascal A.\ Oesch}{7,8,9}
\orcidauthor{0000-0001-5586-6950}{Alberto Torralba}{2}
\orcidauthor{0000-0002-4684-9005}{Raphael E. Hviding}{3}
\orcidauthor{0000-0003-2680-005X}{Gabriel Brammer}{8,9}
\orcidauthor{0000-0003-3769-9559}{Robert A. Simcoe}{1}
\orcidauthor{0000-0002-0974-5266}{Sownak Bose}{10}
\orcidauthor{0000-0002-4989-2471}{Rychard Bouwens}{11}
\orcidauthor{0000-0001-8460-1564}{Pratika Dayal}{12}
\orcidauthor{0000-0003-2895-6218}{Anna-Christina Eilers}{1}
\orcidauthor{0000-0001-7232-5355}{Qinyue Fei}{12}
\orcidauthor{0000-0001-6278-032X}{Lukas J. Furtak}{4,5}
\orcidauthor{0000-0003-0205-9826}{Rashmi Gottumukkala}{8,9}
\orcidauthor{0000-0003-4700-663X}{Andy Goulding}{6}
\orcidauthor{0000-0002-9389-7413}{Kasper E. Heintz}{8,9}
\orcidauthor{0000-0002-3301-3321}{Michaela Hirschmann}{13}
\orcidauthor{0000-0002-5588-9156}{Vasily Kokorev}{4,5}
\orcidauthor{0000-0001-6755-1315}{Joel Leja}{14,15,16}
\orcidauthor{0009-0002-8965-1303}{Zhaoran Liu}{1}
\orcidauthor{0000-0002-5554-8896}{Priyamvada Natarajan}{17,18,19}
\orcidauthor{0009-0007-4394-3366}{Andrew D. Santarelli}{17}
\orcidauthor{0000-0003-4075-7393}{David J. Setton}{6}
\orcidauthor{0000-0002-2838-9033}{Aaron Smith}{20}
\orcidauthor{0000-0002-8224-4505}{Sandro Tacchella}{21,22}
\orcidauthor{0000-0002-3216-1322}{Marta Volonteri}{23}
\orcidauthor{0000-0003-4793-7880}{Fabian Walter}{3}
\orcidauthor{0000-0001-8928-4465}{Andrea Weibel}{7}
\orcidauthor{0000-0003-2919-7495}{Christina C. Williams}{24}
\affiliation{$^1$ MIT Kavli Institute for Astrophysics and Space Research, 70 Vassar Street, Cambridge, MA 02139, USA}
\affiliation{$^2$ Institute of Science and Technology Austria (ISTA), Am Campus 1, 3400 Klosterneuburg, Austria}
\affiliation{$^3$ Max-Planck-Institut f\"ur Astronomie, K\"onigstuhl 17, D-69117 Heidelberg, Germany}
\affiliation{$^4$ Department of Astronomy, The University of Texas at Austin, Austin, TX, USA}
\affiliation{$^5$ Cosmic Frontier Center, The University of Texas at Austin, Austin, TX 78712, USA}
\affiliation{$^6$ Department of Astrophysical Sciences, Princeton University, Princeton, NJ 08544, USA}
\affiliation{$^7$ Department of Astronomy, University of Geneva, Chemin Pegasi 51, 1290 Versoix, Switzerland}
\affiliation{$^8$ Cosmic Dawn Center (DAWN), Copenhagen, Denmark}
\affiliation{$^9$ Niels Bohr Institute, University of Copenhagen, Jagtvej 128, K{\o}benhavn N, DK-2200, Denmark}
\affiliation{$^{10}$ Institute for Computational Cosmology, Department of Physics, Durham University, South Road, Durham DH1 3LE, UK}
\affiliation{$^{11}$ Leiden Observatory, Leiden University, P.O. Box 9513, NL-2300 RA Leiden, the Netherlands}
\affiliation{$^{12}$ David A. Dunlap Department of Astronomy \& Astrophysics, University of Toronto, 50 St. George St., Toronto, ON M5S 3H4, Canada}
\affiliation{$^{13}$ Institute for Physics, Laboratory for Galaxy Evolution and Spectral Modelling, Ecole Polytechnique Federale de Lausanne, Observatoire de Sauverny, Chemin Pegasi 51, 1290 Versoix, Switzerland}
\affiliation{$^{14}$ Department of Astronomy \& Astrophysics, The Pennsylvania State University, University Park, PA 16802, USA}
\affiliation{$^{15}$ Institute for Gravitation and the Cosmos, The Pennsylvania State University, University Park, PA 16802, USA}
\affiliation{$^{16}$ Institute for Computational \& Data Sciences, The Pennsylvania State University, University Park, PA 16802, USA}
\affiliation{$^{17}$ Department of Astronomy, Yale University, New Haven, CT 06511, USA}
\affiliation{$^{18}$ Department of Physics, Yale University, New Haven, CT 06511, USA}
\affiliation{$^{19}$ Yale Center for Astronomy \& Astrophysics, Yale University, New Haven, CT 06520, USA}
\affiliation{$^{20}$ Department of Physics, The University of Texas at Dallas, 800 W Campbell Rd, Richardson, TX 75080, USA}
\affiliation{$^{21}$ The Kavli Institute for Cosmology (KICC), University of Cambridge, Madingley Road, Cambridge, CB3 0HA, UK}
\affiliation{$^{22}$ Cavendish Laboratory, University of Cambridge, 19 JJ Thomson Avenue, Cambridge, CB3 0HE, UK}
\affiliation{$^{23}$ Institut d'Astrophysique de Paris, CNRS, Sorbonne Universit\'e, 98bis Boulevard Arago, 75014, Paris, France}
\affiliation{$^{24}$ NSF National Optical-Infrared Astronomy Research Laboratory, 950 North Cherry Avenue, Tucson, AZ 85719, USA}

\thanks{$^*$E-mail: \href{mailto:wendysun@mit.edu}{wendysun@mit.edu}, \href{mailto:rnaidu@mit.edu}{rnaidu@mit.edu}}
\thanks{$\ddagger$ NASA Hubble Fellow, Pappalardo Fellow}

\begin{abstract}
    The central engines of Little Red Dots (LRDs) may be ``black hole stars" (BH*s), early stages of black hole growth characterized by dense gas envelopes. So far, the most direct evidence for BH*s comes from a handful of sources where the host galaxy is completely outshone as suggested by their remarkably steep Balmer breaks. Here we present a novel scheme to disentangle BH*s from their host galaxies assuming that the [\ion{O}{3}]5008\AA\ line arises exclusively from the host. Using a sample of 98 LRDs ($z\approx2-9$) with high quality NIRSpec/PRISM spectra, we demonstrate that the host-subtracted median stack displays a Balmer break $>2\times$ stronger than massive quiescent galaxies, with the rest-optical continuum resembling a blackbody-like SED ($T_{\rm{eff}}\approx4050$ K, $\log(L_{\rm{bol}})\approx43.9$ erg s$^{-1}$, $R_{\rm{eff}}\approx1300$ au). We measure a steep Balmer decrement (H$\alpha$/H$\beta>10$) and numerous density-sensitive features (e.g., \ion{Fe}{2}, \ion{He}{1}, \ion{O}{1}). These are hallmark signatures of dense gas envelopes, providing population-level evidence that BH*s indeed power LRDs. In the median LRD, BH*s account for $\sim20\%$ of the UV emission, $\sim50\%$ at the Balmer break, and $\sim90\%$ at wavelengths longer than H$\alpha$ with the remainder arising from the host. BH*s preferentially reside in low-mass galaxies ($M_{\rm{\star}}\approx 10^{8}\,{\rm M}_{\rm{\odot}}$) undergoing recent starbursts, as evidenced by extreme emission line EWs (e.g., [\ion{O}{3}]5008\AA$\approx1100$\AA, \ion{C}{3}]$\approx12$\AA), thereby favoring BH* origins linked to star-formation. We show V-shaped LRD selections are biased to high BH*/host fractions ($\gtrsim60\%$ at 5500\AA) -- less dominant BH*s may be powering JWST's blue broad-line AGN. We find BH*s are so commonplace and transient (duty cycle $\sim1\%$, lifetime $\sim10$ Myrs) that every massive black hole may have once shone as a BH*.
\end{abstract}

\section{Introduction}
\label{sec:intro}

One of the most discussed JWST discoveries is a new class of compact, red, point-sources with unprecedented SEDs -- the so-called ``Little Red Dots" (LRDs; \citealt{Matthee24}). Their sheer numbers ($\approx10^{-5}$cMpc$^{-3}$; e.g., \citealt{Greene24, Kocevski24, Kokorev24, Akins25, Zhang25GNLRDs, Lin25cosmos3d}) mean that any satisfying theory of the early Universe must address their origins, and that they must be incorporated in generic models of galaxy formation. While initially thought to be dusty AGN \citep[e.g.,][]{Kocevski23, Matthee24, Harikane23}, massive galaxies \citep[e.g.,][]{Labbe23Nature, Baggen23, Baggen24, PerezGonzalez24}, or some combination thereof \citep[e.g.,][]{Wang24evolved, Ma25, Ronayne25}, it has become clear that LRDs cannot be explained with the traditional building blocks of galaxy or AGN SEDs.

LRDs are characterized by a constellation of peculiar features. Their optical to UV spectra show a characteristic ``V-shape", with blue UV slopes ( ``\texttt{\textbackslash}'') paired with a rising, red optical component (``/") \citep[e.g.,][]{Setton24}. Similarly, the morphology is often strongly wavelength-dependent -- typically point-sources in the rest-optical, even at JWST's exquisite resolution, even when sheared by gravitational lensing down to $<30$ pc \citep[e.g.,][]{Furtak23QSO1, Golubchik25}, whereas often extended and clumpy in the rest-UV \citep[e.g.,][]{Zhang25,Chen25hostifany, Rinaldi25}. In contrast to typical AGN, they are remarkably X-ray faint for their optical luminosity \citep[e.g.,][]{Yue24, Ananna24, Sacchi25} and show no variability, even across decades \citep[e.g.,][]{Zhang25var, Tee25, Burke25}, with only a handful exceptions \citep[e.g.,][]{Furtak25, Ji25BT, Zhang25cepheid}. Despite the extreme Balmer decrements ($\approx10$; e.g., \citealt{Brooks25Balmer, degraaff25pop, Torralba25IFU}), stringent non-detections in the FIR and the faint mid-IR leave little room for dust \citep[e.g.,][]{Casey25, Xiao25, Setton25}, raising questions about where the redness of LRDs comes from. 

One of the most perplexing features of LRDs is the deep, smooth Balmer break seen in a subset of these objects \citep[e.g.,][]{Labbe24, Wang24, Wang24evolved, Kokorev24giant,Ji25BT,Torralba25IFU}. In the first few billion years, Balmer breaks are typically observed in massive, quiescent galaxies, with the breaks arising due to absorption in the stellar atmospheres of A-type stars \citep[e.g.,][]{degraaff24, Carnall24, Weibel24QG}. However, if this was indeed the case in LRDs \citep[e.g.,][]{Labbe23Nature}, their prolific numbers would imply a cosmic stellar mass density at odds with the baryonic fraction in $\Lambda$CDM \citep[e.g.,][]{Boylan-Kolchin23}. Furthermore, a massive galaxy interpretation ($\approx10^{9-11}\,{\rm M}_{\rm{\odot}}$) is also in tension with clustering studies that show LRDs reside in environments similar to low-mass ($\approx10^{7-8}$) dwarf galaxies \citep[e.g.,][]{Matthee25LRDclustering, Lin25clustering, Pizzati25}.

A turning point in the LRD puzzle arrived with the discovery of two objects -- MoM-BH*-1 \citep[][]{Naidu25BHstar} and The Cliff \citep[][]{degraaff25}. While all previously known LRDs showed Balmer breaks with strength ($f^{\nu}(4050\rm{\AA})$/$f^{\nu}(3670\rm{\AA})$) comparable to massive galaxies \citep[e.g.,][]{Furtak24, Wang24evolved, Labbe24,Williams24, Ma25, Ji25BT}, these two sources have breaks far stronger ($>2\times$) than the theoretical maximum expected for a standard dust-free stellar population \citep[e.g.,][]{degraaff25}. This rules out stars as the origin of the break, and instead points to a hitherto unobserved physical phenomenon. A key clue to the origin of the break lay in the deep Balmer absorption observed in both objects, a feature seen in virtually every LRD examined with deep high-resolution ($R
\gtrsim1500$) spectroscopy \citep[e.g.,][]{Matthee24, Lin24, deugenio25z5lrd, deugenio25irony, Torralba25IFU}. Gas that is dense enough to produce Balmer absorption is also likely to produce a Balmer break \citep[][]{IM25, Ji25BT}. Building on this link, and noting the simultaneous occurrence of black hole-like and star-like features, \citet[][]{Naidu25BHstar,degraaff25} proposed MoM-BH*-1 and The Cliff were a new type of astrophysical object, so-called ``black hole stars" (BH*s) with $M_{\rm{BH}}\approx10^{6} \rm{M}_{\rm{\odot}}$. As such, the term ``BH*" as used in this paper denotes an empirical phenomenon -- black holes with emergent spectra that are variations on these two archetypal sources. 

BH*s have been modeled to varying degrees of success as black holes that are thoroughly enshrouded in cocoons of dense gas \citep[e.g.,][]{Naidu25BHstar, degraaff25, Rusakov25, Liu25, Kido25, Torralba25IFU, Ji25BT, Begelman25, Santarelli25}. Accretion onto the black hole is the energy source, while the gas cocoon processes the emergent radiation. The resultant SED displays classic signatures of black holes such as broad Balmer lines (albeit with different origins) and immense luminosity (e.g., $L_{\rm{bol}}\approx10^{10-12} L_{\rm{\odot}}$; e.g., \citealt{Greene25}). Simultaneously, they also show star-like features such as deep Balmer breaks, ``stellar" absorption features (e.g., the Balmer series with increasing absorption strength down the series; \citealt{deugenio25irony,deugenio25twice,Torralba25IFU, Naidu25BHstar}), and SEDs peaking in the optical with fall-offs to both longer and shorter wavelengths \citep[e.g.,][]{Greene25, degraaff25pop, Umeda25}. Emission lines in BH*s may not only be produced by an accretion disk (if it exists) but also via collisional excitation \citep[e.g.,][]{Torralba25IFU, degraaff25pop}. The lines escape the dense cocoon via scattering (e.g., electron scattering, resonance scattering), as commonly seen in stellar phenomena such as luminous blue variables (LBVs) and Type-IIn (i.e., cocooned) supernovae \citep[e.g.,][]{Dessart09, Rusakov25, Naidu25BHstar, Chang25, Sneppen26}.

Excitingly, BH*s share many of their key properties with early proposed theoretical models of BH formation \citep[e.g.,][]{Rees84}, particularly those linked to the accelerated growth of BHs in gas-rich incubating environments such as nuclear star clusters \citep[e.g.,][]{Alexander14, Inayoshi16, Inayoshi20} and the direct collapse of gas in star-forming halos \citep[e.g.,][]{Begelman06, Lodato07, Natarajan11} that may result in objects such as the failed supernovae of massive stars known as ``quasistars" \citep[e.g.,][]{Begelman08,Volonteri10}. We note here again that the term BH* as used in this paper (and as coined in \citealt{Naidu25BHstar, degraaff25}) is an empirical phenomenon agnostic to formation channel and how exactly the BH* ends up in a gas-enshrouded state (e.g., via nuclear clusters, quasistars, direct collapse).

It is not just the objects with large Balmer breaks, but \textit{all} LRDs that may be powered by BH*s. \citet[][]{Naidu25BHstar} combined the spectrum of MoM-BH*-1 with a neighboring UV-bright galaxy to argue that LRDs may be described as combinations of host galaxies and BH*s. Indeed, \citet{degraaff25pop,Barro25} recently built on this approach by using The Cliff instead of MoM-BH*-1 as their BH* template to explore the diversity of LRDs. As per this composite picture, the host galaxies in the ``pure" BH*s \citep[e.g.,][]{Naidu25BHstar, degraaff24} are being thoroughly outshone, whereas in V-shaped LRDs the host is relatively brighter, contributing the blue UV component and diluting the Balmer break (see e.g., \citealt{Taylor25, Golubchik25, Umeda25} for similar-spirited host+BH* modeling of LRDs).

The compact optical morphology vs. extended UV is then a difference between BH* light vs. galaxy light \citep[e.g.,][]{Zhang25} as opposed to scattered AGN light \citep[e.g.,][]{Labbe23}. The other perplexing properties of LRDs also begin to make sense as a consequence of the physics of BH*s. The BH* SED is weak in X-rays \citep[e.g.,][]{Kido25}, the redness of LRDs is due to absorption by gas and so the FIR remains weak \citep[e.g.,][]{Setton25, Xiao25, Casey25}, the apparently thermal SED accounts for the flat MIR \citep[e.g.,][]{Williams24}, and finally variability arises not from accretion disk physics \citep[e.g.,][]{Secunda25} but due to e.g., ``stellar pulsation" in the extended photospheres that produces fluctuations on longer timescales \citep[e.g.,][]{Zhang25var, Cantiello25}. The emission lines are broadened not by virial kinematics, but emerge from the optically thick cocoon of gas via various scattering (e.g., electron scattering, resonance) and collisional processes implying traditional BH mass indicators may not apply \citep[e.g.,][]{Naidu25BHstar, Rusakov25, Chang25}. The pseudo-blackbody-like SED that peaks in the optical translates to a far lower bolometric luminosity \citep[e.g.,][]{Greene25, degraaff25pop, Umeda25}. These lower masses and luminosities fit comfortably within the paradigm of galaxy/BH co-evolution versus earlier inferences of apparently ``overmassive" black holes with $M_{\rm{BH}}/M_{\rm{\star}}\approx1-100\%$ \citep[e.g.,][]{Furtak24,Ji25BT, Matthee25LRDclustering} that were a few orders of magnitude higher than the local $M_{\rm{BH}}/M_{\rm{\star}}\approx0.01\%$ \citep[e.g.,][]{RV15}.

While all these lines of evidence are extremely promising for the ``LRD = BH* + Host" picture, we are still lacking the most direct test of this hypothesis. If all LRDs are truly BH*s embedded within host galaxies, carefully subtracting the galaxy light must leave behind a spectrum resembling the ``pure" BH*s such as MoM-BH*-1 and The Cliff with all the hallmark signatures of radiative transfer in a dense gas ``photosphere" such as sharp Balmer breaks \citep[e.g.,][]{IM25, Ji25BT, Naidu25BHstar} and steep Balmer decrements \citep[e.g.,][]{Chang25, Torralba25IFU, degraaff25pop}. Morphological decomposition -- e.g., with a point source representing the BH* plus a S\'ersic component for the host -- has proven extremely challenging owing to the compact sizes of the hosts \citep[e.g.,][]{Matthee24, Zhang25, Chen25hostifany, Baggen25MI, Rinaldi25}, and even when extended \citep[e.g.,][]{Kokorev25glimpsed}, the resolution of the photometry is too coarse to e.g., measure the strength of the Balmer break or infer line ratios. 

In this paper we develop a novel empirical approach for LRD/host decomposition directly on the spectra to test the nature of their central engines. Some of the cleanest clues to the formation and assembly of BHs are available via the relationship of the accreting BH with that of the stellar populations in their hosts. We exploit the vast JWST spectroscopic archive of galaxies at similar redshifts as LRDs to approximate their hosts. We then test whether the host-subtracted spectra truly resemble BH*s, thereby providing a stringent test of this paradigm. The resulting BH* and host spectra then allow us to derive population-level constraints on their properties. A preliminary version of this method was piloted in \citet[][]{Naidu25BHstar} to explain the spectrum of A2744-QSO1 \citep[][]{Furtak24} by subtracting a stack of observed galaxies to show the residual spectrum closely resembled MoM-BH*-1.

Throughout this work, we adopt a flat $\Lambda$CDM cosmology with parameters as per \citet[][]{Planck18}. Magnitudes are in the AB system \citep[e.g.,][]{Oke83}. For summary statistics, we report medians with uncertainties on the median from bootstrapping (16$^{\rm{th}}$ and 84$^{\rm{th}}$ percentiles). We often reference the Balmer break strength for various objects -- this is the ratio in $f_{\rm{\nu}}$ in two spectral windows ([3620--3720]\AA, [4000--4100]\AA) motivated by \citet[][]{Wang24evolved} and used across the LRD literature \citep[e.g.,][]{degraaff25,degraaff25pop, Naidu25BHstar, Taylor25}.

\section{Data}
\label{sec:data}

\begin{figure*}
    \centering
    \includegraphics[width=0.95\linewidth]{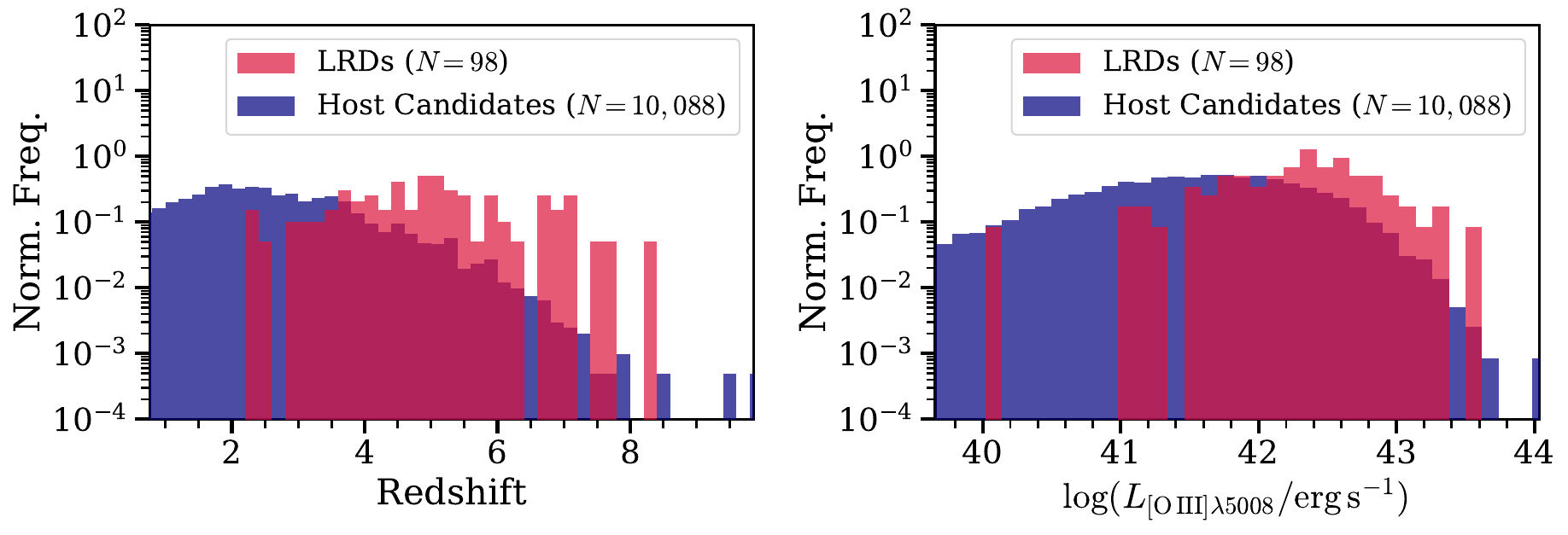}
    \caption{\textbf{Distributions of redshift (left) and [\ion{O}{3}] luminosity (right) of the LRD sample and the candidate host galaxy sample.} Our empirical host subtraction method relies on matching LRDs with peer galaxies that lie at similar redshift and that display similar [\ion{O}{3}] luminosity (\S\ref{sec:methods}). The vastness of the DAWN JWST archive ensures that NIRSpec/PRISM spectra of peer galaxies are available for every LRD in our sample.}
\label{fig:sample_distribution}
\end{figure*}

The data used in this work is drawn from \texttt{v4.4} of the public DAWN JWST Archive\footnote{\url{https://dawn-cph.github.io/dja/index.html}} \citep[DJA;][]{DJA}. We make use of the $\approx17,000$ high-quality (\texttt{grade}$=$3, i.e., visually inspected robust redshifts) NIRSpec/PRISM spectra uniformly processed with the \texttt{msaexp} software \citep[][]{msaexp}. The data processing is described in detail in \citet[][]{Heintz25,degraaff25rubies}. Relevant to this work, a key feature of the \texttt{v4} reductions is an extension of the nominal wavelength range of the PRISM spectra to $\approx5.5\mu$m (versus the $<5.3\mu$m in early standard JWST reductions), improved bar-shadow corrections, and updated flux calibration \citep{Valentino25,Pollock25}.

In addition to the public \texttt{v4.4} DJA release, we include 13 LRD PRISM spectra from the Mirage or Miracle Survey (MoM, GO-5224; PIs: Oesch \& Naidu; \citealt{Naidu25BHstar,Naidu25z14}) processed with the exact same pipeline and recently analyzed in \citet[][]{degraaff25pop}. We also study the five LRDs discovered with the NIRCam grism in the FRESCO Survey \citep[][]{Oesch23, Matthee24}, which were recently followed up with NIRSpec/IFU (PRISM+G395H) in GO-5664 (PI: Matthee). These are among the most luminous LRDs known, e.g., FRESCO-GN-9771 \citep[][]{Matthee24,Torralba25IFU} is the most luminous object ($L_{\rm{bol}}\approx10^{45}$ erg s$^{-1}$) in our sample.

\subsection{LRD Sample}

We begin with a sample of 119 LRDs selected by \citet[][]{degraaff25pop} from \texttt{v4.4} of the DJA and the aforementioned datasets, complemented by the 5 additional sources from GO-5664. These LRDs are selected to display a characteristic V-shaped continuum in their PRISM spectrum along with point-source morphology in the NIRCam imaging (F444W; \texttt{v7} DJA mosaics). In detail, the V-shaped criterion is implemented on PRISM spectra based on the UV slope ($\beta_{\rm{UV}}$) and optical slope ($\beta_{\rm{opt.}}$) measured at wavelengths on either side of the Balmer break ($0.12-0.36\mu$m and $0.36-0.70\mu$m) as follows: $\beta_{\rm{UV}}<-0.2$, $\beta_{\rm{opt}}>0$ and $\beta_{\rm{UV}}-\beta_{\rm{opt}}<0.5$. \citet[][]{Hviding25} demonstrate that selecting for V-shaped point-sources implicitly also selects for broad Balmer lines.  When NIRSpec M-grating spectra are available for these prism spectra, the broad-line fraction is $98\%$ \citep[][]{degraaff25pop}. This sample therefore has all the defining characteristics of LRDs \citep[e.g.,][]{Matthee24, Kokorev24, Labbe25LRD, Akins25}. We refer readers to \citet[][]{degraaff25pop} for a thorough overview of the properties of this sample.

To ensure we are robustly probing the key features of interest (e.g., the Balmer break, Balmer lines) and cleanly measuring the metrics used in our decomposition procedure ([\ion{O}{3}], continuum level) we require every $1000\rm{\AA}$ window at $\lambda_{\rm{rest}} = 3000 - 6000 \rm{\AA}$ to have a median SNR per pixel of $>2$. This SNR threshold is based on experimentation with the mock test framework we will discuss in \S\ref{sec:mocks}, pushing to as low SNR as possible to include the maximum number of sources without affecting the quality of reconstruction (i.e., the reconstructed LRD stack is within 1$\sigma$ of the simulated LRD stack). We also require sources to have $>90\%$ of their observed spectral pixels to be flagged as ``valid" (i.e., no data quality issues) in the DJA reductions. This leaves us with a sample of 98 LRDs with high quality spectra that form the basis of this paper.

\subsection{Candidate Host Galaxy Sample}
\label{sec:hostsample}

After applying similar quality cuts to the full DJA sample (median SNR per pix $>2$ in 1000\AA\ windows, $>95\%$ valid pixels) and excluding LRDs we are left with $10,088$ sources with high-quality PRISM spectra. We emphasize that for this large parent pool of candidate hosts no other cuts are made on galaxy properties (line fluxes, EWs, star-forming vs. quiescent, redshift). As such, this sample represents the full diversity of the high-redshift Universe observed with JWST/NIRSpec. In what follows, candidate hosts are selected from this sample based on each LRD's properties, in particular their proximity in redshift and [\ion{O}{3}] luminosity (see \S\ref{sec:methods} for motivation and details). Note that we do not explicitly exclude non-LRD AGN from the candidate host sample. Identifying AGN from prism spectra alone (whose large statistics and continuous wavelength coverage our method relies on) is challenging without supporting higher resolution observations \citep[e.g.,][]{Chisholm24, Hviding25, Scholtz25}. However, such objects are expected to comprise only a few percent of a blind [\ion{O}{3}]-selected sample as seen in the non-detection of broad H$\beta$ in stacks even at the highest [\ion{O}{3}] luminosities ($\approx10^{43-44}$ erg s$^{-1}$; e.g., \citealt{Meyer24}) and as per their incidence fractions relative to LRDs in the [\ion{O}{3}] luminosity range studied in this paper ($\approx1:1$; e.g., \citealt{Hviding25}) .

\subsection{Emission Line Fluxes}

We use the \texttt{UNITE} emission line fitting package \citep[][]{Hviding25} to measure line fluxes ([\ion{O}{3}], H$\alpha$, H$\beta$) self-consistently for all individual objects and stacks. Briefly, \texttt{UNITE} translates model spectra to observed spectra while accounting for all relevant instrumental effects (e.g., NIRSpec's wavelength dependent resolution; \citealt{degraaff24rubies}). All emission lines are fit as combinations of narrow/emission (the better-fit between FWHM $=$ 0-750 km s$^{-1}$ or FWHM $=$ 0-1,000 km s$^{-1}$) and broad (FWHM $=$ 250-2,500 km s$^{-1}$) Gaussian line profiles. The key measurement for host/galaxy decomposition is the [\ion{O}{3}] luminosity -- we have verified that our derived values are in excellent agreement (in almost all cases within 20\%) with the spline fits released with DJA \texttt{v4.4}. As discussed in \S\ref{sec:data}, our sample includes sources that are not part of DJA \texttt{v4.4} and so for consistency we derive fluxes for all sources with \texttt{UNITE}.

\section{Methods: LRD Decomposition}
\label{sec:methods}

Here we describe our approach to separate LRD hosts from their central engines. The high-level idea is that each LRD is matched with an ensemble of real, observed galaxies that are plausible approximations of the LRD's host galaxy. By subtracting the SEDs of each of these candidate hosts from the LRD SED, we build an empirically motivated posterior for the SED of the central engine, thereby testing whether LRDs truly host BH*s.

The key advantage of this data-driven approach vs. traditional SED fitting (as with galaxy+AGN models; e.g., \citealt{Lyu24, Chavez25, Taylor25}) is that we make minimal assumptions about the nature of the central engines. SED fitting depends on the template sets used. At the moment, the development of ab initio templates (of e.g., BH* models) that are able to match the key features of LRDs is still in the early stages \citep[e.g.,][]{Liu25BB,Santarelli25, Ji25BT,Naidu25BHstar,Taylor25}.

\begin{figure*}
    \centering
    \includegraphics[width=0.8\linewidth]{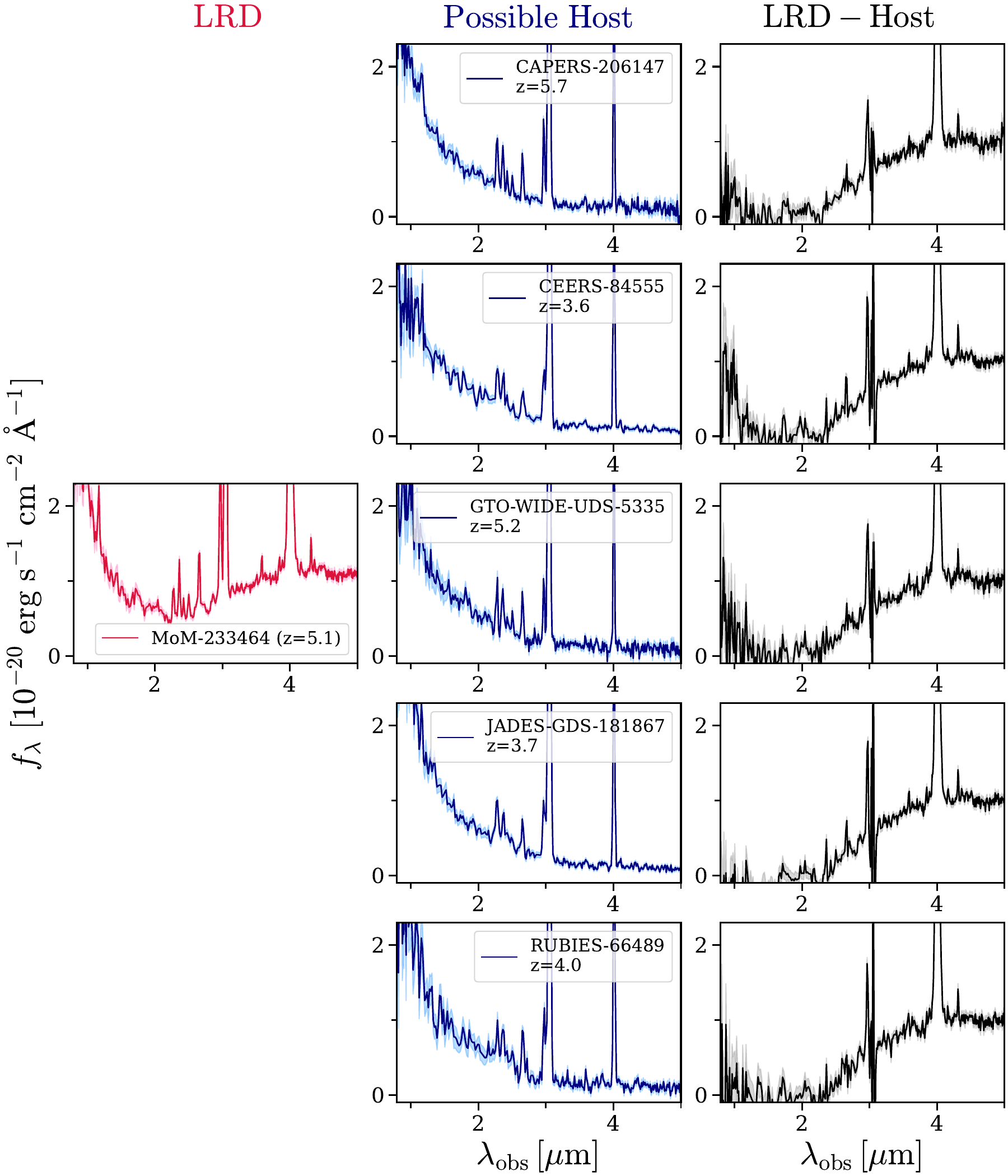}
    \caption{\textbf{Illustration of our empirical LRD decomposition strategy.} Every LRD (red; left) is matched with candidate host galaxies (navy; center) at similar redshift ($|\Delta z|<1.5$) and of comparable [\ion{O}{3}] luminosity ($\pm0.5$ dex) from the DAWN JWST Archive (DJA). By subtracting each candidate host from the LRD after scaling to match the [\ion{O}{3}] luminosity we derive a set of possible $\text{LRD}-\text{Host}$ SEDs (black; right). We show five candidate hosts here for simplicity, whereas in practice each LRD is matched to a median of 81 possible hosts -- e.g., the LRD shown above has 95 matches. The $\text{LRD}-\text{Host}$ SEDs display over/under-subtraction around the [\ion{O}{3}] wavelength due to subtle differences in the [\ion{O}{3}] line width, but note that every $\text{LRD}-\text{Host}$ spectrum has zero [\ion{O}{3}] luminosity by construction.}
    \label{fig:method}
\end{figure*}

\subsection{Assumptions}
\label{sec:assumptions}

Our method is built on three key assumptions.

\begin{enumerate}
    \item \textit{Every LRD is a combination of a host galaxy and a central engine.} This assumption is motivated by LRDs whose rest-optical SEDs cannot be explained by any known stellar population and are dominated essentially entirely by a single point-source \citep[][]{Naidu25BHstar,degraaff25}. Combining a component resembling BH* models with blue star-forming galaxies results in a V-shape SED as seen in typical LRDs \citep[e.g.,][]{Naidu25BHstar,Ji25BT,Taylor25, degraaff25pop}. In what follows, we make no assumptions about the properties of the central engines (except \#3 below) and instead seek to infer the host-subtracted LRD spectrum.     
    
    \item \textit{The hosts can be approximated by peer galaxies at similar redshifts.} In other words, if the central engine did not exist, we assume the host would resemble some subset of the galaxy population cataloged in the DJA. This is motivated by the finding that typical LRDs resemble generic galaxies in the rest-UV, the wavelengths where the host is expected to dominate in the composite picture involving a red BH* \citep[e.g.,][]{Naidu25BHstar,Taylor25, Ji25BT}. Evidence for this picture comes from the extended UV morphology \citep[e.g.,][]{Rinaldi25, Chen25hostifany, Torralba25LyA}, blue UV slopes \citep[e.g.,][]{Matthee24, degraaff25pop, Barro25}, and lack of UV variability with all hints of variability till date reported in the rest-optical \citep[e.g.,][]{Furtak25, Ji25BT, Zhang25cepheid, Naidu25BHstar, deugenio25z5lrd}. Furthermore, clustering studies show LRDs reside in unremarkable environments indistinguishable from generic low-mass galaxies at similar redshift \citep[e.g.,][]{Lin25clustering, Matthee25LRDclustering}. Intriguingly, \citet[][]{Matthee25LRDclustering} observed that the LRD host mass derived from clustering is similar to the mass inferred by assuming all the UV light arises from the host -- as we will show later (\S\ref{sec:bhstarfrac}), this works well for the least luminous LRDs, the kind studied in their work.
    
    \item\textit{[\ion{O}{3}] emission arises purely from the host galaxy.} [\ion{O}{3}]$5008$\AA\ is the one feature accessible in virtually every LRD PRISM spectrum that appears to be powered essentially entirely by the host galaxy, and so it forms the foundation of our method. [\ion{O}{3}] is always observed to be very narrow in LRDs ($\approx50-200$ km s$^{-1}$) even with JWST's highest resolution grating ($R\gtrsim3000$) \citep[e.g.,][]{Torralba25IFU,deugenio25z5lrd,Ji25BT}. Furthermore, the line-width is consistent with the stellar/dynamical masses inferred for the hosts \citep[e.g.,][]{Maiolino24, Ji25BT, Torralba25IFU} tying its origin to ionized gas in the ISM. Indeed, on a population level, the [\ion{O}{3}] luminosity in LRDs is tightly correlated with $M_{\rm{UV}}$ \citep[][]{degraaff25pop}. As discussed above (\#2), the UV SEDs of LRDs are likely dominated by the host galaxy and so these correlations tie [\ion{O}{3}] to the host. Finally, the two LRDs with the strongest Balmer breaks \citep[][]{Naidu25BHstar,degraaff25} -- interpreted to be almost pure BH*s outshining extremely faint hosts -- are accompanied by almost non-existent [\ion{O}{3}] emission. Indeed, \citet{degraaff25pop, Barro25} report population trends between [\ion{O}{3}] EWs and Balmer break strengths -- the stronger the break, the weaker the [\ion{O}{3}] EW. Indeed, theoretical models that successfully reproduce such deep Balmer breaks and the accompanying Balmer absorption invoke gas densities that are a few orders of magnitude higher than the critical density of the [\ion{O}{3}] doublet leading to collisional de-excitation ($n_{\rm{H}}/\rm{cm}^{-3}\approx10^{9-12}$ vs. $\approx10^{6}$; \citealt{IM25,Liu25BB,Chang25}). For all these reasons we assume the [\ion{O}{3}] observed in LRDs arises due to the host galaxy. We acknowledge that this is a simplifying assumption, which based on current evidence, likely holds for a large fraction of the LRD population. We investigate the consequences of breaking this assumption by testing with a $0 - 100\%$ [\ion{O}{3}] contribution from the central engine in \S\ref{sec:host}, though note the ``pure'' BH*s (MoM-BH*-1 and The Cliff) imply a negligible contribution ($<<1\%$).
\end{enumerate}

Overall, our approach may be summarized as empirical SED fitting by using galaxies at similar redshifts. Matching in [\ion{O}{3}] luminosity selects for a combination of the instantaneous star-formation, ionizing photon efficiency, and metallicity \citep[e.g.,][]{Matthee23, Meyer24, Endsley24}. We also place constraints on the shape of the allowed SED (\S\ref{sec:implementation}), which in turn acts as a constraint on the allowed star-formation history and stellar mass \citep[e.g.,][]{Leja19, Iyer19, Tacchella22}. A challenge for this empirical-spirited method is that some LRD hosts appear to occupy poorly populated regions of parameter space in the DJA such that we may find few peer galaxies (e.g., ultra metal-poor $z>6$ galaxies; \citealt{Maiolino25}). However, as we will discuss in \S\ref{sec:implementation}, this situation occurs only in a handful of LRDs. Some considerations for future work include exploring departures from our main assumptions such as the hosts being drawn from the general galaxy population. As an example, we do not account for the possibility of differential dust extinction between LRD hosts and candidate galaxies --  if such a systematic difference exists, it could bias the inferred host $M_{\rm{UV}}$ and the relative contribution of host to the total light \citep[e.g.,][]{Nikopoulos25, Brooks25Balmer, Killi23, Shivaei25, Woodrum25}. As another example, if LRD hosts possess higher gas densities than typical galaxies, [\ion{O}{3}] emissivity may be suppressed, which would lead our matching procedure to select biased peer galaxies.

\begin{figure*}
    \centering    \includegraphics[width=0.85\linewidth]{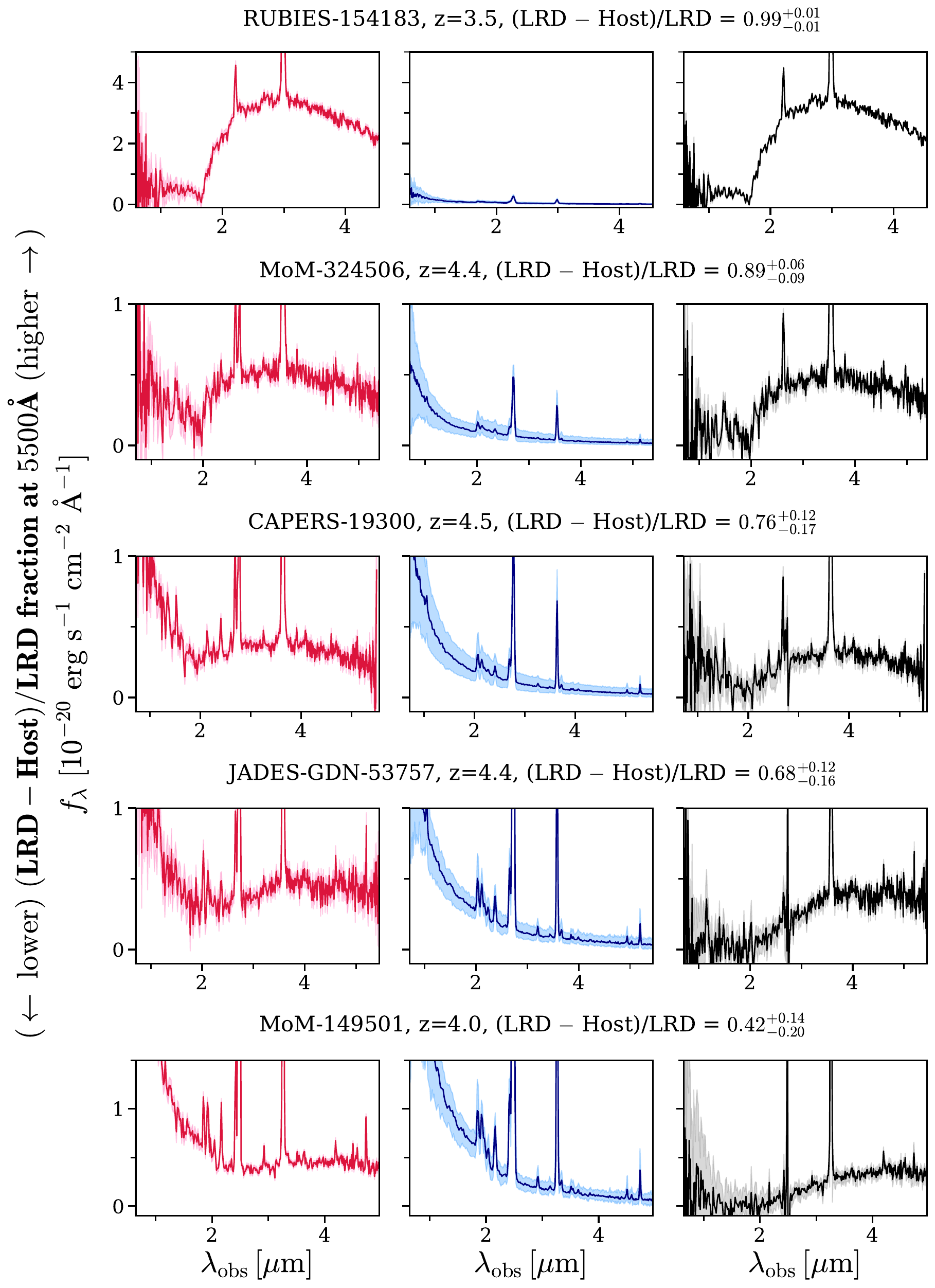}
    \caption{ \textbf{Examples of LRDs spanning the full range of ($\text{LRD}- \text{Host}$)/$\text{LRD}$ fractions inferred for our sample.} Median stacks for the inferred hosts (center) and $\text{LRD}-\text{Host}$ spectra (i.e., the central engine; right) are shown, sorted by the ($\text{LRD} - \text{Host}$ )/LRD fraction at 5500\AA. Errors on the LRD are observed uncertainties, while on the Host and $\text{LRD}-\text{Host}$ spectra are posteriors based on all the possible host candidates. While the central engine produces almost all the optical light in the LRD in the top row (The Cliff from \citealt{degraaff25}), the host galaxy is dominant in the LRD in the bottom row (MoM-149501). Note the diversity in the $\text{LRD}-\text{Host}$ SEDs -- they are not carbon copies of The Cliff (top row), with the peak wavelength of the optical SED occurring at comparable or redder wavelengths. In our framework, this diversity of central engines (see Appendix \ref{fig:appendix_gallery} for decomposition of the full sample), along with the diversity of hosts and their relative fractions accounts for the diversity in LRD SED shapes.}
    \label{fig:workedexamples}
\end{figure*}

\subsection{Implementation}
\label{sec:implementation}

To construct a set of candidate host galaxies for a given LRD we select sources from the pool of all possible 10,088 hosts (\S\ref{sec:hostsample}) that satisfy:
\begin{enumerate}
    \item Redshift within $\Delta(z) =\pm1.5$.
    \item $L_{\rm{[OIII]}}$ within $\pm 0.5$ dex.
    \item Host SED$\times$($L_{\rm{[OIII], LRD}}/L_{\rm{[OIII], host}}$) $\lesssim$ the LRD SED across all $\lambda_{\rm{obs}}>0.8\mu$m.
\end{enumerate}

The first two criteria select for galaxies with properties matched to the LRD host. These galaxy's SEDs are then rescaled by a factor of $L_{\rm{[OIII], LRD}}/L_{\rm{[OIII], host}}$ such that the candidate host accounts for all the [\ion{O}{3}] in the LRD. We then require the rescaled host galaxy SED must not exceed the LRD's SED at all observed wavelengths. This is a simple sanity check to ensure that a contributing component (the host galaxy) does not exceed the sum (the LRD). In particular, we compare spline fits to the LRD and host continua at $\lambda_{\rm{obs}}>0.8\mu$m to ensure the host does not exceed the LRD by more than $25\%$ over any contiguous region of $\lambda_{\rm{obs}} = 0.3\mu$m, failing which it is removed from the set of candidates. The $25\%$ tolerance ensures that any minor differences due to e.g., stray noise pixels do not rule out otherwise reasonable candidates. We set the thresholds for all three criteria via experimentation, balancing finding close matches with finding enough matches for robust statistics (see \S\ref{sec:mocks} for mock tests that validate these choices). 

Across the 98 LRDs we study here, after filtering by the first condition requiring proximity in redshift a median of $23.4^{+29.7}_{-17.3}\,\%$ candidates survive. $9.9^{+12.5}_{-8.0}\,\%$ candidates remain after the second step matching in [\ion{O}{3}] luminosity. And finally, $0.8^{+2.0}_{-0.7}\,\%$ candidates remain after the third step requiring SEDs consistent with the LRD. Overall, the median LRD is matched to 81 candidate host galaxies. Further details of the host subtraction procedure are fleshed out below using worked examples. In the handful of cases ($N=5$) where the LRD has less than $5$ host matches (for heterogeneous reasons -- e.g., occurring at high redshifts with few peers, extreme [\ion{O}{3}] values), we iteratively relax the $\Delta(z)$ constraint to $2.0$ and to $2.5$, and the $L_{\rm{[OIII]}}$ constraint to 0.6 and 0.7 dex until five matches are found.

\subsection{Examples}
\label{sec:worked}

In Figure \ref{fig:method} we illustrate our procedure using a single LRD, MoM-233464 at $z_{\rm{spec}}=5.1$. Based on the criteria listed in \S\ref{sec:implementation} we find 95 galaxies with similar redshift and [\ion{O}{3}] luminosity. Each of these sources matched in $L_{\rm{[OIII]}}$ and redshift is treated as an equally likely candidate host and subtracted from the LRD spectrum. To subtract the spectra, all candidate hosts are resampled onto the observed-frame wavelength grid of the LRD with flux conservation using the \texttt{spectres} software \citep[][]{spectres}.

The resulting difference spectra ($\text{LRD} - \text{Host}$) represent draws from the empirically constructed posterior of the SED of the central engine. Note that these difference spectra, by construction, have zero [\ion{O}{3}] luminosity. However, residuals around the [\ion{O}{3}] line are apparent due to differences in e.g., the kinematic width of the line between the LRD and the candidate hosts. We characterize the posterior distribution of each LRD's difference spectra (median and percentiles) to report properties for every individual LRD (e.g., the fraction of light contributed by the host as a function of wavelength). Later, we stack these difference spectra across the full sample (median stack with bootstrap resampling) to represent the median central engine of LRD. Specifically, for each bootstrap iteration, we randomly draw one BH* for each LRD (which gives $98$ BH*s corresponding to the $98$ LRDs), and then we resample with replacement over 10,000 iterations to report the median spectrum with associated uncertainties. We validate every step of this pipeline through realistic mock tests in \S\ref{sec:mocks} that account for e.g., the impact of redshift-dependent differences in resolution.

In Figure \ref{fig:workedexamples} we show median stacked host galaxy and $\text{LRD}-\text{Host}$ spectra for five representative LRDs spanning the full dynamic range of ($\text{LRD}-\text{Host}$)/$\text{LRD}$ ratios ($\approx40-99\%$) in the rest-optical ($\lambda_{\rm{0}}=5500$\AA) in our sample. One of the known BH*-dominated LRDs, The Cliff at $z=3.5$ (RUBIES-154183; \citealt{degraaff25}) has room only for an extremely faint host in its SED, such that the ($\text{LRD}-\text{Host}$)/LRD fraction we derive for it at 5500\AA\ is $99\%$. At lower ratios, the characteristic V-shape LRD SED begins to emerge. 

A key finding of our work is already apparent in this figure -- the SED of the central engine (i.e., $\text{LRD}-\text{Host}$) is remarkably similar across various LRDs, with a continuum that is very weak at wavelengths bluer than the rest-optical ($\approx3000-4000$\AA). At the same time, it is notable that these SEDs do not have a single universal shape (e.g., exactly mirroring The Cliff or MoM-BH*-1), but vary from source to source with the continuum peaking at a range of wavelengths. For example, MoM-149501 and JADES-GDN-53757 in the bottom two panels have ``cold" central engines, with much redder SEDs than the sources powering The Cliff and MoM-324506 shown in the top two rows. We remind readers that these findings follow from our simple assumptions (and their associated caveats) discussed in \S\ref{sec:assumptions}. We will further quantify this diversity in \S\ref{sec:bhstarfrac}.

\subsection{Mock Tests}
\label{sec:mocks}

\begin{figure*}
    \centering
\includegraphics[width=\linewidth]{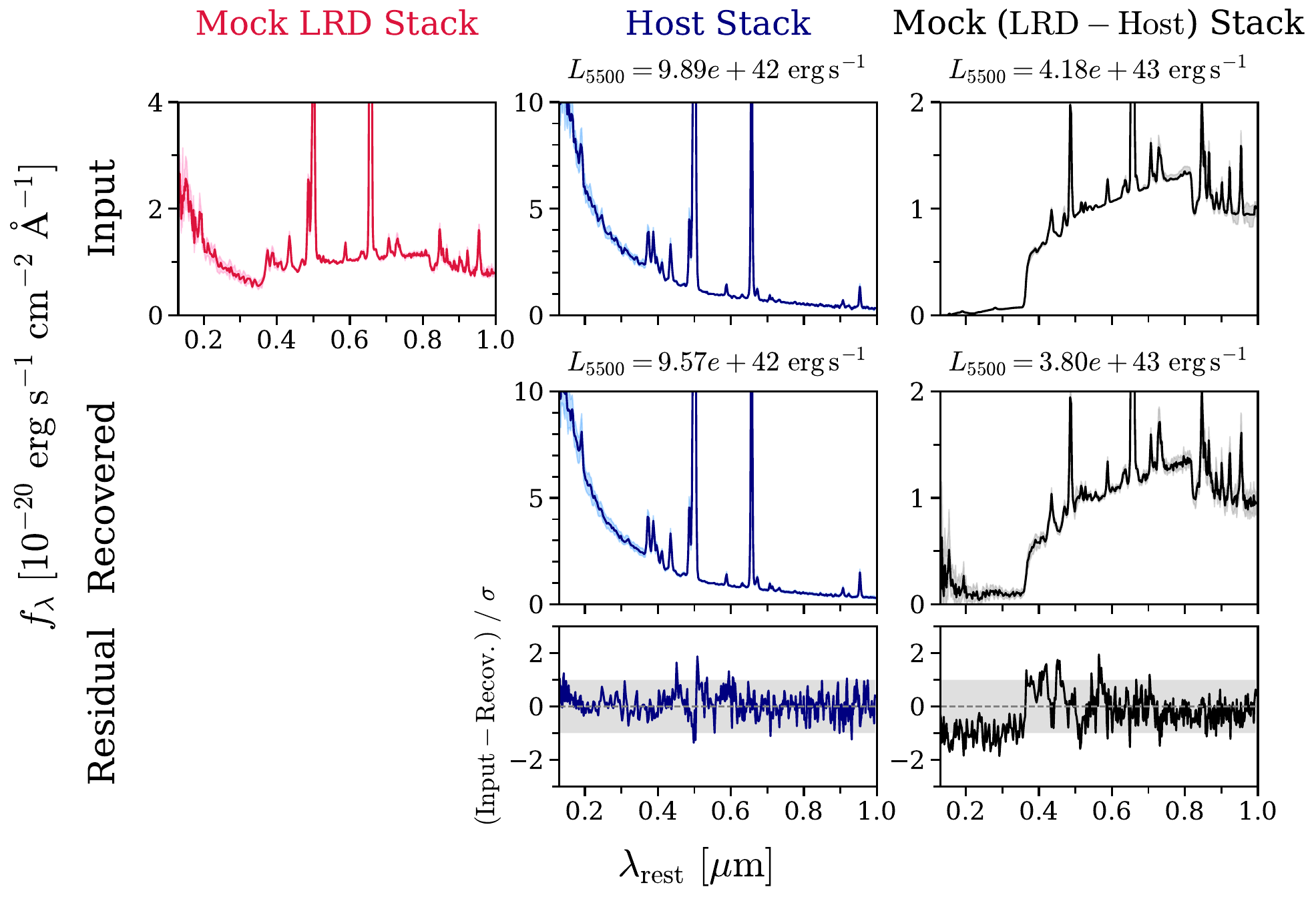}
    \caption{\textbf{Mock test results to validate our host/BH* decomposition strategy.} \textbf{Top:} We produce a sample of synthetic LRDs by combining spectra of observed galaxies (top-center) and \texttt{CLOUDY} models (top-right). Every LRD in our sample has a corresponding synthetic LRD with matched BH*/LRD fraction and noise properties (see \S\ref{sec:mocks} for details). The median stack of the synthetic LRDs is shown in the top-left (red). While we show median stacks here (top-row) as the injected input we are attempting to reconstruct, our decomposition procedure is run on individual synthetic LRDs (exactly as in Fig. \ref{fig:method}). \textbf{Center, Right:} The reconstructed host and BH* stacks are generally in excellent agreement with the input stacks as evidenced by the residuals that agree within $1\sigma$ (bottom panels). The uncertainty in the recovered BH* stack is most pronounced in the rest-UV below $\lesssim2000$\AA\ where spurious emission features appear as the subtracted hosts are not exactly the same as the true hosts -- in what follows, we do not interpret this portion of the reconstructed BH* spectrum.}
    \label{fig:mock_test}
\end{figure*}

\begin{deluxetable*}{lrrrr}
\tabletypesize{\footnotesize}
\tablecaption{Summary of Mock Test Median Stacks}
\tablehead{
\colhead{} & \colhead{Input BH*} & \colhead{Recovered BH*} & \colhead{Input Host} & \colhead{Recovered Host}}
\startdata
\label{tab:mock}
\vspace{-0.3cm}\\
$\log(L_{5500})$ & $43.62^{+0.06}_{-0.10}$ & $43.58^{+0.08}_{-0.09}$ & $43.00^{+0.05}_{-0.04}$ & $42.98^{+0.04}_{-0.06}$ \\
Balmer break & $1.94^{+0.11}_{-0.10}$ & $2.24^{+0.34}_{-0.26}$ & $0.96^{+0.05}_{-0.05}$ & $0.94^{+0.06}_{-0.06}$ \\
$\beta_{\mathrm{UV}}$ & $6.55^{+0.24}_{-0.22}$ & $-1.95^{+0.81}_{-0.74}$ & $-1.68^{+0.08}_{-0.07}$ & $-1.60^{+0.10}_{-0.09}$ \\
H$\alpha\lambda6564$\AA \, EW$_{\rm{0}}$ & $845.7^{+116.5}_{-99.8}$ & $872.5^{+48.8}_{-44.9}$ & $910.3^{+148.4}_{-141.2}$ & $896.3^{+137.5}_{-141.7}$ \\
H$\beta\lambda4862$\AA \, EW$_{\rm{0}}$ & $81.3^{+10.9}_{-9.6}$ & $75.2^{+8.4}_{-7.9}$ & $150.0^{+18.6}_{-21.4}$ & $164.9^{+15.9}_{-15.7}$ \\
\nion{O}{iii}$\lambda5008$\AA \, EW$_{\rm{0}}$ & -- & -- & $1018.7^{+108.8}_{-116.2}$ & $1209.8^{+87.6}_{-96.6}$
\enddata
\tablecomments{Luminosities are in units of erg s$^{-1}$, and rest-frame EWs in \AA. The input and recovered quantities are in excellent agreement. While the luminosities, Balmer break, and optical line EWs are recovered well, the derived BH* at $\lesssim2000$\AA\ is noisy, and quantities in this regime such as $\beta_{\rm{UV}}$ should be treated with care (see Fig. \ref{fig:mock_test}, \S\ref{sec:mocks}).}
\end{deluxetable*}

Before we begin interpreting results from the LRD decomposition, we perform a realistic end-to-end mock test of our pipeline to test the robustness of our results. We produce a synthetic dataset of LRDs with properties matched to our sample. Each synthetic LRD is composed of a model BH* spectrum and the observed spectrum of a host galaxy. What we seek to test here is whether a median injected host and BH* SED is recovered self-consistently by our procedure -- this is not a test of e.g., the underlying physical assumptions outlined in \S\ref{sec:assumptions}.

For realistic model BH* spectra, we draw from the \texttt{CLOUDY} grid presented in \citet[][]{Naidu25BHstar}. We note that this family of models \citep[e.g.,][]{IM25,Ji25BT,Naidu25BHstar,Taylor25, Torralba25IFU,Golubchik25,Yan25} is currently the only one able to produce a match to both the continuum as well as Balmer lines of LRDs. For the exercise at hand, we use a subset of 154 BH* models that match the observed properties of LRDs. These models display a Balmer break strength of $>3$ and have L(H$\alpha$)/L($5100$\AA) within 0.5 dex of the tight relation ($<0.25$ dex scatter) reported in \citet[][]{degraaff25pop}. While we are using the current best models for BH* spectra, we note that there are some surprising features that are not well-known among LRDs yet -- e.g., there are deep Paschen breaks in some of these models, which match hints of this feature in CAPERS-9226, MACS-J0647-1045, and MoM-233464 displayed in the Appendix as well as in our faintest substack in Fig. \ref{fig:bhstarfrac} (see Paschen jumps reported in some LRDs in \citealt{Sneppen26Paschen}; see also the discussion of BH* data/model differences in \citealt{degraaff25}). Nonetheless, what matters for the test at hand is whether the recovered BH* stack matches the injected BH* stack.

To compose a realistic sample closely paralleling the data, we produce one synthetic LRD per observed LRD. The BH* is chosen at random from the \texttt{CLOUDY} models. For the host component, we randomly select a single individual host from the high-SNR ([\ion{O}{3}] SNR$>$5) subset of matched hosts identified for that specific LRD (as per the criteria listed in \S\ref{sec:implementation}). By using one host per observed LRD, we maintain a one-to-one correspondence between the observed LRD and mock LRD populations. Because the \texttt{CLOUDY} models are noiseless, we prioritize high-SNR host templates to ensure that our initial synthetic LRD is relatively clean before noise is injected. The BH* template is scaled and combined with the host spectrum such that the resulting synthetic LRD has the same BH*/LRD fraction at 5500\AA\ as the observed LRD (estimated from its median stack; e.g., right-most panels of Fig. \ref{fig:workedexamples}). To transform the clean synthetic LRD into a realistic mock observation, the SNR of the synthetic spectrum is matched pixel by pixel with the observed spectrum, and we perturb the spectrum based on these errors to mimic the noise in our sample.

Once we have built the mock LRD dataset, we run exactly the same decomposition procedure that we use on observed LRDs, and obtain the results shown in Figure \ref{fig:mock_test} and Table \ref{tab:mock}. As shown in the bottom panels of Fig. \ref{fig:mock_test}, the reconstructed BH* and host median stacks closely mirror the injected stacks, differing only at the $\approx10\%$ level and with residuals generally within $1\sigma$. Rerunning the reconstruction with a different set of mock LRDs (randomly selecting hosts and \texttt{CLOUDY} models) produces very similar results. At the same time, areas of caution are highlighted. The reconstructed BH* shows spurious emission lines at $\lesssim2000$\AA\ and is systematically uncertain at these wavelengths (e.g., with an artificially bluer $\beta_{\rm{UV}}$). This is because the injected BH*s are extremely faint at UV wavelengths, and so the subtle mismatches between the true host and the candidate host (e.g., in UV emission line strengths) are brought into sharper relief than in the rest optical. Informed by this test, we interpret the noisy $<2000$\AA\ region of our recovered BH* spectrum with caution.

We end this section by noting that the mock test results demonstrate the overall robustness of our approach. Various areas of concern (e.g., the fidelity of stacked EWs, the varying PRISM spectral resolution with redshift) are addressed in a realistic manner. All the key quantities discussed in this work -- e.g., the continuum luminosities, Balmer breaks, line EWs -- are recovered successfully (Table \ref{tab:mock}). Furthermore, we understand the wavelengths ($\lesssim2000$\AA) where results must be treated with care.

\section{Results}
\label{sec:results}

\subsection{Little Red Dot - Host Galaxy = Black Hole Star}
\label{sec:reveal}

\begin{figure*}
    \centering
    \includegraphics[width=\linewidth]{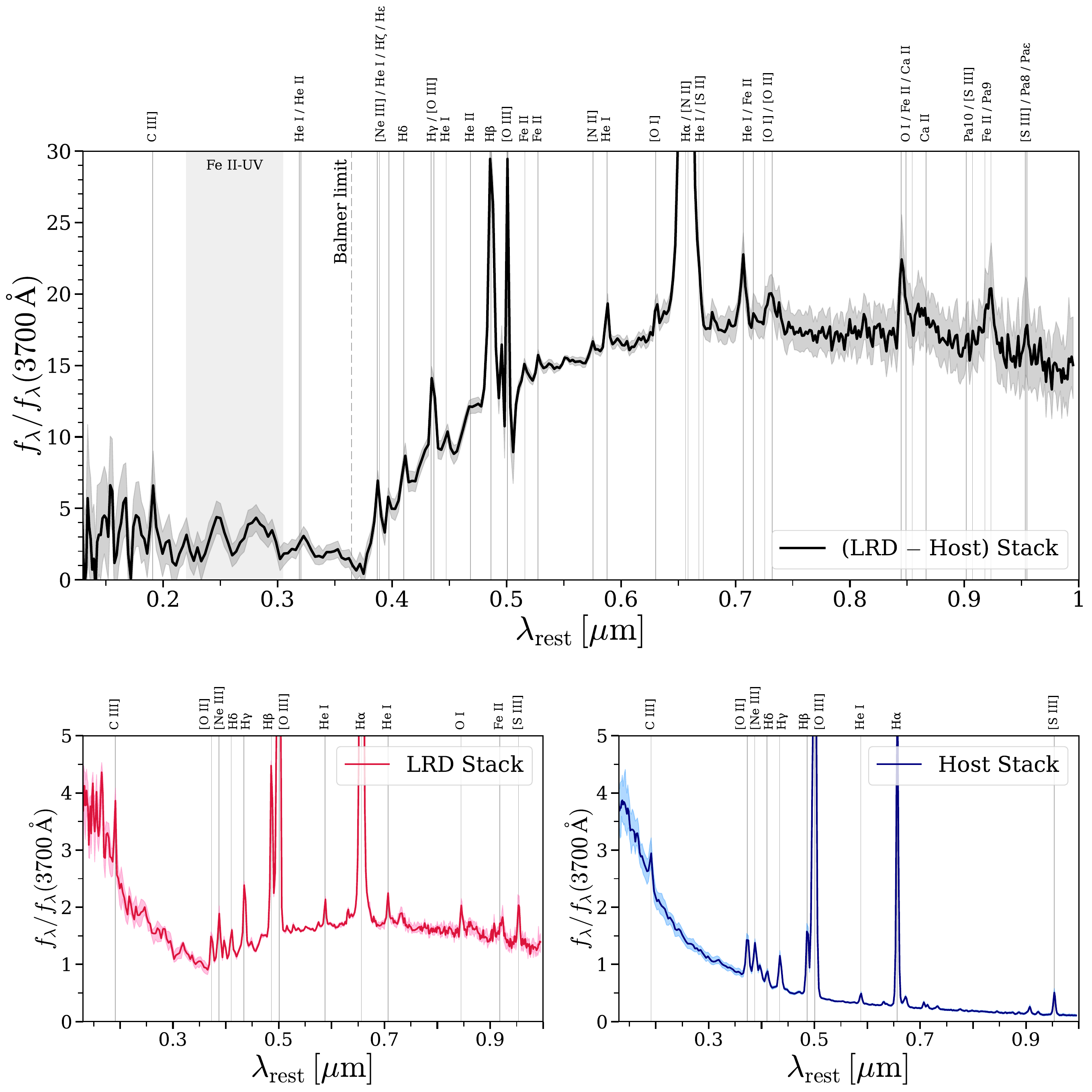}
    \caption{\textbf{Stacks of $\text{LRD} - \text{Host}$  difference spectra (top) and host galaxies (bottom-right) constructed by applying our decomposition procedure to each of the 98 LRDs in our sample (stacked in the bottom-left).} \textbf{Top:} The central engine of the \textit{median} LRD shows a steep Balmer break ($f^{\nu}(4050\rm{\AA})$/$f^{\nu}(3670\rm{\AA})\approx7$), extremely strong H$\alpha$ emission ($\approx850$\AA), a steep Balmer decrement (H$\alpha$/H$\beta\approx16$), and strong \ion{O}{1} and Fe emission. These are hallmark signatures of radiative transfer in a dense gas envelope, strongly supporting the picture that the central engine of the typical LRD is indeed a BH*. Note that the [\ion{O}{3}] line by construction has zero luminosity, but appears over/under-subtracted because the kinematic width of the line is never exactly the same across the LRDs and the candidate hosts. \textbf{Bottom-Left:} In LRDs, these hallmark BH* features are mixed with galaxy light and are challenging to discern (e.g., there is only a weak Balmer break), underscoring the need for our procedure. \textbf{Bottom-Right:} The median host galaxy is highly star-forming, as evidenced by strong rest-optical emission lines (EW(\ion{O}{3}+H$\beta$)$\approx1600$\AA) that are $\approx2-3\times$ stronger than the typical galaxy at these redshifts \citep[e.g.,][]{Endsley24}. The detection of strong \ion{C}{3}] ($\approx12$\AA, $>2\times$ the typical EW at $z\approx5$) points to the hosts having undergone a recent burst \citep[e.g.,][]{Roberts-Borsani25}.}
    \label{fig:main_stack}
\end{figure*}

\begin{figure*}
    \centering
\includegraphics[width=\linewidth]{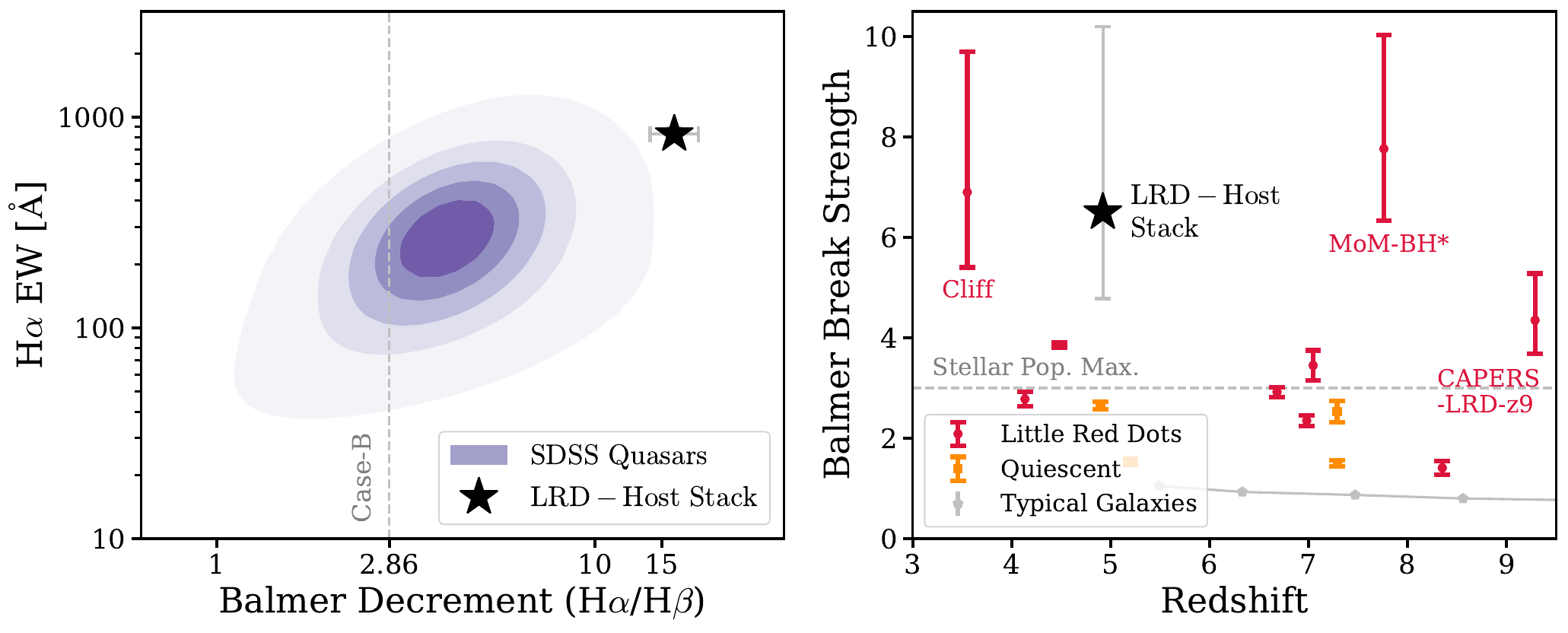}
    \caption{\textbf{The $\text{LRD}-\text{Host}$ difference stack has unusual properties relative to typical AGN and galaxies.} \textbf{Left:} We compare the exceptional H$\alpha$ EW and Balmer decrement of the $\text{LRD}-\text{Host}$ stack against a compilation of $z\approx0-0.6$ SDSS quasars \citep[][]{Wu22SDSS}. Our stack has few peers among these local quasars, hinting that the mechanisms powering the Balmer lines in LRDs are distinct. Indeed, the ubiquity of Balmer absorption \citep[e.g.,][]{Lin24} points to resonant scattering \citep[e.g.,][]{Chang25} and collisional excitation \citep[e.g,][]{Torralba25IFU} as opposed to e.g., photo-ionization and dust attenuation that set H$\alpha$ and H$\alpha$/H$\beta$ among classical AGN. \textbf{Right:} The Balmer break strength of our stack is among the largest observed with JWST, comparable only to The Cliff \citep[][]{degraaff25} and MoM-BH* \citep[][]{Naidu25BHstar} that are the archetypal BH*s. We highlight that the break strength of the stack lies well above the maximum expected from a standard stellar population assuming a dust-free \citet[][]{Chabrier03} IMF (gray dashed line; \citealt{Wang24evolved, degraaff25}), LRDs with Balmer breaks (red; e.g., \citealt{Labbe24,Wang24evolved, Furtak24, Kokorev24giant, Taylor25}), quiescent and mini-quenched galaxies (orange; e.g., \citealt{Strait23,degraaff24, Weibel24QG,Looser24}), and star-forming galaxies (gray points; compiled in \citealt{Roberts-Borsani24}).
    }
    \label{fig:context}
\end{figure*}

\begin{figure*}
    \centering
    \includegraphics[width=0.75\linewidth]{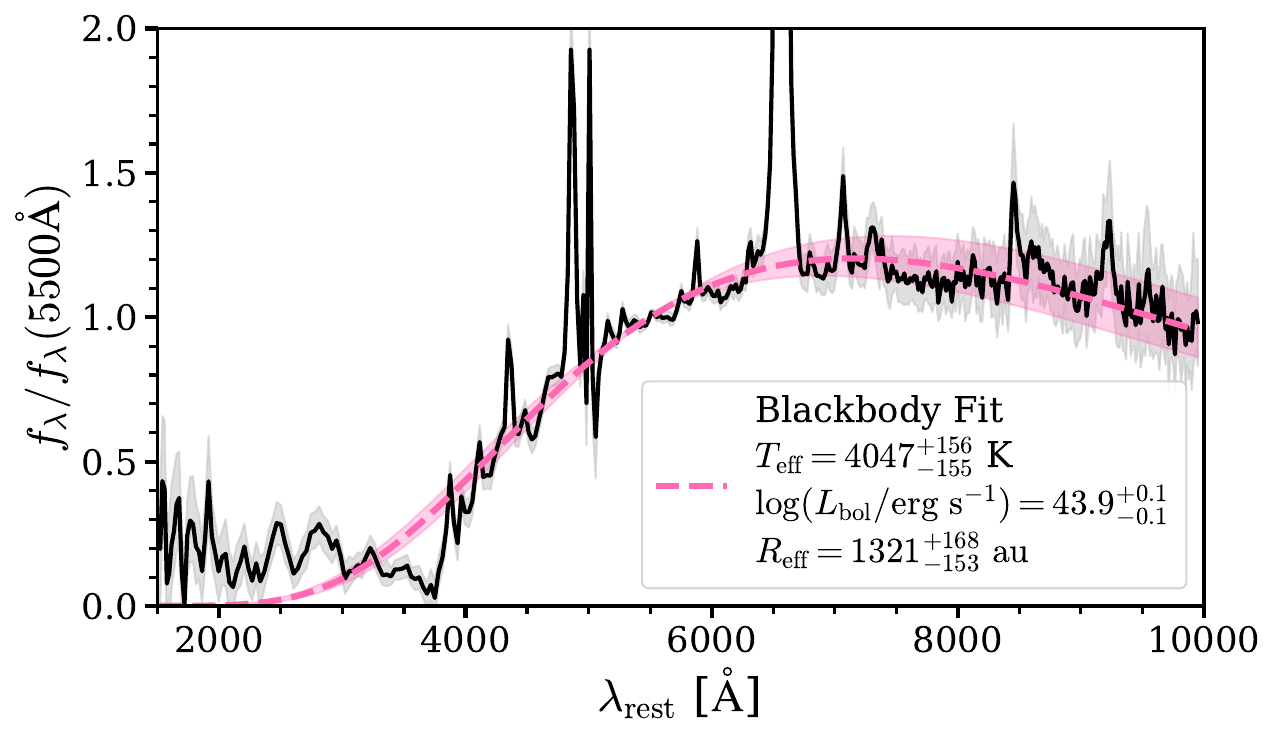}
    \caption{\textbf{A single simple blackbody (pink) provides an empirical description of the rest-optical continuum of the median $\text{LRD} - \text{Host}$ stack (black).} Emission lines are masked and we fit $\lambda_{\rm{rest}}>4200$\AA. The integral of the blackbody function provides an approximation of the bolometric luminosity as empirically demonstrated in  \citet{Greene25,degraaff25pop}. The temperature and luminosity imply a radius of $\approx1300$ au as per the Stefan-Boltzmann law, $\approx100\times$ larger than the largest known stars (red hypergiants reaching $\approx10$ au; e.g., \citealt{Arroyo-Torres13}) but comparable to local broad-line regions \citep[e.g.,][]{Bentz13}.}
    \label{fig:blackbody}
\end{figure*}

In Fig. \ref{fig:main_stack} we display a stacked spectrum representing the median central engine of the 98 LRDs. Measurements based on this stack are summarized in Tables \ref{tab:mainstacklines}, \ref{tab:lrdlines}, \ref{tab:control}, and \ref{tab:properties}. We construct the median stack in a Monte Carlo fashion across 10,000 trials by randomly selecting one $\text{LRD} - \text{Host}$  spectrum for each of the 98 LRDs, bootstrap resampling these difference spectra, and then computing the median spectrum for each trial. Each difference spectrum is normalized at 5500\AA\ before being added to the stack. The spectra are all resampled onto the rest-wavelength grid of the LRD at the median redshift of the sample. For visualization, in Fig. \ref{fig:main_stack} we normalize the median spectrum at 3700\AA\ to emphasize the large Balmer break. Missing wavelength coverage is masked in the stacking, e.g., the redder wavelengths for high-$z$ sources where only H$\beta$ is detected; bluer wavelengths for low-$z$ sources where there is no rest-UV coverage -- as a result the $\lesssim2000$\AA\ and $\gtrsim7000$\AA\ portions are noisier than the rest of the stack. 

The stacked $\text{LRD} - \text{Host}$  spectrum of the full sample of 98 sources in Figure \ref{fig:main_stack} shows a constellation of features that are rare among typical galaxies and AGN (Fig. \ref{fig:context}), but are seen in the handful of known BH*s and expected in models of BH*s, thereby underscoring that the typical LRD is indeed powered by a BH*. In what follows, we provide context for and discuss each of these features:

\afterpage{
\clearpage

\begin{figure}
    \centering
\includegraphics[width=\linewidth]{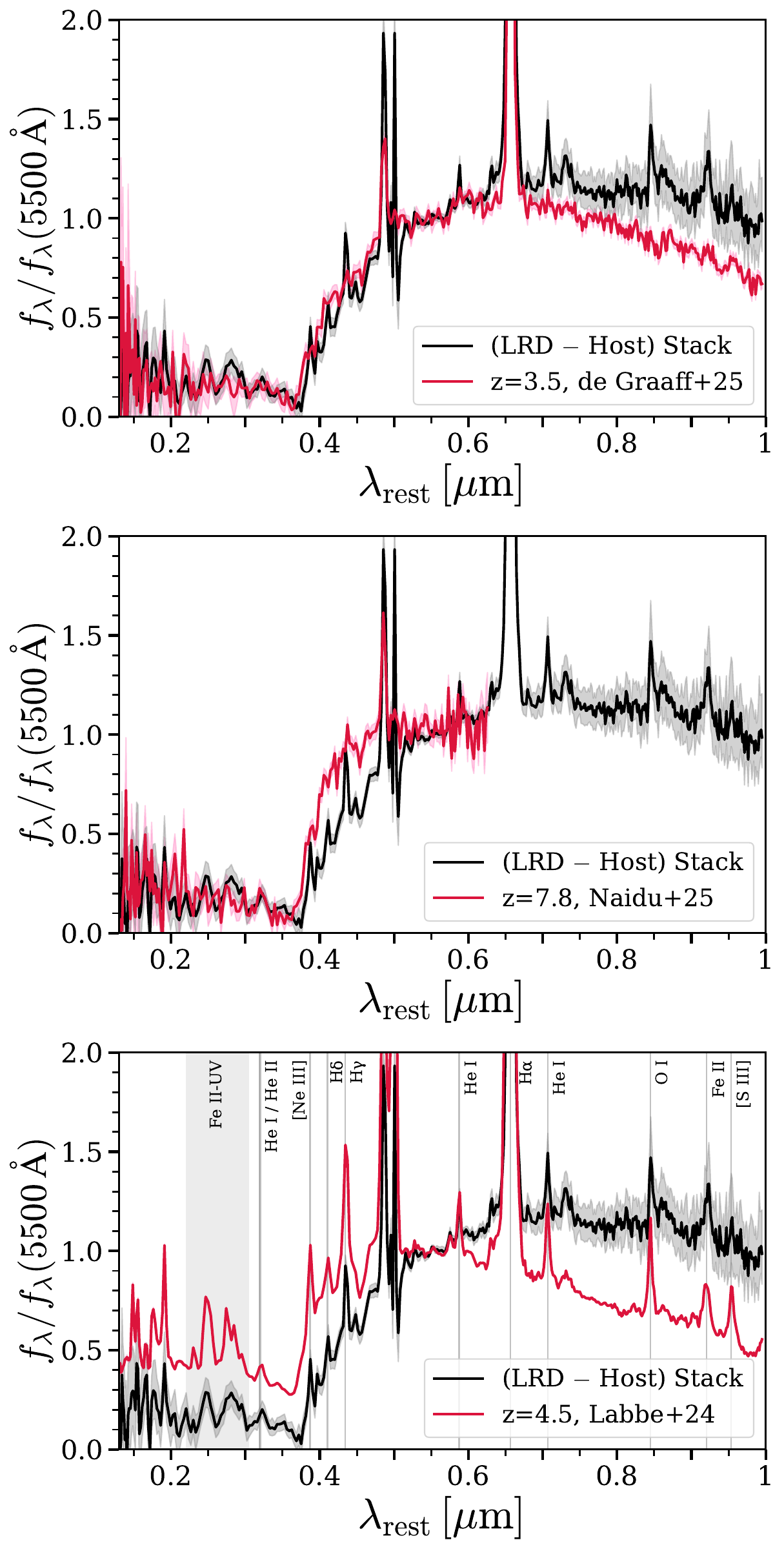}
    \caption{\textbf{Comparison of the $\text{LRD}-\text{Host}$ difference stack (black) with prominent LRDs.} The Cliff (top; BH* fraction $\approx100\%$; \citealt{degraaff25}) and MoM-BH* (center; BH* fraction $\approx100\%$; \citealt{Naidu25BHstar}) are effectively ``pure" BH*s that are outshining their extremely faint host galaxies. These sources bear remarkable similarities to the median stack, showing sharp Balmer breaks and similar continuua, although with ``hotter" $T_{\rm{eff}}\approx4600-5600$ K as evidenced by their spectra peaking at bluer wavelengths than our stack. In the final panel we show A2744-45924 for comparison (BH* fraction $\approx85\%$; \citealt{Labbe24}). This source shows several emission features rare among star-forming galaxies (e.g., \ion{Fe}{2} UV). Such lines have been robustly detected only in extremely bright LRDs so far but are apparent here in the \textit{median} $\text{LRD} - \text{Host}$  stack, underscoring the ubiquity of these features and the common underlying physics across all LRDs. Synthesizing the arguments presented in Figs. \ref{fig:main_stack}, \ref{fig:context}, \ref{fig:blackbody}, and \ref{fig:cliff_monster} we henceforth refer to the ``$\text{LRD}-\text{Host}$ stack" as the ``BH* stack".}
    \label{fig:cliff_monster}
\end{figure}

\begin{deluxetable}{lrr}[h!]
\tabletypesize{\footnotesize}
\tablecaption{Emission Line Measurements for BH* Stack}
\tablehead{
\colhead{Line} & \colhead{$\log(L)$} & \colhead{EW$_{\rm{0}}$} }
\startdata
\label{tab:mainstacklines}
\vspace{-0.3cm}\\
\nion{He}{i}$\lambda3889$\AA & $40.82^{+0.12}_{-0.14}$ & $57.6^{+19.6}_{-16.9}$ \\
H$\gamma\lambda4341$\AA + [\nion{O}{iii}]$\lambda4363$\AA & $41.08^{+0.06}_{-0.07}$ & $37.4^{+6.6}_{-6.1}$ \\
H$\beta\lambda4862$\AA & $41.54^{+0.05}_{-0.04}$ & $73.5^{+8.6}_{-7.4}$ \\
{[\nion{O}{iii}]}$\lambda5008$\AA & -- & -- \\
\nion{He}{i}$\lambda5877$\AA & $40.68^{+0.06}_{-0.07}$ & $8.1^{+1.3}_{-1.3}$ \\
H$\alpha\lambda6564$\AA & $42.75^{+0.03}_{-0.03}$ & $827.6^{+62.8}_{-64.7}$ \\
\nion{He}{i}$\lambda7067$\AA & $40.89^{+0.06}_{-0.06}$ & $11.4^{+1.7}_{-1.6}$ \\
\nion{O}{i}$\lambda8446$\AA\ + \nion{Fe}{ii} & $41.08^{+0.05}_{-0.06}$ & $18.5^{+2.4}_{-2.4}$ \\
$\mathrm{Pa}9\lambda9231.6$\AA\ + \nion{Fe}{ii} & $41.06^{+0.06}_{-0.07}$ & $18.7^{+2.8}_{-3.0}$
\enddata
\tablecomments{Luminosities are in units of erg s$^{-1}$, and rest-frame EWs in \AA. For \texttt{UNITE} measurements, the stack is shifted to observed frame grid of the close-to-median redshift among the LRDs, $z = 4.54$. The [\ion{O}{3}] luminosity is zero by construction (\S\ref{sec:methods}).}
\end{deluxetable}

\begin{deluxetable}{lrr}[h!]
\tabletypesize{\footnotesize}
\tablecaption{Emission Line Measurements for LRD Stack}
\tablehead{
\colhead{Line} & \colhead{$\log(L)$} & \colhead{EW$_{\rm{0}}$} }
\startdata
\label{tab:lrdlines}
\vspace{-0.3cm}\\
\nion{He}{i}$\lambda3889$\AA & $41.23^{+0.14}_{-0.23}$ & $30.4^{+12.2}_{-13.0}$ \\
H$\gamma\lambda4341$\AA + [\nion{O}{iii}]$\lambda4363$\AA & $41.51^{+0.05}_{-0.06}$ & $53.9^{+7.5}_{-7.3}$ \\
H$\beta\lambda4862$\AA & $41.90^{+0.03}_{-0.03}$ & $109.8^{+6.8}_{-6.9}$ \\
{[\nion{O}{iii}]}$\lambda5008$\AA & $42.35^{+0.04}_{-0.05}$ & $298.7^{+30.8}_{-30.5}$ \\
\nion{He}{i}$\lambda5877$\AA & $40.99^{+0.05}_{-0.06}$ & $12.4^{+1.5}_{-1.5}$ \\
H$\alpha\lambda6564$\AA & $42.84^{+0.04}_{-0.05}$ & $817.2^{+80.9}_{-82.7}$ \\
\nion{He}{i}$\lambda7067$\AA & $41.02^{+0.06}_{-0.07}$ & $12.5^{+2.0}_{-1.8}$
\enddata
\tablecomments{Luminosities are in units of erg s$^{-1}$, and rest-frame EWs in \AA. The EWs are generally consistent with \citet[][]{degraaff25pop}, for e.g., EW(H$\alpha$)$=762^{+43}_{-60}$, accounting for slight differences in the studied sample arising from our more restrictive SNR cuts and the additional luminous sources from \citet[][]{Matthee24, Torralba25IFU} studied here (see \S\ref{sec:data}).}
\end{deluxetable}

\begin{deluxetable}{lrr}[h!]
\tabletypesize{\footnotesize}
\tablecaption{Properties of LRD Host Stack vs. Control Stack}
\tablehead{
\colhead{} & \colhead{LRD Host Stack} & \colhead{Control Stack}}
\startdata
\label{tab:control}
\vspace{-0.3cm}\\
$\log(L_{5500})$ & $42.90^{+0.06}_{-0.06}$ & $43.44^{+0.08}_{-0.09}$ \\
Balmer break & $0.98^{+0.06}_{-0.06}$ & $1.16^{+0.06}_{-0.06}$ \\
$\beta_{\mathrm{UV}}$ & $-1.63^{+0.10}_{-0.09}$ & $-1.83^{+0.11}_{-0.10}$ \\
\nion{C}{iii}]$\lambda1908$\AA \, EW$_{\rm{0}}$ & $11.6^{+5.0}_{-4.7}$ & $3.8^{+1.4}_{-1.4}$ \\
H$\beta\lambda4862$\AA \, EW$_{\rm{0}}$ & $147.4^{+14.8}_{-13.6}$ & $47.9^{+6.4}_{-6.6}$\\ 
{[\nion{O}{iii}]}$\lambda5008$\AA \, EW$_{\rm{0}}$ & $1106.4^{+86.7}_{-79.6}$ & $326.4^{+31.8}_{-31.3}$ \\
\nion{He}{i}$\lambda5877$\AA \, EW$_{\rm{0}}$ & $29.3^{+1.5}_{-1.2}$ & $9.7^{+0.4}_{-0.4}$ \\
H$\alpha\lambda6564$\AA \, EW$_{\rm{0}}$ & $939.0^{+54.6}_{-62.5}$ & $321.0^{+20.3}_{-19.7}$ \\
$\log{L_{\text{\nion{C}{iii}]}\lambda1908\text{\AA}}}$ & $41.07^{+0.14}_{-0.22}$ & $41.06^{+0.13}_{-0.19}$ \\
$\log{L_{\text{H}\beta\lambda4862\text{\AA}}}$ & $41.46^{+0.04}_{-0.04}$ & $41.50^{+0.05}_{-0.06}$ \\
$\log{L_{\text{[\nion{O}{iii}]}\lambda5008\text{\AA}}}$ & $42.30^{+0.03}_{-0.03}$ & $42.30^{+0.04}_{-0.04}$ \\
$\log{L_{\text{\nion{He}{i}}\lambda5877\text{\AA}}}$ & $40.58^{+0.02}_{-0.02}$ & $40.64^{+0.02}_{-0.02}$ \\
$\log{L_{\text{H}\alpha\lambda6564\text{\AA}}}$ & $42.04^{+0.02}_{-0.03}$ & $42.08^{+0.03}_{-0.03}$
\enddata
\tablecomments{Luminosities are in units of erg s$^{-1}$, and rest-frame EWs in \AA.}
\end{deluxetable}

\clearpage
}

\begin{enumerate}
    \item \textit{Balmer Break:} The stack shows a strong Balmer break ($f^{\nu}(4050\rm{\AA})$/$f^{\nu}(3670\rm{\AA}) = 6.50^{+3.70}_{-1.72}$; see Figs. \ref{fig:main_stack}, \ref{fig:context}), which far exceeds the strongest breaks of $\approx2.5$ seen in quiescent galaxies at similar redshifts \citep[e.g.,][]{Weibel24QG, Carnall24, degraaff24} as well as the maximum theoretical break strength expected for a stellar population ($\approx3$) assuming a standard \citet[][]{Chabrier03} IMF with no dust \citep[e.g.,][]{Wang24evolved, degraaff25}. Such extreme breaks are a key feature of BH* models \citep[e.g.,][]{Liu25BB,Naidu25BHstar,Ji25BT}. In Fig. \ref{fig:context}, we show that the break of the stack is also stronger than all known LRDs, with the two exceptions being sources thought to be almost pure BH*s \citep[][]{Naidu25BHstar,degraaff25}. In our framework this is because the host galaxy is diluting the strength of the BH* Balmer break in the typical LRD in our sample. 
    \item \textit{H$\alpha$ strength and Balmer decrement:} The EW(H$\alpha$) of our stack is $\approx850$\AA\, which is far stronger than the typical line strength observed in local AGN (few 100 \AA, e.g., \citealt{Burke25}) as well as quasars at similar redshifts as our sample \citep[e.g.,][]{Liu25}. Simultaneously, the Balmer decrement is extremely steep ($16\pm2$; Fig. \ref{fig:context}) -- this is highly unlikely to be due to dust owing to the strict constraints on the FIR emission of LRDs \citep[e.g.,][]{Casey25, Setton25, Xiao25}. In BH* models, both the extreme line strengths and Balmer decrement are a natural product of collisional excitation \citep[e.g.,][]{Torralba25IFU} and resonant scattering \citep[e.g.,][]{Chang25} expected from the same conditions that produce the deep Balmer break \citep[e.g.,][]{Yan25}. Also note that the absorption strength \textit{increasing} along the Balmer series (thereby boosting the observed Balmer decrement) is typical for BH*s \citep[e.g.,][]{Naidu25BHstar,deugenio25irony, Torralba25IFU}, analogous to regular stars \citep[e.g.,][]{Gray05} and gas-cocooned Type IIn supernovae \citep[e.g.,][]{Dessart09}. Note that this is distinct from the case of absorbing gas lying completely outside the line formation region, where oscillator strengths would set the absorption depth (i.e., H$\alpha$ would be absorbed more than H$\beta$; e.g., \citealt{Draine11}). 
    \item \textit{Fluorescent, Pumped \ion{Fe}{2} and \ion{O}{1} emission:} We observe several strong \ion{Fe}{2} lines (e.g., UV complexes at $\approx2200-3000$\AA, $\approx9000-9200$\AA, $\approx8200-8500$\AA) and \ion{O}{1} (e.g., $8446$\AA) lines. The [\ion{Fe}{2}] forest at some wavelengths is expected to arise in the outer layers of BH* models \citep[e.g.,][]{Torralba25IFU} and some of the complexes may be powered by Ly$\alpha$ and Ly$\beta$ fluorescence in the strong radiation field and dense medium \citep[e.g.,][]{Sigut98, Kokorev25glimpsed}. Strong \ion{O}{1} is also expected in such an environment as a result of Ly$\beta$ fluorescence \citep[e.g.,][]{degraaff25pop, Tripodi25, Juodzbalis24Rosetta}.
    \item \textit{Blackbody-like optical continuum SED:} In Fig. \ref{fig:blackbody} we show that a simple blackbody with $T_{\rm{eff}}=4047^{+156}_{-155}$ K provides an \textit{empirical} description of the optical continuum at $\lambda_{\rm{rest}}>4000$\AA. Similar SED shapes in the rest-optical have been noted previously in fits to individual LRDs, albeit without this kind of BH* decomposition \citep[e.g.,][]{Lin25,degraaff25pop,Umeda25}. Interestingly, various BH* models predict SEDs peaking in the rest-optical with fall-offs to shorter as well as longer wavelengths consistent with our median stack \citep[e.g.,][]{Naidu25BHstar, Liu25BB, Kido25}. Also note that we are able to fit the spectrum with a simple blackbody versus a modified blackbody \citep[][]{degraaff25pop} whose additional parameters may be compensating for the host contribution that we account for here via subtraction. Finally, we note that the physical interpretation of this empirical SED shape is under development -- it has been interpreted as arising from photosphere-like emission \citep[e.g.,][]{Liu26, Kido25, Umeda25, Torralba26, degraaff25pop} as well as absorbed and reprocessed AGN continuum \citep[e.g.,][]{Sneppen26, Pacucci26, Naidu25BHstar, degraaff25}.
\end{enumerate}

\begin{figure*}
    \centering
    \includegraphics[width=0.9\linewidth]{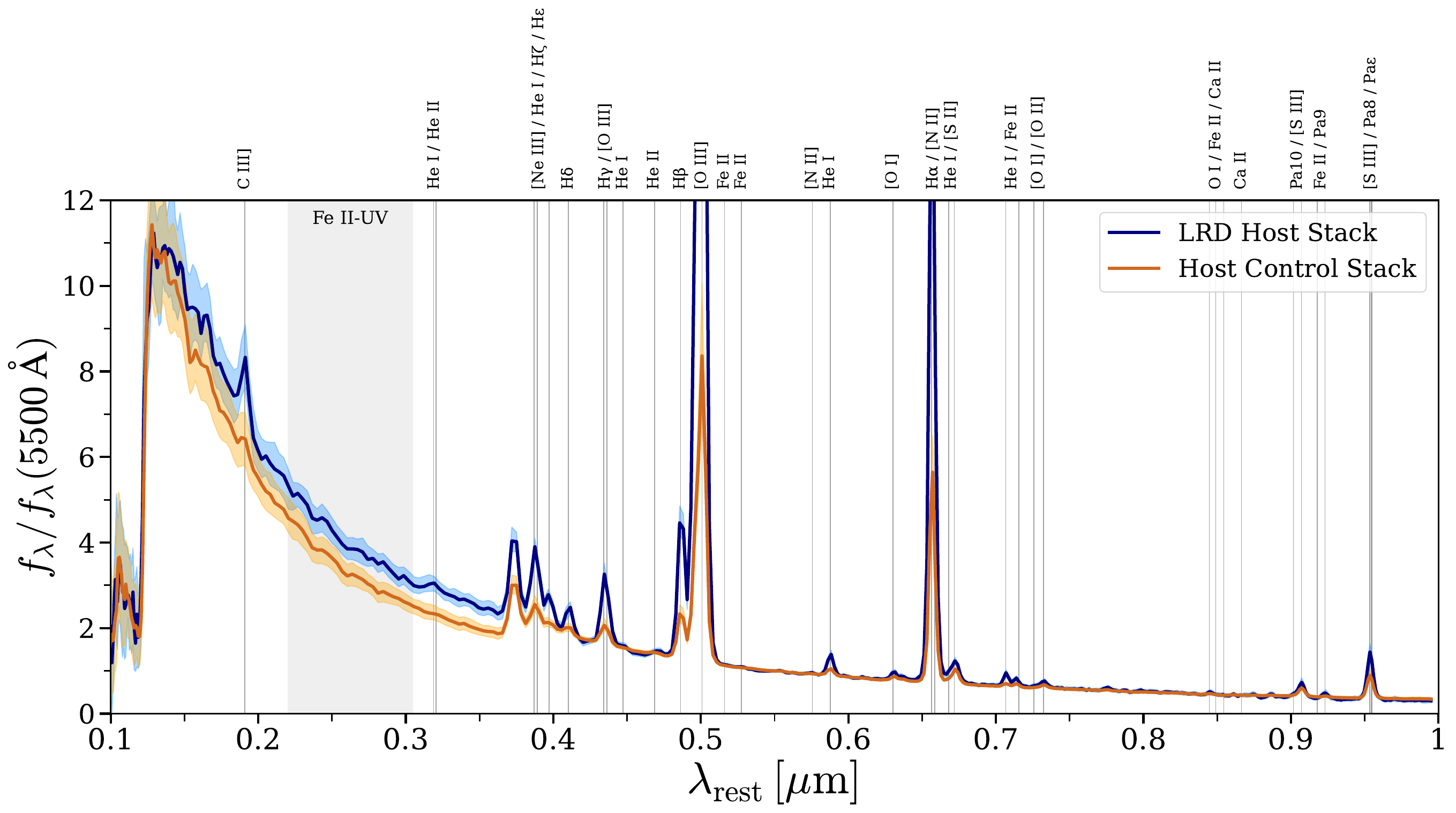}
    \caption{\textbf{Comparison of LRD host stack (blue) with the control stack (yellow).} The galaxies in the control stack match the LRDs' [\ion{O}{3}] luminosities and lie at similar redshift -- the only difference is that they are not required to match the LRDs' SED shapes (i.e., the third condition in \S\ref{sec:implementation}). The LRD hosts display brighter rest-optical lines than the control sample, and show clearly elevated UV features such as \ion{C}{3}] tracing harder ionizing photons and thereby massive stars produced in recent bursts. Another clear difference is in the \ion{He}{1} lines (e.g., 5877\AA, 7067\AA) signaling a denser and hotter ISM in the LRD hosts \citep[e.g.,][]{Berg26}.}
    \label{fig:sfg_control}
\end{figure*}

\begin{figure*}
    \centering \includegraphics[width=0.95\linewidth]{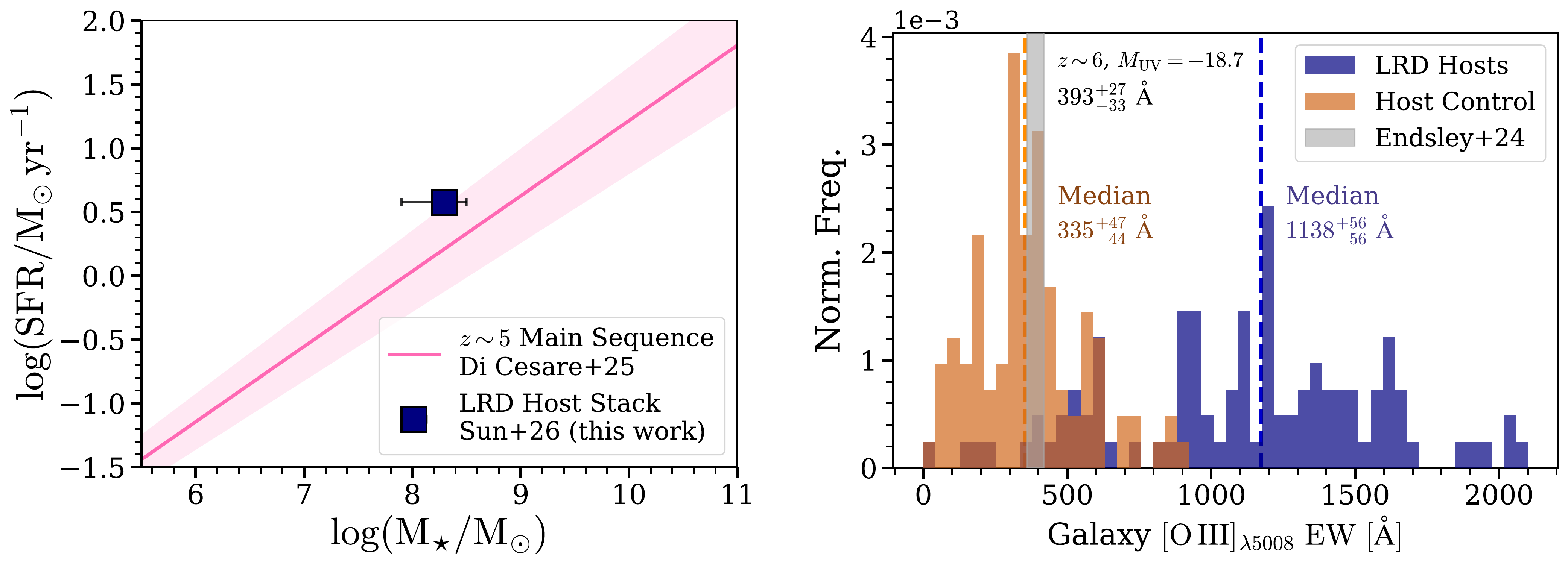}
    \caption{\textbf{The host galaxies of LRDs are starbursts with stronger emission lines than the general population.} \textit{Left:} LRD hosts lie slightly above the star-forming main sequence at the median redshift ($z\approx5$) of our sample. We compare the H$\alpha$-based star-formation rate and \texttt{prospector} SED-fitting-based stellar mass (pink) from \citet[][]{DiCesare25SFMS} with a similarly derived estimate for the LRD hosts (navy blue). \textit{Right:} We compare the [\ion{O}{3}] EW distribution of the LRD hosts to a carefully constructed control sample (\S\ref{sec:host}; dark orange) as well as the population average expected for matched $M_{\rm{UV}}$ and at similar redshift (gray band; \citealt[][]{Endsley24bursty}). LRD hosts display $\approx3\times$ stronger emission lines than both the control stack as well as the population average.}
    \label{fig:host_sfr}
\end{figure*}

\begin{figure*}
    \centering \includegraphics[width=0.9\linewidth]{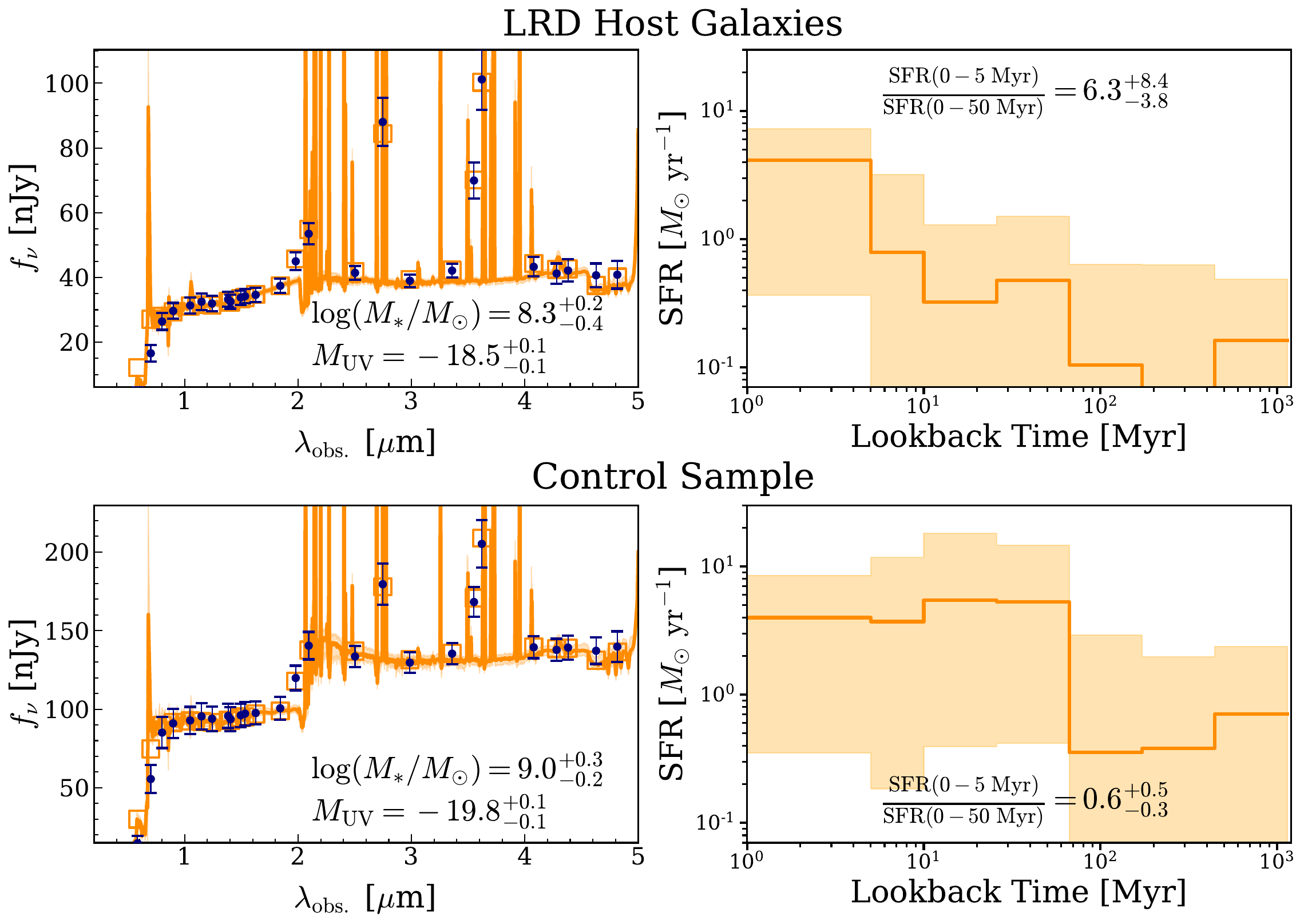}
    \caption{\textbf{LRD host galaxies (top) are observed during a burst whereas a control sample shows a relatively flat recent star-formation history (bottom).} Posterior draws in orange are based on \texttt{prospector} fits to the stacks shown in Fig. \ref{fig:sfg_control}. The stacks are fit in the form of synthesized JWST NIRCam photometry in all medium-band and broad-bands (navy) along with fluxes of strong lines (H$\alpha$, H$\beta$, [\ion{O}{3}]). The recovered SFH of the LRD hosts favors a recent burst such that the SFR has risen by $\approx10\times$ in the last $\approx10$ Myrs, consistent with the $\approx2-3\times$ stronger optical and UV emission lines versus the control sample and typical galaxies at similar redshift (see Fig. \ref{fig:host_sfr}).}
    \label{fig:prospector}
\end{figure*}

Finally, in Fig. \ref{fig:cliff_monster} we compare the median $\text{LRD}-\text{Host}$ stack against LRDs that have been modeled as effectively pure BH*s -- MoM-BH*-1 \citep[][]{Naidu25BHstar} and The Cliff \citep[][]{degraaff25}. The stack bears a striking resemblance to these objects, showing a similar continuum SED shape and sharp Balmer break. The key difference is that these sources are somewhat ``hotter" than our stack, with their SEDs peaking at bluer wavelengths and also resulting in sharper breaks. In the same figure we also show A2744-45924 \citep[][]{Labbe24} for comparison -- while this source is not a pure BH* like the other two objects (BH* fraction at 5500\AA\ of $\approx92\%$), it displays extremely strong emission lines  characteristic of BH*s (e.g., \ion{Fe}{2} UV, \ion{He}{1}, \ion{O}{1}), which line up with several lines in our stack, albeit at different strengths, as this source is much hotter and more luminous than the median stack.

The median stack presented in this section shows the central engines of LRDs display a strikingly uniform SED. This SED matches numerous predictions of BH* models and resembles the two ``pure" BH*s. All these lines of evidence combined lead us to conclude that BH*s are indeed powering LRDs. That is, $\text{LRD} - \text{Host}$  = BH*. Henceforth, we refer to the ``$\text{LRD} - \text{Host}$'' difference spectra as BH* spectra.

{\setlength{\tabcolsep}{0pt}
\begin{deluxetable}{lr}
\tabletypesize{\footnotesize}
\tablewidth{0pt}
\tablecaption{Summary of Median BH* Stack}
\tablehead{
\multicolumn{2}{c}{Empirical Properties}
\label{tab:properties}
}
Sample Size & 98\\
Median Redshift $\langle z_{\rm{spec}}\rangle$ (range) & 4.92 (2.26 to 8.36)\\
Optical Luminosity [$\log(L(5500\rm{\AA})/\rm{erg\ s^{-1}})$] & $43.49^{+0.05}_{-0.06}$ \\
Balmer Break [$f^{\nu}(4050\rm{\AA})$/$f^{\nu}(3670\rm{\AA})$] & $6.50^{+3.70}_{-1.72}$ \\
Balmer Decrement [H$\alpha$/H$\beta$] & $16.2^{+2.2}_{-2.2}$ \\
\\[-1ex]
\hline\vspace{-0.2cm}\\
\multicolumn{2}{c}{Blackbody Fitting (\S\ref{sec:reveal})}\vspace{0.05cm} \\
\hline \vspace{-0.15cm}\\
Effective Temperature [$T_{\rm{eff}}$/K] & $4047^{+156}_{-155}$\\
Bolometric Luminosity [$\log(L_{\rm{bol}}$/erg s$^{-1})$] & $43.9^{+0.1}_{-0.1}$\\
Radius [$R_{\rm{eff}}$/au] & $1321^{+168}_{-153}$\\ \\[-1ex]
\hline\vspace{-0.2cm}\\
\multicolumn{2}{c}{Host Galaxy Properties (\texttt{Prospector} SED fitting; \S\ref{sec:host})}\vspace{0.05cm} \\
\hline \vspace{-0.15cm}\\
UV Luminosity [$M_{\rm{UV}}$] & $-18.5^{+0.1}_{-0.1}$\\
Stellar Mass [log($M_{\rm{*}}/\rm{M}_{\rm{\odot}}$)] & $8.3^{+0.2}_{-0.4}$\\
Star-Formation Rate (${\rm{10\ Myr}}$) [$\rm{M}_{\rm{\odot}}$ yr$^{-1}$] & $3.1^{+1.5}_{-2.6}$
\enddata
\end{deluxetable}

\begin{figure*}
    \centering
    \includegraphics[width=\linewidth]{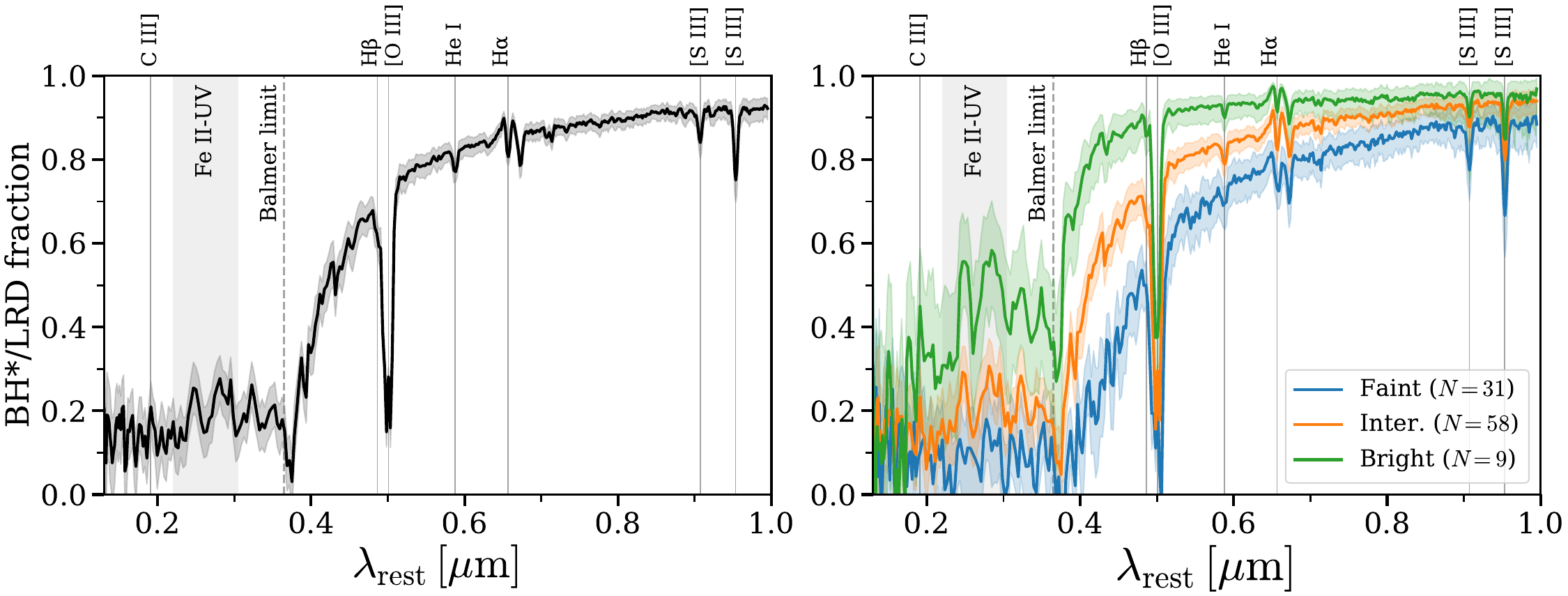}
    \caption{\textbf{Wavelength-dependent BH*/LRD fractions of the BH* stack (left) and three substacks divided by $L_{5500}$ (right).} The faint substack has $42.2 < \log\left(L_{\rm{5500}} / \text{erg}\,\text{s}^{-1}\right) < 43.2$; the intermediate substack has $43.2 < \log\left(L_{\rm{5500}} / \text{erg}\,\text{s}^{-1}\right) < 44.2$; the bright substack has $44.2 < \log\left(L_{\rm{5500}} / \text{erg}\,\text{s}^{-1}\right) < 45.2$. Generally, the BH* provides relatively modest contributions in the rest-UV, but there is a sharp increase around $\approx3600-4000$\AA\ (not necessarily at the Balmer limit, but redwards for the faintest). At redder wavelengths the BH* begins to dominate, accounting for almost all the light around $\approx1\mu$m. There is a strong luminosity dependence to the BH* contribution -- the brightest BH*s (green curve) are also luminous in the UV, contributing $\approx50\%$ around the region where the Fe UV forest is observed (corresponding to the prominent peaks in the UV). On the other hand, fainter BH*s (blue) make their presence known only in the rest-optical and are effectively invisible at bluer wavelengths. The sharp, almost vertical transition in BH* fraction redwards of the Balmer limit is luminosity dependent, and occurs only in the most luminous sources, whereas in fainter BH*s we witness a more gradual transition. Emission lines show expected trends -- forbidden lines such as [\ion{O}{3}] (by construction) and the [\ion{S}{3}] doublet are prominent in hosts relative to the BH*. 
    }
    \label{fig:bhstarfrac}
\end{figure*}

\subsection{Host Galaxy Properties -- LRD Hosts are Dwarf Galaxies Undergoing Starbursts }
\label{sec:host}

In Fig. \ref{fig:sfg_control} and Table \ref{tab:control} we compare the stack of LRD host galaxies against a redshift- and [\ion{O}{3}] luminosity-matched control sample. We generate the control sample by following only the first two steps of the host matching procedure described in \S\ref{sec:implementation} -- i.e., we do not impose any constraints on the SED shape of the host galaxy to match the SED shape of the LRD. The control sources are scaled such that their [\ion{O}{3}] luminosities match the corresponding LRDs before stacking. Again, in a Monte Carlo fashion over 10,000 trials we draw one control sample galaxy per LRD, and bootstrap resample these sources to produce the control stack.

Remarkably, despite being fully matched in $L_{\rm{[OIII]}}$, there are clear differences between the LRD host and control stacks. LRD hosts are fainter ($M_{\rm{UV}}=-18.5$ vs. $-19.8$) and lower-mass ($\log(M_{\rm{\star}}/\rm{M}_{\rm{\odot}})=8.3$ vs. 9.0) galaxies relative to the control sample. Further, in the hosts the rest-optical emission lines (e.g., H$\alpha$, [\ion{O}{3}]) are far stronger ($\approx2-3\times$ in EW) than the control (see Fig. \ref{fig:host_sfr}). Indeed, the rest-optical lines (e.g., [\ion{O}{3}]+H$\beta$) are even stronger than the average galaxy at comparable $M_{\rm{UV}}\approx-18.5$ at $z\approx6$ \citep[e.g.,][]{Endsley24, Endsley24bursty}. Moreover, \ion{C}{3}] emission is apparent in the rest-UV -- these are telling signs of recent starbursts and young ages \citep[e.g.,][]{Roberts-Borsani24}. Further, the \ion{He}{1} lines ($5877$\AA, $7067$\AA) sensitive to the density and temperature of the ISM \citep[e.g.,][]{Berg26} are more prominent among LRD hosts, pointing to a hotter, denser ISM. Taken together these findings tie the emergence of BH*s -- perhaps their formation \citep[e.g.,][]{Begelman25}, or their entrapment in gas \citep[e.g.,][]{Alexander14} -- to recent star-formation and a hot, dense ISM.

Indeed, this is borne out also by the star-formation histories of the stacks that we present in Figure \ref{fig:prospector} using the \texttt{Prospector} SED-fitting framework \citep{Leja17, Leja19, Johnson21}. The LRD host stacks shows a rising star-formation history, with a $\approx10\times$ increase between the epoch of observation and $\approx10-25$ Myrs prior. On the other hand, the control stack has a relatively steady SFH over the same time period.

In detail, to fit these stacks with \texttt{prospector} we follow the modeling choices from \citet[][]{NM24} (in turn adapted from \citealt[][]{Tacchella22,Tacchella23SMACS}). We use \texttt{FSPS} \citep{FSPS1,FSPS2,FSPS3} with the \texttt{MIST} stellar models \citep[][]{Choi17} and \texttt{MILES} spectral library \citep[][]{MILES2011} and assume a \citet{Chabrier03} IMF. Nebular emission is modeled with the \texttt{CLOUDY} \citep{Ferland17} grid introduced in \citet{Byler17}. The free parameters include seven bins describing the non-parametric star-formation history, the total stellar mass, stellar and gas-phase metallicities, nebular emission parameters, and a flexible dust model \citep{KriekConroy13}. We adopt a ``bursty continuity" prior for the star-formation history \citep{Tacchella22} with the time bins logarithmically spaced following \citep[][]{Leja19} up to a formation redshift of $z=20$. We hold the first two bins fixed at lookback times of 0-5 and 5-10 Myrs to capture bursts that power strong emission lines \citep[e.g.,][]{Whitler22} and that are expected to grow ubiquitous with increasing redshift. We synthesize NIRCam photometry from the stacks in all JWST bands and pass the measured emission line fluxes to the model.

We emphasize that our derived host properties hinge on the assumption that all [\ion{O}{3}] flux arises from the host (see S\ref{sec:methods} for motivation). One could posit scenarios wherein the host contribution to [\ion{O}{3}] is not exactly $\approx100\%$. For example, just as UV photons are leaking from the BH* (e.g., \citealt{Labbe24,Torralba25IFU}; Figs. \ref{fig:main_stack}, \ref{fig:bhstarfrac}), ionizing photons that are able to power [\ion{O}{3}] in surrounding gas may also be escaping, thereby boosting the EWs and leading us to overestimate the host galaxy EWs. For all the reasons discussed in S\ref{sec:methods}, we deem this unlikely to be the case in the typical LRD. It is also worth noting that LRDs are inferred to have soft massive star-like ionizing spectra \citep[][]{Wang25missing} and lack extreme ionizing conditions, as evidenced by their lack of \ion{He}{2} emission and that \ion{Fe}{2} and not higher ionization Fe lines are observed \citep[][]{Torralba25IFU, deugenio25irony, Lin25, tang26}.

Current empirical constraints suggest that the actual contribution of [\ion{O}{3}] from the BH* is likely negligible. The ``pure'' BH* archetype MoM-BH*-1 has the ratio $L_{\text{H}\beta\lambda4862\text{\AA}} / L_{\text{[\nion{O}{III}]}\lambda5008\text{\AA}} \gtrsim 15$ \citep[][]{Naidu25BHstar}, with The Cliff showing a similar ratio \citep[][]{Matthee26}. Applying this ratio to our LRD sample (Table \ref{tab:lrdlines}) -- and conservatively assuming that the totality of H$\beta$ flux arises from the BH* yields a predicted BH* contribution of $L_{\text{[\nion{O}{III}]}\lambda5008\text{\AA}} < 10^{40.7}$ erg s$^{-1}$, which is almost two orders of magnitude fainter than the observed [\ion{O}{3}] in our LRD sample, suggesting that [\ion{O}{3}] remains a robust tracer of the host. While we deem a significant [\ion{O}{3}] contribution from the BH* unlikely, we cannot completely rule out that there is absolutely no intrinsic [\ion{O}{3}] component powered by accretion that varies with other properties of the LRD, which may contribute in part to this apparent large difference in EWs between the control sample and LRD hosts. We therefore explore an extreme case for what BH*-driven contamination in [\ion{O}{3}] might mean for our results. 

We re-run our stacking procedure assuming a $0 - 100\%$ [\ion{O}{3}] contribution arising from the BH*. Note that the host-matching step ($L_{\rm{[OIII]}}$ within $3\times$) is relatively flexible and resistant to any degree of contamination, and so the difference in results is from the SED scaling step (\S\ref{sec:implementation}). We find that even when the BH* contributes $50\%$ of [\ion{O}{3}], the recovered BH* Balmer break strength is still stronger than the maximum expected from a standard stellar population assuming a dust-free \citet[][]{Chabrier03} IMF.

\subsection{Wavelength-Dependent BH*/LRD Fraction}
\label{sec:bhstarfrac}

A signature feature of LRDs is the remarkable diversity across the population, as well as their strikingly different appearance as a function of wavelength. On average, LRDs show extended morphologies in the rest-UV and are completely dominated by point-sources at the H$\alpha$ wavelength \citep[e.g.,][]{Matthee24,Akins25, Zhang25, Kokorev25glimpsed, Juodzbalis24Rosetta}. This transition in morphology coincides with a transition from the blue UV continuum to the characteristic red optical continuum . There is also perplexing diversity, with some sources remaining unresolved point-sources with BH* signatures (e.g., the \ion{Fe}{2} UV complex) far into the UV \citep[e.g.,][]{Furtak24, Labbe24,Torralba25IFU, Torralba25LyA, Tang25, Akins25restUVlines}.

In Fig. \ref{fig:bhstarfrac} we are able to explain these observations in terms of variations in the BH*/LRD fraction with wavelength. In the median LRD, the BH* fraction remains modest (but non-zero) in the rest-UV ($\approx20\%$). There is an abrupt transition at $\approx3600-4000$\AA, redwards of which the BH* begins to dominate the total light. By $\approx1\mu$m almost the entire SED ($>90\%$) is explained by the BH*.

Clear minima appear at the positions of forbidden lines that cannot be produced in the dense BH*: [\ion{O}{3}] by construction, and at [\ion{S}{3}] as an outcome of our procedure. Intriguingly, the BH*/LRD fraction is relatively flat at the H$\alpha$ wavelength -- the broad H$\alpha$ emission of the BH* is balanced by similarly strong narrow H$\alpha$ emission from the hosts  ($\approx830$\AA vs. $\approx940$\AA; Tables \ref{tab:mainstacklines}, \ref{tab:control}). Note that we lack the resolution to discern narrow and broad components in the Balmer lines, or else we might expect a sharp minima at the location of the narrow line of the host.  Strikingly, the FeII-UV feature is seen in the \textit{median} stack of all sources -- this feature is extremely prominent in the most luminous sources (right panel, Fig. \ref{fig:bhstarfrac}) and has previously been reported in such objects \citep[e.g.,][]{Labbe24, Torralba25IFU}, but here we see this is a feature occurring in the typical LRD.

\begin{figure*}
    \centering
    \includegraphics[width=\linewidth]{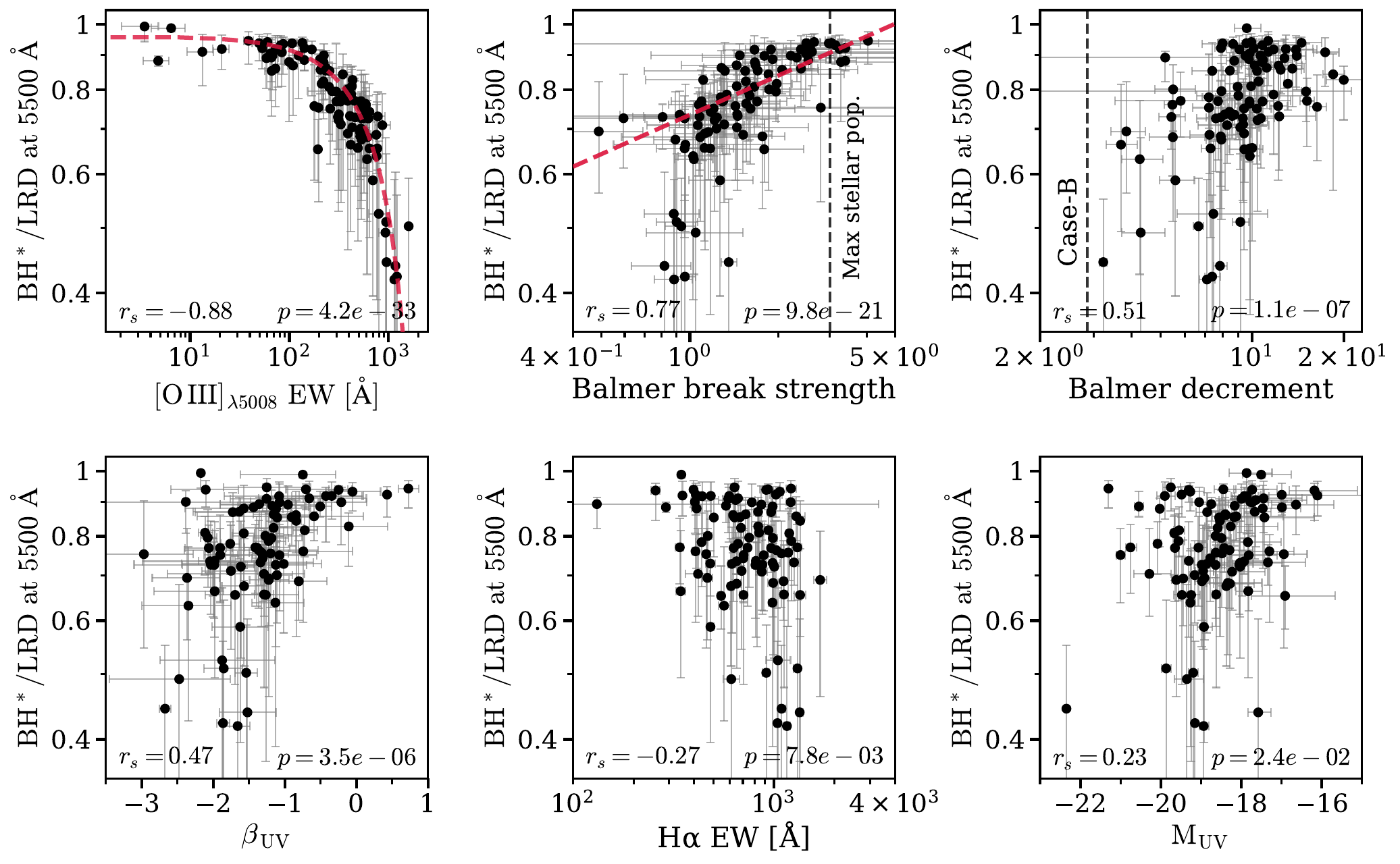}
    \caption{\textbf{Correlations of median BH*/LRD fraction at $5500\rm{\AA}$ (rest-frame) for each LRD with LRD observables.} Spearman's rank correlation coefficient $r_s$ and its associated $p$-value are provided for each LRD observable. The key correlations we notice for the BH*/LRD fraction are with LRD [\ion{O}{3}] EW and Balmer break strength, for which we provide fits in Eq. ~\ref{eq:bhstar_frac_vs_lrd_o3_ew} - ~\ref{eq:bhstar_frac_vs_lrd_bb}. These strong correlations are likely because the continuum of the BH* overwhelms the host, thereby diluting the [\ion{O}{3}] and producing strong Balmer break. There are also weak trends with Balmer decrement (consistent with the scenario where pure BH*s with extreme decrements of $> 10$ are diluted by low-mass galaxies with close-to-Case-B decrements of $\approx 3.8$) and $\beta_{\text{UV}}$ (corresponding to the fact that typical hosts have blue UV slopes whereas BH*s have red UV slopes).}
\label{fig:bhstar_frac_correlation}
\end{figure*}

To study the luminosity dependence of the BH*/LRD fraction, we construct substacks following the exact same procedure as the main stack (\S\ref{sec:methods}), but by splitting the sample in three bins by the optical luminosity, $L_{\rm{5500}}$ (Fig. \ref{fig:bhstarfrac}). Such substacks will be analyzed in detail in Naidu \& Sun et al. in prep., while here we focus on the strong luminosity dependence in the BH*/LRD fraction across these substacks (Fig. \ref{fig:bhstar_frac_correlation}). In particular, the most luminous sources tend to remain bright into the UV, contributing almost $\approx50\%$ of the light at $\approx2000-3000$\AA\ as seen via their prominent UV \ion{Fe}{2} emission. On the other hand, fainter BH*s tend to ramp up gradually with wavelength and only begin to dominate the SED past $\approx5000$\AA. Nonetheless, by $\approx1\mu$m, irrespective of how luminous an LRD is, almost all light ($\approx90\%$) arises from the BH*. 

The transition wavelength, where we witness an abrupt change in BH* fraction is interestingly not exactly at the Balmer limit \citep[e.g.,][]{Setton24}. The sharp, almost vertical edge around the Balmer limit is seen most prominently among the most luminous sources -- the sharpness is akin to the almost vertical Lyman-$\alpha$ break seen in high-redshift galaxies. Fainter sources instead have a more gradual transition and reach a $\approx50\%$ BH* fraction at redder wavelengths. In a companion paper (Naidu \& Sun+26), we show that this may be understood in terms of a correlation between the luminosity and effective temperature of BH*s (akin to the main-sequence of stars). Luminous, hot BH*s emit strongly at bluer wavelengths whereas faint, cold BH*s have SEDs that peak at redder wavelengths, with little flux around H$_{\infty}$ (i.e., little continuum to ``break" at the Balmer limit) and therefore show more gradual transitions.

\begin{figure*}
    \centering
    \includegraphics[width=\linewidth]
    {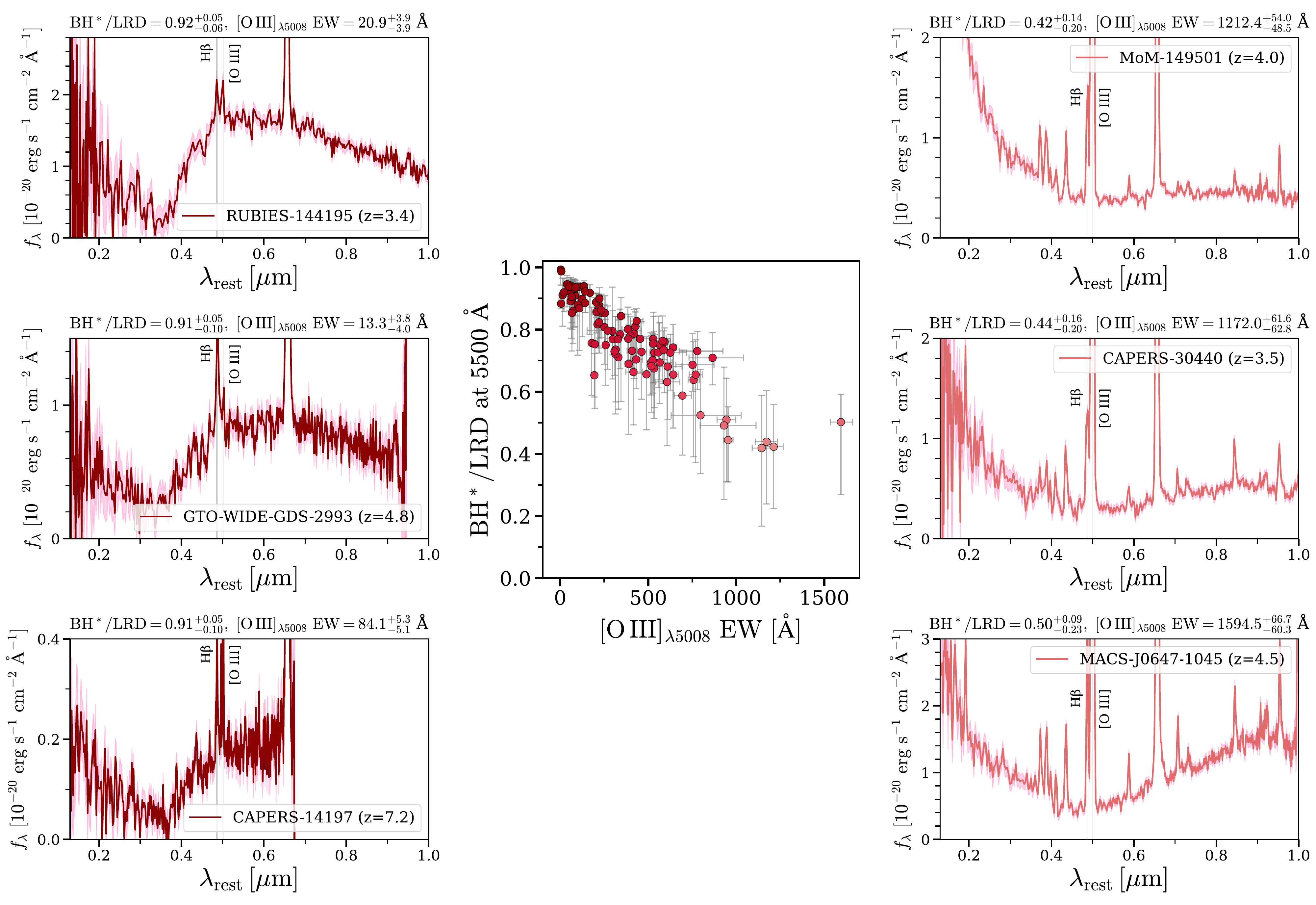}
    \caption{\textbf{Extreme sources in the BH*/LRD fraction vs. LRD [\ion{O}{3}] EW$_{\rm{0}}$ space.} On the high BH* fraction (low LRD [\ion{O}{3}] EW$_{\rm{0}}$) end -- sources useful for studying ``pure" BH*s with minimal host infilling -- we show objects with BH* fractions $> 90\%$, including RUBIES-144195, GTO-WIDE-GDS-2993, and CAPERS-14197. On the low BH* fraction (high LRD [\ion{O}{3}] EW$_{\rm{0}}$) end -- some of the most extreme starbursts in the JWST archive -- we show objects with BH* fraction $\leq 50\%$ whose spectra are heavily diluted by host-galaxy light, including MoM-149501, CAPERS-30440, and MACS-J0647-1045. Intriguingly, we find no LRDs with BH* fraction $<40\%$, with all but a handful of sources lying at $>60\%$ (discussed further in \S\ref{sec:vshaped}, Fig. \ref{fig:vshaped}).}
    \label{fig:bhstar_frac_vs_o3}
\end{figure*}

These trends are independently supported by morphological evidence -- e.g., the most luminous LRDs often display point-source morphology even in the UV \citep[e.g.,][]{Torralba25LyA}. Indeed, \citet[][]{Zhang25} recently performed morphological decomposition of 4-band photometry for a stack of 217 LRDs based on COSMOS-Web \citep[][]{Casey23} and found a BH* fraction of $\approx80-90\%$ in the rest-optical and $\approx60\%$ in the rest-UV, in excellent agreement with our luminous stack that is likely the best subset to compare against an LRD sample selected from relatively shallow wide-area imaging (Fig. \ref{fig:bhstarfrac}). Further, modeling of the rest-optical spectra with modified blackbodies \citep[][]{degraaff25pop} or by using The Cliff as an empirical template \citep[][]{Barro25} yields consistent results -- e.g., the strongest Balmer breaks (tracking high BH* fractions; see Fig. \ref{fig:bhstar_frac_vs_o3}) are seen in sources that peak at the bluest wavelengths (i.e., the hottest and brightest sources).

The stellar mass of the LRD hosts for our median stack is $\log(M_{\rm{\star}}/ {\rm M}_{\rm{\odot}})=8.3^{+0.2}_{-0.4}$, in excellent agreement with independent estimates from clustering that span $\log(M_{\rm{\star}}/ {\rm M}_{\rm{\odot}})=7.5-8.5$ \citep[e.g.,][]{Matthee25LRDclustering, Lin25clustering}. Splitting the BH*s by luminosity (as in Fig. \ref{fig:bhstarfrac}), we find more luminous BH*s reside in more massive hosts. In particular, BH*s with $\log(L_{\rm{5500}}/\rm{erg\ s^{-1}}) =$ [$<43.2$, $43.2-44.2$, $>44.2$] reside in galaxies with $\log(M_{\rm{\star}}/ {\rm M}_{\rm{\odot}}) =$ [$7.5^{+0.3}_{-0.2}$, $8.0^{+0.3}_{-0.3}$, $8.7^{+0.5}_{-0.5}$]. This is again supported by clustering studies, with the fainter, lensed LRDs from \citet[][]{Matthee25LRDclustering} residing in $\approx10^{7.5}\,{\rm M}_{\rm{\odot}}$ galaxies with the brighter blank field LRDs from \citet[][]{Lin25} inhabiting $\approx 10^{8.5}\,{\rm M}_{\rm{\odot}}$ systems.

\subsection{BH* Fractions from Simple Observables}
\label{sec:predictbhstarfrac}

In Fig. \ref{fig:bhstar_frac_correlation} we explore whether the fraction of light arising from the BH* (at 5500\AA) correlates with easily accessible observables in LRDs, even when full PRISM coverage is unavailable (e.g., grism-selected line emitters  \citealt{NM24}; medium-band excesses \citealt{Williams23}). This may help accelerate the search for BH*-dominated objects so that we may study the physics of BH*s with minimal interference from host light. For instance, while studying absorption features, one would like to target sources with the highest BH* fractions to avoid host infilling.

The key correlations we notice for the BH* fraction are with the LRD [\ion{O}{3}] EW and LRD Balmer break strength consistent with the the semi-empirical models in \citet[][]{degraaff25pop,Barro25}. Indeed, the few known ``pure" BH*s show extremely weak [\ion{O}{3}] EWs alongside deep Balmer breaks \citep[e.g.,][]{Naidu25BHstar, degraaff25}. Fundamentally, these relationships are likely because the continuum of the BH* overwhelms the host and thereby dilutes the [\ion{O}{3}] and produces extreme Balmer breaks. For [\ion{O}{3}], alternatively, some of the brightest BH*s may emerge in extremely metal-poor hosts where the line is intrinsically weak \citep[e.g.,][]{Maiolino25}. Note that since our [\ion{O}{3}]-matching procedure relies on luminosity and not EWs, the correlation we find is not (entirely) by construction as the relative continuum levels that set the BH*/LRD fraction at $5500$\AA\ are an outcome of our procedure.

We quantify trends against the BH*/LRD fraction by reporting their Spearman's rank correlation coefficient and its associated $p$-value in Fig. \ref{fig:bhstar_frac_correlation}. The best fit curves to the two most significant relations are as follows:
\begin{equation}
\begin{split}
    \frac{\text{BH*}}{\text{LRD}}\bigg|_{5500\text{\AA}}
    &= \left[(-4.40 \pm 0.25)\right] \left(\frac{\text{LRD [\ion{O}{3}] EW$_{\rm{0}}$ }}{ 10^{4}\text{\AA}}\right) \\
    &\quad + \left[(9.60 \pm 0.06) \times 10^{-1}\right] \, ,
\end{split}
\label{eq:bhstar_frac_vs_lrd_o3_ew}
\end{equation}
\begin{equation}
    \frac{\text{BH*}}{\text{LRD}}\bigg|_{5500\text{\AA}} = 10^{-0.13 \pm 0.01} \, (\text{LRD Balmer Break})^{0.19 \pm 0.02} \, .
\label{eq:bhstar_frac_vs_lrd_bb}
\end{equation}

We notice a weak trend with Balmer decrement, consistent with the scenario where pure BH*s with extreme decrements of $> 10$ are diluted by low-mass host galaxies with close-to-Case-B decrements of $\approx 3.8$. There is also a weak trend with $\beta_{\rm{UV}}$, which is reflective of the fact that typical hosts have blue UV slopes whereas BH*s have red UV slopes, and so this provides some handle on the BH* fraction, particularly among the reddest slopes. It is also noteworthy that H$\alpha$ does not trace the BH* fraction. The BH* and the host galaxy contribute comparable H$\alpha$ line strengths ($830 \rm{\AA}$ and $940 \rm{\AA}$; Tables \ref{tab:mainstacklines}, \ref{tab:control}), so the LRD EW(H$\alpha$) remains high across all BH*/LRD fractions and is insensitive to the mixing ratio. Similarly, $M_{\rm{UV}}$ may be understood as largely reflecting the host mass and star-formation history, and is therefore not an excellent diagnostic.

\subsection{Remarkable Individual Sources}
\label{sec:remarkable}

It is interesting to consider outlier sources in the BH*/LRD fraction. In our sample only two sources have BH* fractions of $>90\%$ at $>1\sigma$ confidence -- MoM-BH*-1 and The Cliff. Sixteen other sources have median BH* fractions $>90\%$, but with much wider posteriors. Of these, we consider those sources secure that are also outliers in the empirical metrics discussed in the previous section (Balmer break strength, EW([\ion{O}{3}])).

In terms of Balmer breaks, there are five objects with breaks stronger than the stellar population maximum for a \citet[][]{Chabrier03} IMF ($\gtrsim3$; \citealt{Wang24evolved}) and with BH* fraction $>90\%$. These objects are some of the most well-studied LRDs including: The Cliff \citep[][]{degraaff25},  MoM-BH*-1 \citep[][]{Naidu25BHstar}, A2744-45924 \citep[][]{Labbe24}, and A2744-QSO1 \citep[][]{Furtak24}. A source that may belong with these objects but is yet to be followed up in detail is CAPERS-EGS-14197 (see Fig. \ref{fig:bhstar_frac_vs_o3}) lying at $z=7.15$ with a strong Balmer break ($5.3^{+6.0}_{-2.1}$), a BH* fraction of $\approx91^{+5}_{-10}\%$, and a very weak [\ion{O}{3}] EW of 84$^{+5}_{-5}$\AA.

Next, we select $>90\%$ BH* fraction sources with the lowest [\ion{O}{3}] EWs. Two such objects are shown in Fig. \ref{fig:bhstar_frac_vs_o3} -- RUBIES-144195 ($z=3.35$, EW=$21^{+4}_{-4}$\AA) and GTO-WIDE-GDS-2993 ($z=4.81$, EW=$13^{+4}_{-4}$\AA). The only LRDs with lower EWs and higher BH* fractions are MoM-BH*-1 and The Cliff, marking these sources as promising targets for high-resolution spectroscopy.

We now turn to sources with the lowest BH* fractions. Some of these objects have exceptional host galaxies, with remarkably strong [\ion{O}{3}] EWs. Accounting for the rest-optical continuum from the BH*, the EWs of lines from the host galaxy must be even higher -- [\ion{O}{3}]5008\AA\ EWs of $\approx2500-3500$\AA. These objects rank among the strongest line emitters in the entire JWST archive \citep[e.g.,][]{Endsley24bursty}. Sources in this category we highlight here include MoM-149501, CAPERS-30440, and MACS-J0647-1045 \citep[][]{Killi23}. These objects represent the extremes of star-formation. While the Balmer lines are still difficult to interpret given the BH* contribution, even from the [\ion{O}{3}] lines we estimate a $\xi_{\rm{ion}}\approx10^{26}\ \rm{Hz}\ erg^{-1}$ \citep[e.g.,][]{Chevallard18} and likely a significant Lyman continuum escape fraction, $f_{\rm{esc}}>20\%$, given the synchrony of ionizing photon production and ionizing photon escape \citep[e.g.,][]{NM22}. We speculate that these objects might represent the later stages of BH* evolution, wherein star formation picks up and stellar mass assembly in the host begins to dominate the emerging emission.

\section{Discussion}
\label{sec:discussion}

\subsection{LRD = Host + BH*: A Simple Composite Model for LRDs}
\label{sec:unification}

In this work we have shown the picture of $\text{LRD} - \text{Host}$  = BH* holds not only for the most extreme sources with large Balmer breaks, but is a defining characteristic for the entire LRD population. The full diversity of LRD SEDs may be described by the diversity of BH*s themselves in various stages of mass assembly (e.g., with varying temperatures, luminosities, surface gravities) combined with the diversity of their host galaxies (e.g., with different masses, star-formation histories, dust content). 

The key insight is that BH*s may be thought of as star-like objects lying at cosmological distances, shining with luminosities of $\approx10^{10} L_{\rm{\odot}}$ and $\approx4000$ K temperatures \citep[e.g.,][]{Begelman08, Begelman25, degraaff25pop, Umeda25, Kido25, Barro25}. Making this connection and appealing to this new astrophysical building block clarifies many of the mysteries of the LRD SED. As seen in radiative transfer models of BH*s, accretion energy is reprocessed into an SED peaking in the rest-optical and falling off towards both longer and shorter wavelengths (Fig. \ref{fig:blackbody}; e.g., \citealt{ Liu25,Naidu25BHstar}), with strong absorption below the Balmer break acting akin to a Lyman break (e.g., \citealt{IM25}). The transition across wavelengths between optical-dominated BH* and UV-dominated galaxy (Fig. \ref{fig:bhstarfrac}) produces the V-shape continuum \citep[e.g.,][]{Labbe23} as well as the sharp contrast between the optical and UV morphology \citep[e.g.,][]{Zhang25}. 

A key open question is the origin of the UV \citep[e.g.,][]{Inayoshi25coevol, degraaff25pop, Asada26}. Our results demonstrate that even in the typical LRD (Fig. \ref{fig:main_stack}), and particularly in the most luminous sources (Fig. \ref{fig:bhstarfrac}; see FRESCO-GN-9771 in Appendix Fig. \ref{fig:appendix_gallery}), there is significant UV emission that cannot be attributed to the host galaxy (e.g., the prominent \ion{Fe}{2} UV in these stacks; \citealt{Labbe24, Torralba25IFU, deugenio25irony}). Where does this UV light come from? One possibility is that this is light leaking from the BH* itself where $n=2$ Hydrogen is unable to block the light -- indeed, \citet[][]{Torralba25IFU} show such a model drawn from the family of \texttt{CLOUDY} models produced in \citet[][]{Naidu25BHstar} to explain the SED shape of FRESCO-GN-9771 (see also \citealt{deugenio25irony}). A caveat is that this region of the predicted \texttt{CLOUDY} spectrum is sensitive to details of the modeling of the Fe atom. Another possibility \citep[e.g.,][]{Inayoshi25coevol, Asada26} is that a compact cluster immediately surrounding the BH* may be responsible for this UV -- however, note that our subtraction method is agnostic to the source (e.g., clusters) of the [\ion{O}{3}] (i.e., we find peer galaxies as long as there are such sources in the DJA), and so for excess UV to remain in the host-subtracted stack we would need these stars to produce a peculiar spectrum (absent in [\ion{O}{3}], and with a unique shape that peaks around the \ion{Fe}{2} wavelength). As challenging as it is, deep H-grating spectroscopy below the Balmer break is the most direct path to confirm if this is indeed BH* light (with e.g., line widths commensurate to the optical) in these sources (e.g., JWST-GO-8047; PI: B. Wang).

\subsection{V-Shaped LRDs may be the tip of the iceberg}
\label{sec:vshaped}

\begin{figure*}
    \centering \includegraphics[width=0.90\linewidth]{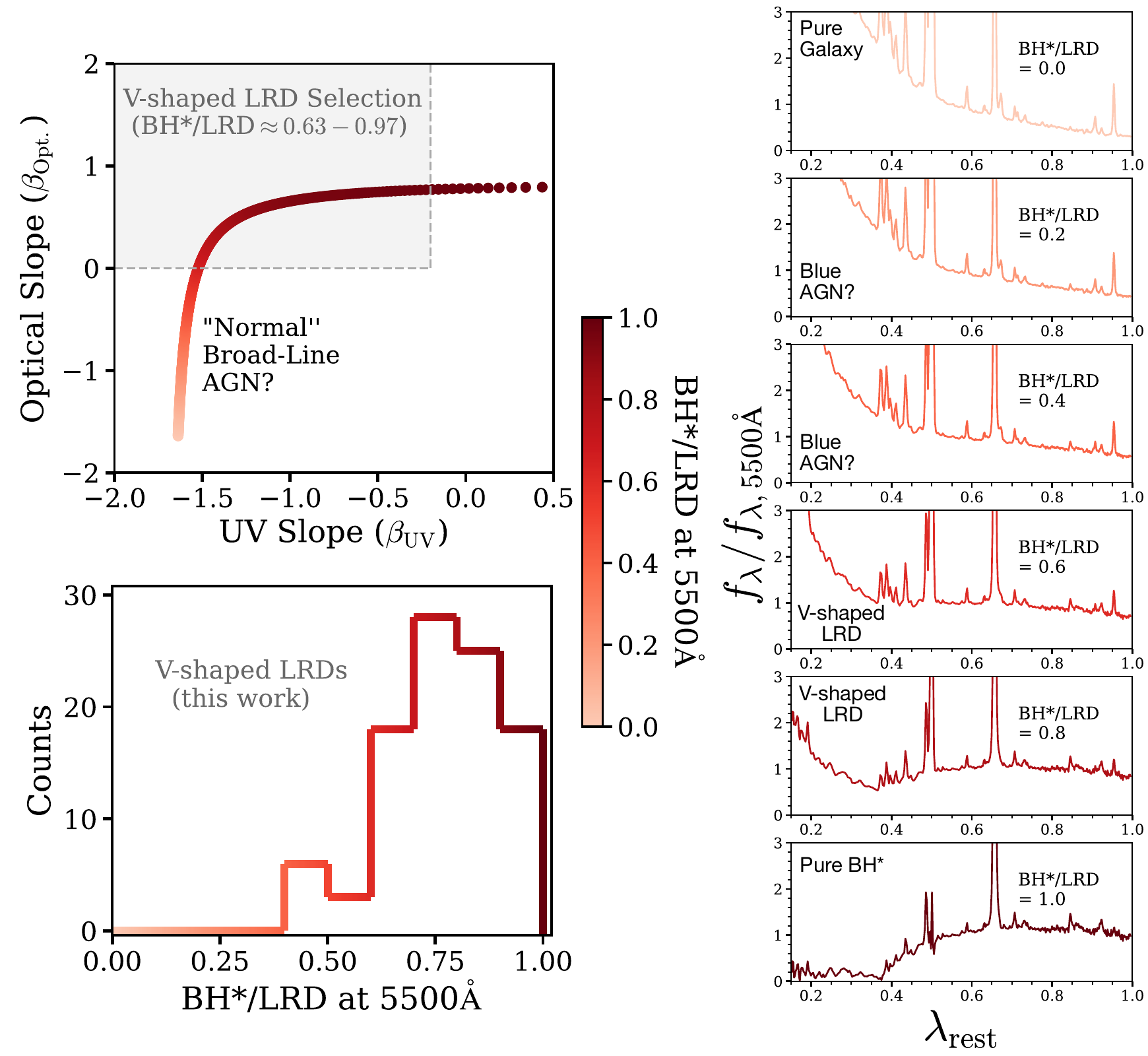}
    \caption{\textbf{BH*s may not only be powering V-shaped LRDs, but a larger population of broad-line AGN.} \textbf{Top-Left:} By combining the median BH* stack and host stacks with varying ratios, we demonstrate that V-shaped selections depicted in shaded gray \citep[e.g.,][]{Labbe23,Kokorev24,Hviding25} are preferentially selecting sources with high BH* fractions of $\gtrsim60\%$. \textbf{Bottom-Left:} Distribution of BH*/LRD fractions in the rest-optical for the sample studied in the work. Notice the steep drop-off at $\lesssim60\%$ as suggested by the exercise in the top-left panel. The handful sources at lower BH* fractions are in particularly low-mass hosts or have very cold BH*s (see Fig. \ref{fig:bhstar_frac_vs_o3}). 
    \textbf{Right:} Sequence of BH* + host galaxy combinations ranging from $0\%$ BH* (top) to $100\%$ BH* (bottom). The characteristic V-shaped SED appears only among sources where the BH* dominates the rest-optical ($\gtrsim60\%$). Sources with relatively high BH* fractions ($\approx20-40\%$) show broad H$\alpha$ but do not satisfy LRD selection criteria. Therefore, JWST's broad-line AGN samples -- which share many of the LRDs' puzzling properties such as X-ray weakness \citep[e.g.,][]{Maiolino24Chandra, Yue24, Ananna24} -- may also be hosting BH*s.}
    \label{fig:vshaped}
\end{figure*}

What is the relationship between LRDs, BH*s and the overall population of bluer, broad-line sources that JWST is revealing at high-$z$? Comparing number densities, V-shaped LRDs comprise $\approx10-50\%$ of these samples depending on the luminosity studied \citep[e.g.,][]{Maiolino23, Matthee24, Taylor24, Kokorev24, Hviding25}. However, it is noteworthy that several perplexing facets of LRDs -- e.g., X-ray faintness, MIR faintness, lack of outflows, lack of optical variability, strong Nitrogen emission -- are seen in broad-line samples \textit{where no LRD selection is applied} \citep[e.g.,][]{Kokubo25, Maiolino24Chandra, Yue24, Isobe25}. Furthermore, these non-LRD broad-line sources fall on the same tight correlation ($<0.2$ dex scatter) between $L_{\rm{5100}}$ and $L_{\rm{H\alpha}}$ followed by LRDs \citep[][]{degraaff25pop}. Might some of these broad-line sources also be hosting BH*s? 

From our modeling, we infer a sharp drop-off in sources with BH*/LRD fractions (measured at $\approx5500$\AA) below $\approx60\%$ (evident in Figs. \ref{fig:bhstar_frac_correlation}, \ref{fig:bhstar_frac_vs_o3}). In particular, $\approx90\%$ of the sources in our sample are dominated by BH*s in the rest-optical with $>60\%$ BH*/LRD ratios. The lowest ratios we recover are $\approx40\%$, occurring in a handful of sources (examples shown in Fig. \ref{fig:bhstar_frac_vs_o3}). 

It is possible that this is a selection effect -- our sources are selected to be V-shaped (via optical and UV slopes) and point-sources ($>50\%$ from a point-source at 4$\mu$m) \citep[][]{Hviding25, degraaff25pop}, as is standard in LRD definitions \citep[e.g.,][]{Labbe23, Kokorev24, Akins24}. However, in our composite picture, there is no physical significance for the V-shape, which is merely a result of two components with different SED slopes being combined. 

To quantify the effect of V-shape selections, we perform a simple experiment in Fig. \ref{fig:vshaped} where we take our median BH* stack and host stack and vary their BH*/LRD fraction in the rest-optical. We find that the standard V-shaped selections are preferentially selecting sources where the BH* is outshining the host in the rest-optical. BH*s that are embedded in relatively brighter galaxies are currently not being classified as LRDs, and are being selected as ``normal" broad-line AGN with blue UV and optical slopes or being completely missed. The selection bias may be particularly acute for redder BH*s that peak redwards of $5500\rm{\AA}$; in such cases, the host galaxy dominates the continuum at the typical continuum turnover, which potentially causes these redder BH*s to be missed from LRD selections. Note that since BH*s are not outshining the galaxy in the rest-optical, these sources may not have point-source morphologies -- indeed, this is often the case among broad-line samples \citep[e.g.,][]{Harikane23, Maiolino23, Fei25glimpsed}. Recently, \citet[][]{Yanagisawa26} presented a striking validation of this picture with the discovery of two LRDs embedded in a strongly lensed galaxy -- if not for the spatial resolution afforded by lensing, these systems would be completely outshone by their host galaxy.

A key implication is that the``overmassive" black holes being reported among the non-LRD population may also have overestimated masses and bolometric luminosities (though note some sources may resemble more ``standard" accretion disks; e.g., \citealt{Fabian25, Maiolino24GNz11}). Another implication of this exercise is that the already abundant BH* population may be even more widespread than LRD numbers imply \citep[e.g.,][]{Barro25}. What the exact fraction of BH*s among broad-line samples is remains to be seen, and will require developing sensitive proxies to distinguish between BH*-powered AGN from AGN truly resembling $z=0$ samples (Fig. \ref{fig:context}).

Another important implication is that the dramatic redshift evolution in the number density of V-shaped LRDs --  $\approx10\times$ between $z\approx6$ to $z\approx2$ \citep[e.g.,][]{Ma25countingLRDs, Kocevski24} -- may not be as straightforward to interpret as BH*s being a predominantly early Universe phenomenon \citep[e.g.,][]{Inayoshi25coevol}. It is possible that BH*s may be forming at a relatively steady rate, but the fraction of host galaxies they can outshine to produce a V-shaped spectrum falls with cosmic time, reflecting the stellar mass build up of the Universe (e.g., a $\approx10\times$ increase in the cosmic star-formation mass density at $\gtrsim10^{8} M_{\rm{}\odot}$ between $z=6$ to $z=2$; \citealt{Weibel24}).

For now, we note that the fraction of BH*s among non-LRDs, particularly broad-line sources at high-redshift, is highly likely to be non-zero \citetext{see \citealp{brazzini26, madau26, Matthee26} for similar discussion}. As per our composite picture, there is no physical reason per se for a $\approx60\%$ floor in the BH* fraction, and for BH*s to occur only in V-shaped sources \citep[e.g.,][]{Yanagisawa26}. Some models envision a tight coupling of the fates of BH*s and the stars surrounding them where strong correlations may be expected between the host and BH* components -- but even in such models the optical to UV ratio (proxy for BH* to host ratio) is expected to vary widely \citep[e.g.,][]{Inayoshi25coevol, Zwick25}.

\subsection{The Lifetime and Duty Cycle of BH*s}
\label{sec:duty}

A key open question about BH*s is how long-lived they are, which in turn has implications for how important of a phase they represent in the origin stories of massive black holes and in shaping the course of galaxy formation. The duty cycle and lifetime of BH*s can be derived similar to quasar duty cycles \citep[e.g.,][]{Small92,Martini01,Eilers24} by assuming every galaxy of the characteristic host mass goes through an LRD phase:
\begin{equation}
    \rm{Duty\ Cycle} = \frac{\rm{Lifetime\ of\ BH^*s}}{\rm{Hubble\ Time}} \simeq \frac{\phi(\rm{LRDs)}}{\rm{\phi(Hosts)}}.
\end{equation}

Our unique constraints on the typical host stellar mass ($\approx10^{8}\,{\rm M}_{\rm{\odot}}$) allow us to solve for the lifetime. For the median redshift of our sample ($z\approx5$), $\approx10^{8}\,{\rm M}_{\rm{\odot}}$ galaxies occur at a number density of $\phi/\rm{Mpc^{-3}}\approx 10^{-2}$ \citep[e.g.,][]{Weibel24}. On the other hand, $z\approx5$ LRDs with an $M_{\rm{UV}}\approx-18.5$ (the median for our stack) are found at a rate of $\phi/\rm{Mpc^{-3}}\approx 10^{-4}$ \citep[e.g.,][]{Kokorev24}. This yields a duty cycle of $\approx1\%$ and a BH* lifetime of $\approx10$ Myrs, consistent with constraints from clustering \citep[e.g.,][]{Matthee25LRDclustering, Lin25clustering, Pizzati25}. A caveat is that as shown in \S\ref{sec:vshaped}, V-shaped LRDs likely provide an incomplete census of BH*s -- in particular, other broad-line sources may also be powered by BH*s which would mean LRD luminosity functions (LFs) are under-counting BH*s possibly by $\approx2\times$ (comparing broad-line non-LRD vs. LRD numbers at similar luminosities as our sample; \citealt{Hviding25}), which would imply a duty cycle of $\approx2\%$ and lifetimes of $\approx20$ Myrs.

Excitingly, BH* models predict similarly short lifetimes (e.g., $\approx20$ Myrs in \citealt[][]{Santarelli25}) consistent with our results. It is also interesting to note that models of star-formation burstiness that explain the scatter in the main sequence report excess variability on $\approx10$ Myr timescales relative to 50 Myr timescales \citep[e.g.,][]{Simmonds25,Cole25, Munoz26} that may point to efficient, recurrent fueling of their centers via e.g., compaction events and compact starbursts that may produce conditions ripe for BH* formation or fueling \citep[e.g.,][]{Tacchella16, McClymont26}.

If only $\approx2\%$ of the possible hosts at any moment are shining with a short-lived BH*, they may be thought of as flickering on and off akin to disco lights when viewed across the entire galaxy population (as opposed to long-lived, steady flood-lights). BH*s may therefore represent a crucial phase in the growth of almost every massive black hole. Indeed, assuming similar parameters ($\approx4\%$ duty cycle, lifetime of few $\approx10$s of Myrs), \citet[][]{Begelman25} observe that the descendants of LRDs may account for almost the entire present-day BH mass density.

\subsection{How do BH*s form? Further evidence for Nuclear Star Clusters and Supermassive Stars?}
\label{sec:formation}

BH*s bear a striking resemblance to aspects of various seeding and growth models. The cocoons of gas nourishing BH*s are highly conducive to super-Eddington accretion (e.g., spherical accretion may dissipate radiation pressure via convection; \citealt{Begelman79, Begelman08, Begelman25, Liu25BB}). Their short lifetimes and low duty cycle (\S\ref{sec:duty}) imply this is a commonplace phenomenon that almost every galaxy experiences across diverse environments -- and so their seeding channels may be similarly abundant for instance via globular clusters/nuclear star clusters \citep[e.g.,][]{Alexander14}. The presence of metals in BH* atmospheres (e.g., Fe, O in the median stack) may imply BH* formation does not require pristine conditions \citep[e.g.,][]{Pacucci26} as these sources are detected over several billion years of cosmic time (see e.g., LRD analogs at $z\approx0$; \citealt{Izotov07,Izotov08,Lin25, Ji25lol}), or because metals are perhaps formed in the process. 

Furthermore, we argue in Figs. \ref{fig:sfg_control}, \ref{fig:host_sfr} and \ref{fig:prospector} that BH*s preferentially live in hosts with very recent bursts of star-formation and highly dense ISM conditions. This is intriguing for several proposed channels of BH seeding and growth. For example, the dense, gas-rich nuclear star clusters (NSCs) possibly arising from these bursts may be prime sites for the formation of supermassive stars (SMSs; $\approx10^{5}\,{\rm M}_{\rm{\odot}}$), e.g, via runaway collisions that can build up $\approx10^{5}\,{\rm M}_{\rm{\odot}}$ in a mere $\approx2$ Myrs \citep[e.g.,][]{Denissenkov14, Gieles18, Woods19, Charbonnel23}. The remnants of SMSs may form BH*s, perhaps via a failed supernova scenario where the SMS fails to cast off its outer layers even after collapsing into a black hole at its center \citep[e.g.,][]{Begelman08, Begelman25} or because the BH may be entrained in gas in its dense formation environment \citep[e.g.,][]{Alexander14, Chon26}. It is also intriguing to note an emerging connection between BH*s, SMSs, and globular clusters from chemical abundances, particularly Nitrogen enhancement that is prominent among broad-line AGN as well as LRDs \citep[e.g.,][]{Belokurov23, Charbonnel23, Morel25, Ji25Nitrogen, Isobe25, Schaerer25GCs}.

It has also been proposed that SMSs may in fact be the central engines of LRDs as opposed to BH*s. There is no direct signature yet that rules SMSs out per se (e.g., broad lines could arise from electron scattering in their stellar atmospheres similar to BH*s). However, unlike BH*s we are lacking SMS models at the moment that are able to reproduce the observed emission lines, continuua, temperatures, and luminosities simultaneously \citep[e.g.,][]{Nandal25,Zwick25}. Most significantly, we note that the extremely short lifetimes of SMSs -- $\approx2$ Myr even under optimistic assumptions \citep[][]{Nandal25SMSlifetime} -- are at odds with our derived BH* lifetimes of 10--20 Myrs (\S\ref{sec:duty}).

\section{Summary}
\label{sec:summary}

\noindent $\bigstar$ We present a novel, empirical approach to isolate the central engines of LRDs. The key assumption we make is that the [\ion{O}{3}] luminosity arises exclusively from the host galaxy. We use observed galaxies at similar redshift and with similar $L_{\rm{[OIII]}}$ to approximate and subtract the hosts (\S\ref{sec:methods}, Figs. \ref{fig:method}, \ref{fig:workedexamples}). We validate this approach with detailed mock tests (Fig. \ref{fig:mock_test}, Table \ref{tab:mock}), and then apply it to a sample of 98 LRDs selected from the DAWN JWST Archive with high quality NIRSpec/PRISM spectra \citep[][]{degraaff25pop, Torralba25IFU}.

\vspace{0.2cm}
\noindent $\bigstar$ The host-subtracted median stack of our full sample represents the central engine of LRDs (Figure \ref{fig:main_stack}, \S\ref{sec:reveal}, Table \ref{tab:mainstacklines}). This stack exhibits a constellation of features that are the hallmark signatures of BH*s, leading us to conclude BH*s are indeed the central engines of LRDs. These features include:
\begin{itemize}
    \item \textit{Balmer break:} The stack has a strong Balmer break ($f^{\nu}(4050\rm{\AA})$/$f^{\nu}(3670\rm{\AA}) = 6.50^{+3.70}_{-1.72}$ that far exceeds the strongest breaks seen in quiescent galaxies. The only sources with comparable breaks are the archetypal BH*s, MoM-BH* and The Cliff. [Fig. \ref{fig:context}]
    
    \item \textit{H$\alpha$ strength and Balmer decrement:} The EW(H$\alpha$) $\approx850$\AA\ and  remarkable Balmer decrement (H$\alpha$/H$\beta\approx 16$) are far more extreme than local AGN. The Balmer decrement is unlikely to be due to dust given stringent non-detections of LRDs in the FIR. Instead, these lines are expected as products of collisional excitation, resonant scattering, and star-like absorption ($\tau_{\rm{H\beta}}>\tau_{\rm{H\alpha}}$) in BH* models. [Fig. \ref{fig:context}]
    
    \item \textit{Fluorescent, Pumped \ion{Fe}{2} and \ion{O}{1} emission:} The stack contains several strong \ion{Fe}{2} and \ion{O}{1} lines, including in the UV, which are expected to arise in BH*-like conditions as a result of Ly$\alpha$ and Ly$\beta$ fluorescence. This is also evidence that the BH* envelopes are not metal-free. [Fig. \ref{fig:main_stack}]
    
    \item \textit{Blackbody-like optical SED:} A single blackbody with $T_{\rm{eff}}=4047^{+156}_{-155}$ K provides an empirical description of the rest-optical continuum. An SED peaking in the optical and falling off at longer and shorter wavelengths is a common feature among BH* models. The fit implies an $L_{\rm{bol}}=10^{43.9^{+0.1}_{-0.1}}$ erg s$^{-1}$, while the Stefan-Boltzmann law implies an enormous radius of $1321^{+168}_{-153}$ au, two orders of magnitude larger than the largest known stars but comparable to local broad-line regions. [Fig. \ref{fig:blackbody}]
\end{itemize}

\noindent $\bigstar$ Decomposing the emission of the BH* from the host galaxy allows us to cleanly infer the properties of the two separate components. We find the following host properties:

\begin{itemize}
    \item LRD hosts are faint ($M_{\rm{UV}}\approx-18.5$) dwarf galaxies ($\approx10^{8}\,{\rm M}_{\rm{\odot}}$), in excellent agreement with constraints from clustering. [Fig. \ref{fig:prospector}, Table \ref{tab:properties}]
    
    \item Compared to a control sample carefully matched in $L_{\rm{[OIII]}}$, we find LRD hosts show far stronger ($\approx3\times$) emission line EWs -- in the rest-optical (e.g., EW([\ion{O}{3}])$\approx1100$\AA) as well as in the rest-UV (e.g., EW([\ion{C}{3})$\approx12$\AA) -- implying highly ionizing, young stellar populations. [Fig. \ref{fig:host_sfr}, Table \ref{tab:control}]
    
    \item Indeed, LRD hosts lie above the star-forming main sequence, and display rising star-formation histories. These features hint that recent starbursts may play a role in the formation of BH*s and/or their entrapment in gas. [Fig. \ref{fig:prospector}].
\end{itemize}

\noindent $\bigstar$ We study how combinations of BH*s and hosts produce the remarkable diversity of LRD SEDs. We track the BH*/LRD fraction with wavelength and across various properties.

\begin{itemize}
    \item In the typical LRD, the BH*/LRD fraction remains modest (but non-zero) in the rest-UV ($\approx20\%$ at $2000-3000$\AA); then, there is an abrupt transition around, but not exactly at the Balmer break ($\approx3600-4000$\AA), redwards of which the BH* begins to dominate the light; by $\approx1\mu$m almost the entire SED is explained by the BH*. 
     
     \item The BH*/LRD fraction with wavelength shows strong luminosity dependence -- the brightest (hottest) BH*s account for almost half the light in the UV and show the sharpest, almost vertical Balmer breaks. Fainter (colder) BH*s ramp up gradually with wavelength and do not display such sharp breaks. [Figure \ref{fig:bhstarfrac}, \S\ref{sec:bhstarfrac}]
     
     \item BH*/LRD fractions correlate with easily accessible observables, namely the LRD [\ion{O}{3}] EW and Balmer break strength. This may help accelerate the search for BH*-dominated objects, enabling the study of ``pure" BH*s without interference from hosts. On the other extreme, host-dominated LRDs appear to be some of the most spectacular starbursts at high-redshift with e.g., [OIII]${\rm{5008}\AA}$ EWs reaching $\approx3000$\AA. [Figs. 
     \ref{fig:bhstar_frac_correlation}, \ref{fig:bhstar_frac_vs_o3}, \S\ref{sec:predictbhstarfrac}, \ref{sec:remarkable}]
\end{itemize}

\vspace{0.3cm}
\noindent$\bigstar$ V-shaped LRD selections are preferentially sensitive to high BH*/LRD fractions of $\gtrsim60\%$. The fraction of BH*s among non-LRDs (e.g., ``normal'' broad-line AGN) is highly likely to be non-zero, and the population of BH*s may be even more widespread (up to $>2\times$) than current LRD numbers suggest. [Fig. \ref{fig:vshaped}, \S\ref{sec:vshaped}]

\vspace{0.3cm}
\noindent$\bigstar$ Exploiting our constraints on the characteristic host mass and their corresponding number densities, we infer a BH* duty cycle of $\approx1\%$ and a BH* lifetime of $\approx10$ Myrs. BH*s are a short-lived, and yet ubiquitous phenomenon, implying they are a key phase in the origin story of possibly every massive BH. [\S\ref{sec:duty}]

\vspace{0.3cm}
It has long been anticipated that the early Universe held spectacular, unknown chapters of the origin story of black holes. BH*s may represent one such chapter. With a clear, resolved view of BH*s as well as the galaxies they grow in, we may now begin unraveling the detailed physics of their birth and growth.

\section*{Acknowledgments}

We thank the two anonymous referees for their insightful comments that have strengthened this work.

WQS and RPN acknowledge funding from {\it JWST} programs GO-3516, GO-5224, and the MIT Undergraduate Research Opportunities Program (UROP). Support for this work was provided by NASA through the NASA Hubble Fellowship grant HST-HF2-51515.001-A awarded by the Space Telescope Science Institute, which is operated by the Association of Universities for Research in Astronomy, Incorporated, under NASA contract NAS5-26555. RPN thanks Neil Pappalardo and Jane Pappalardo for their generous support of the MIT Pappalardo Fellowships in Physics, and for their enthusiasm and encouragement for pursuing the earliest galaxies and black holes. JM and AT acknowledge funding from the European Union (ERC, AGENTS,  101076224). KEH acknowledges support from the Independent Research Fund Denmark (DFF) under grant 5251-00009B and co-funding by the European Union (ERC, HEAVYMETAL, 101071865). Views and opinions expressed are, however, those of the authors only and do not necessarily reflect those of the European Union or the European Research Council. Neither the European Union nor the granting authority can be held responsible for them. REH acknowledges support by the German Aerospace Center (DLR) and the Federal Ministry for Economic Affairs and Energy (BMWi) through program 50OR2403 `RUBIES'.

The data products presented herein were retrieved from the Dawn JWST Archive (DJA). DJA is an initiative of the Cosmic Dawn Center (DAWN), which is funded by the Danish National Research Foundation under grant DNRF140. This work is based on observations made with the NASA/ESA/CSA James Webb Space Telescope. The data were obtained from the
Mikulski Archive for Space Telescopes at the Space Telescope Science Institute, which is operated by the Association of Universities for Research in Astronomy, Inc., under NASA contract NAS 5-03127
for JWST. Support for programs
\#3516, \#5224, \#5664 was provided by NASA through grants from the Space
Telescope Science Institute, which is operated by the Association of
Universities for Research in Astronomy, Inc., under NASA contract
NAS 5-03127. 

The spectra used in this paper are associated with programs 1180 \citep[][]{DEugenio25},
1181 (PI: D. Eisenstein), 1208 \citep[][]{Willott22}, 1210 (PI: N. Luetzgendorf), 1211 \citep[][]{Maseda24}, 1212 - 1215 (PI: N. Luetzgendorf), 1228 \citep[][]{Luhman24}, 1229 \citep[][]{Luhman24b}, 1286 (PI: N. Luetzgendorf), 1287 (PI: K. Isaak), 1345 \citep[][]{Finkelstein23}, 1433 \citep[][]{Hsiao24}, 1747 (PI: G. Roberts-Borsani), 2028 \citep[][]{Wang24j0910}, 2073 (PI: J. Hennawi), 2198 \citep[][]{Barrufet25}, 2282 \citep[][]{Bradley23}, 2561 \citep[][]{Bezanson24}, 2565 \citep[][]{Nanayakkara25}, 2640 (PI: W. Best), 2750 \citep[][]{ArrabalHaro23}, 2756 \citep[][]{Mascia24},
2767 \citep[][]{Williams23rxj}, 2770 (PI: M. McCaughrean), 3073 \citep[][]{Castellano24abell}, 3215 \citep[][]{Eisenstein25}, 4106 (PI: E. Nelson), 4233 \citep[][]{degraaff25rubies}, 4446 \citep[][]{Frye24}, 4557 (PI: H. Yan), 5105 \citep[][]{Shen24nexus}, 5224 (PIs: P.A. Oesch \& R.P. Naidu), 6368 (PI: M. Dickinson), 6541 \citep[][]{DeCoursey25}, 6585 (PI: D. Coulter), 6642 (PI: J. Muzerolle Page), and FRESCO IFU \citep[][]{Matthee24, Torralba25IFU}. 

Software used in developing this work includes: \texttt{matplotlib} \citep{matplotlib}, \texttt{jupyter} \citep{jupyter}, \texttt{IPython} \citep{ipython}, \texttt{numpy} \citep{numpy}, \texttt{scipy} \citep{scipy}, \texttt{TOPCAT} \citep{topcat}, \texttt{Astropy} \citep{astropy}, \texttt{msaexp} \citep[][]{msaexp}.

% \end{CJK*}

\bibliography{MasterBiblio}
\bibliographystyle{apj}

\appendix
\section{Host/BH* Decomposition for Individual LRDs}
In Figs. \ref{fig:all_lrd_decomposition} - \ref{fig:final_lrd_decomposition} we summarize the spectral decomposition of every LRD analyzed in this work by presenting its median host galaxy stack and median $\text{LRD}-\text{Host}$ (i.e., BH*) stack following the format in Fig. \ref{fig:workedexamples}. The sources are sorted in descending order of their BH* fraction at 5500\AA, ranging from $\approx100\%$ to $\approx40\%$.

\begin{figure*}
    \centering
    \includegraphics[width=0.75\linewidth]{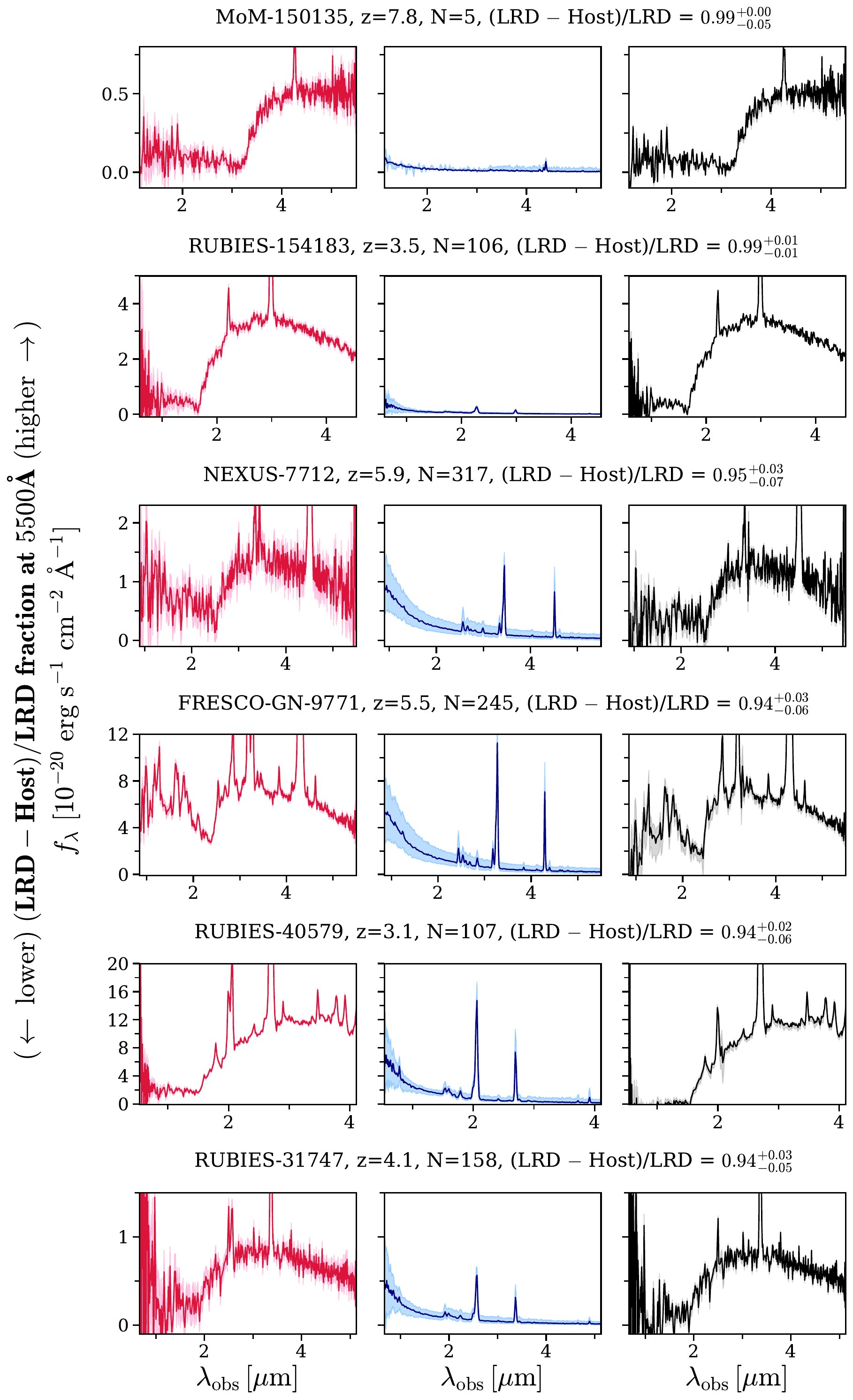}
    \label{fig:appendix_gallery}
    \caption{\textbf{Decomposition for all LRDs in our sample} (to be continued). Median stacks for the inferred hosts (center) and $\text{LRD}-\text{Host}$ spectra (i.e., the central engine; right) are shown, sorted by the ($\text{LRD} - \text{Host}$ )/LRD fraction at 5500\AA. The number of host matches ($N$) for the LRD is indicated in each title.}
    \label{fig:all_lrd_decomposition}
\end{figure*}

\begin{figure*}
    \centering
    \includegraphics[width=0.75\linewidth]{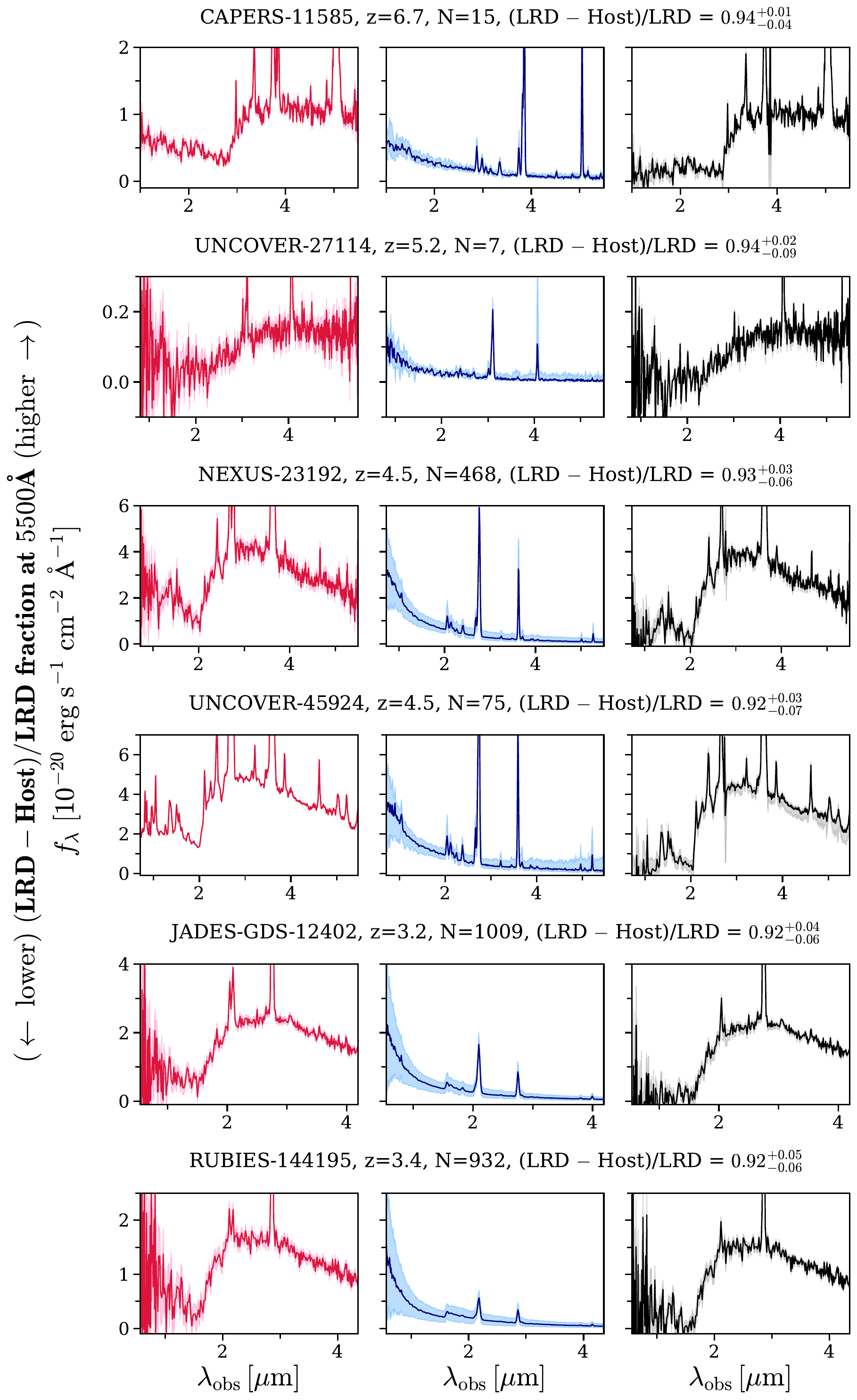}
    \caption{Same as Figure \ref{fig:all_lrd_decomposition}.}
\end{figure*}

\begin{figure*}
    \centering
    \includegraphics[width=0.75\linewidth]{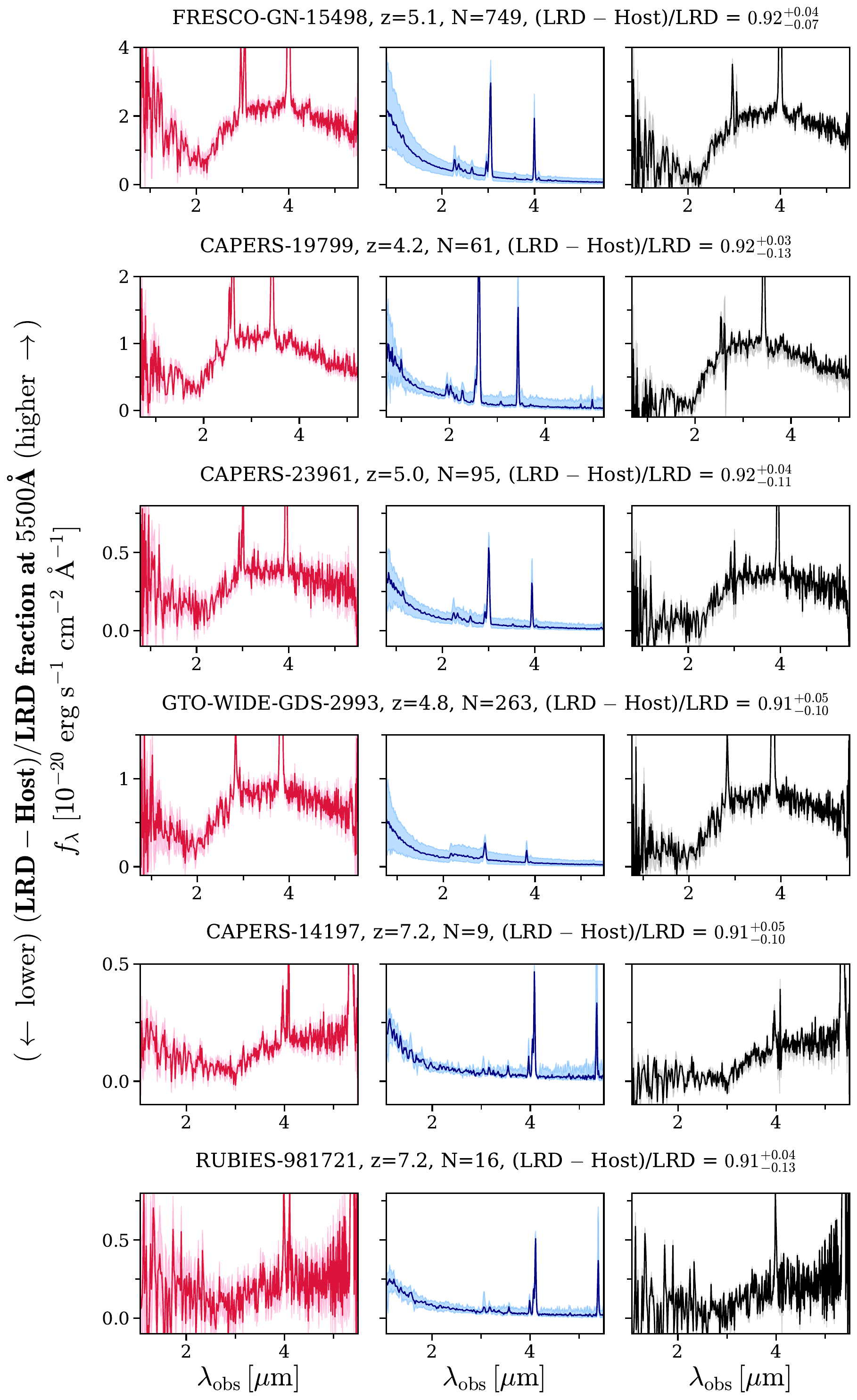}
    \caption{Same as Figure \ref{fig:all_lrd_decomposition}.}
\end{figure*}

\begin{figure*}
    \centering
    \includegraphics[width=0.75\linewidth]{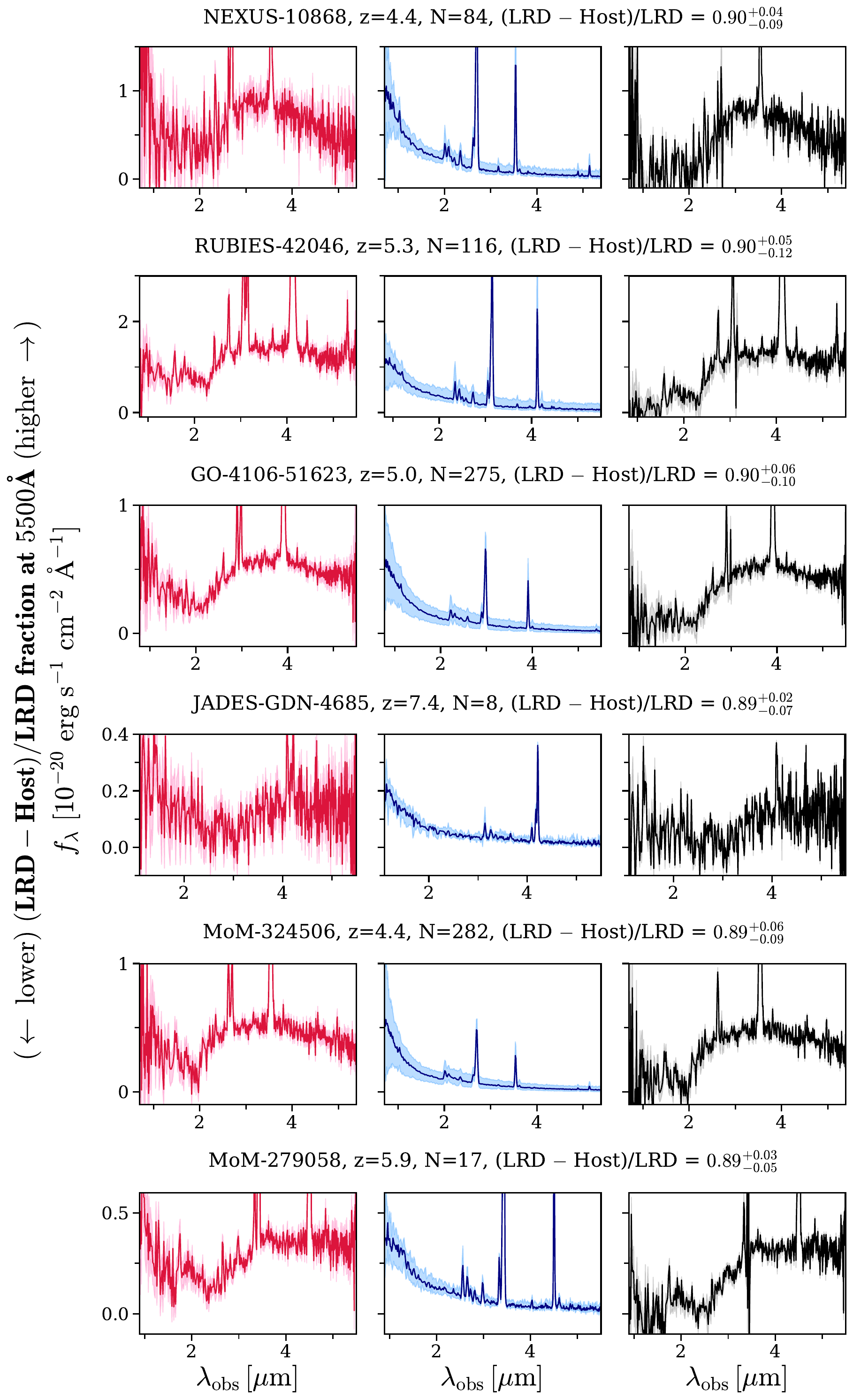}
    \caption{Same as Figure \ref{fig:all_lrd_decomposition}.}
\end{figure*}

\begin{figure*}
    \centering
    \includegraphics[width=0.75\linewidth]{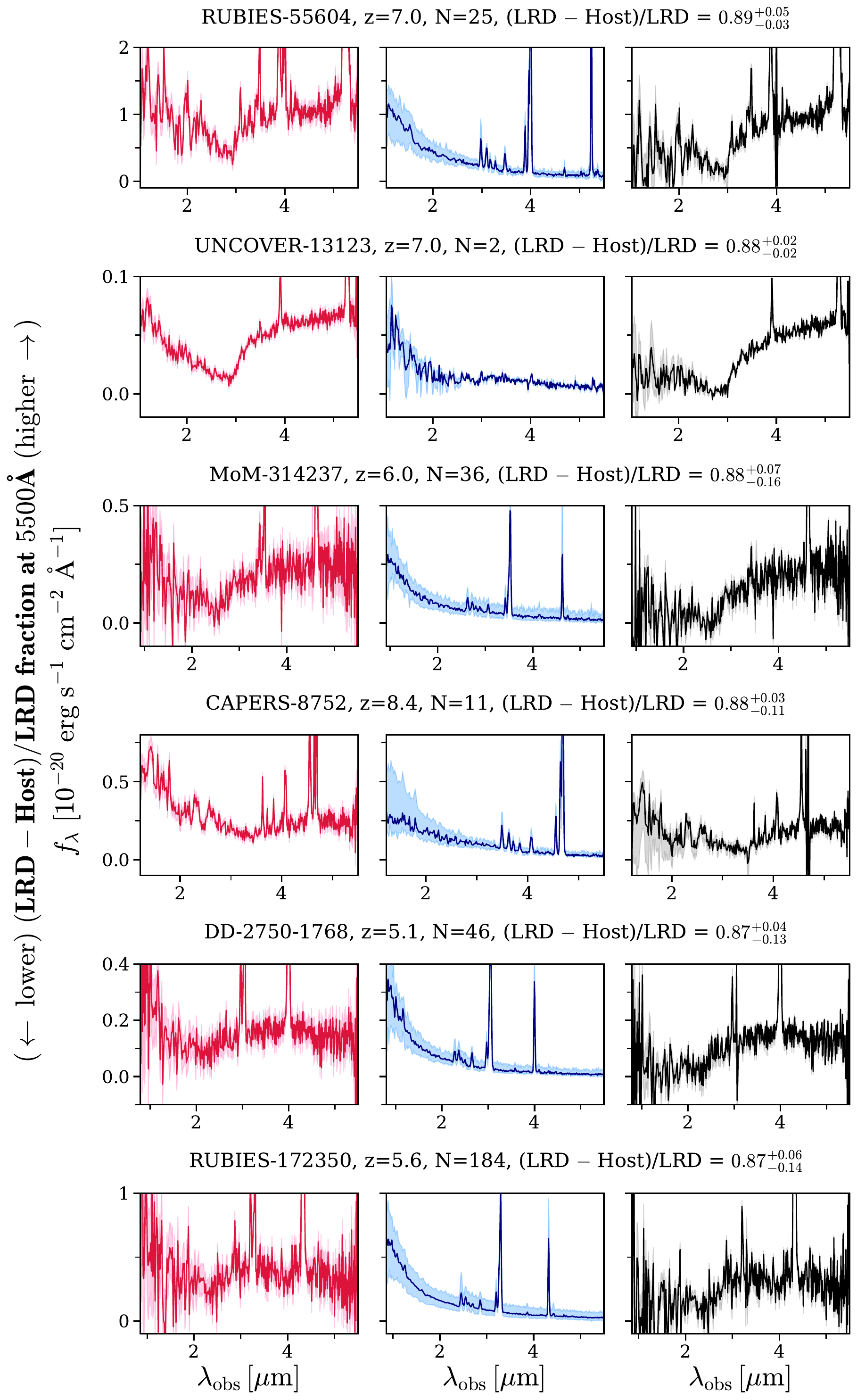}
    \caption{Same as Figure \ref{fig:all_lrd_decomposition}.}
\end{figure*}

\begin{figure*}
    \centering
    \includegraphics[width=0.75\linewidth]{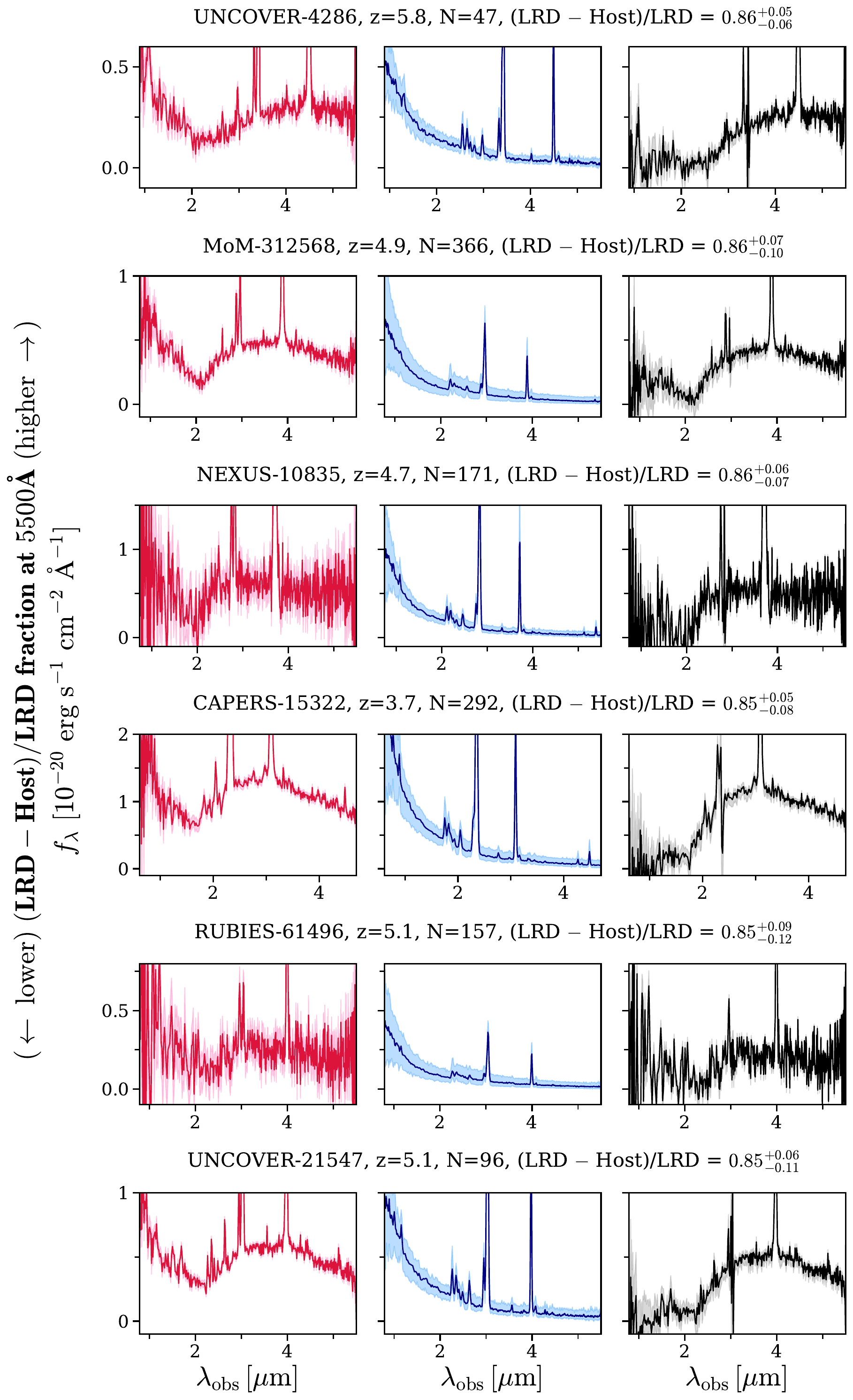}
    \caption{Same as Figure \ref{fig:all_lrd_decomposition}.}
\end{figure*}

\begin{figure*}
    \centering
    \includegraphics[width=0.75\linewidth]{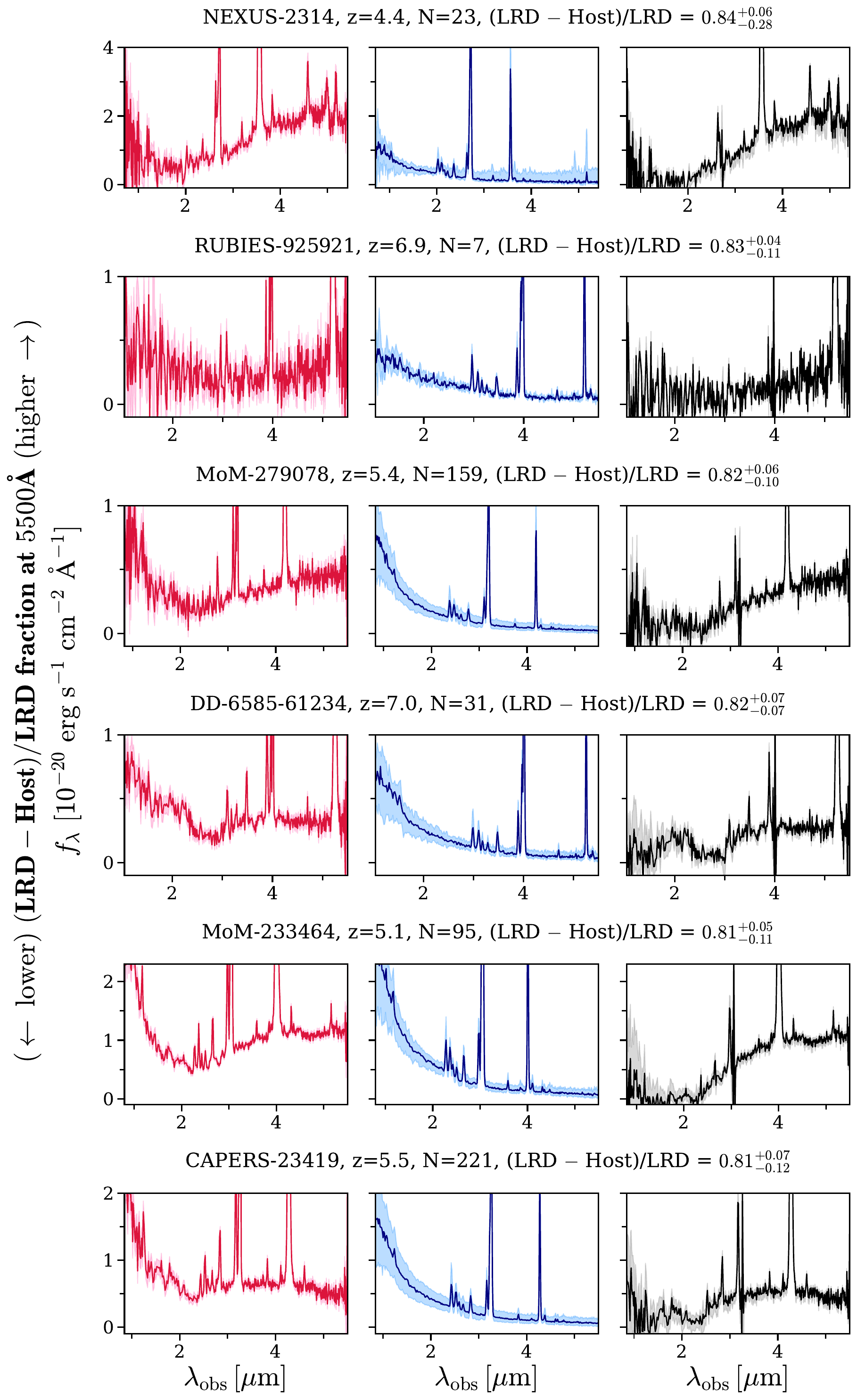}
    \caption{Same as Figure \ref{fig:all_lrd_decomposition}.}
\end{figure*}

\begin{figure*}
    \centering
    \includegraphics[width=0.75\linewidth]{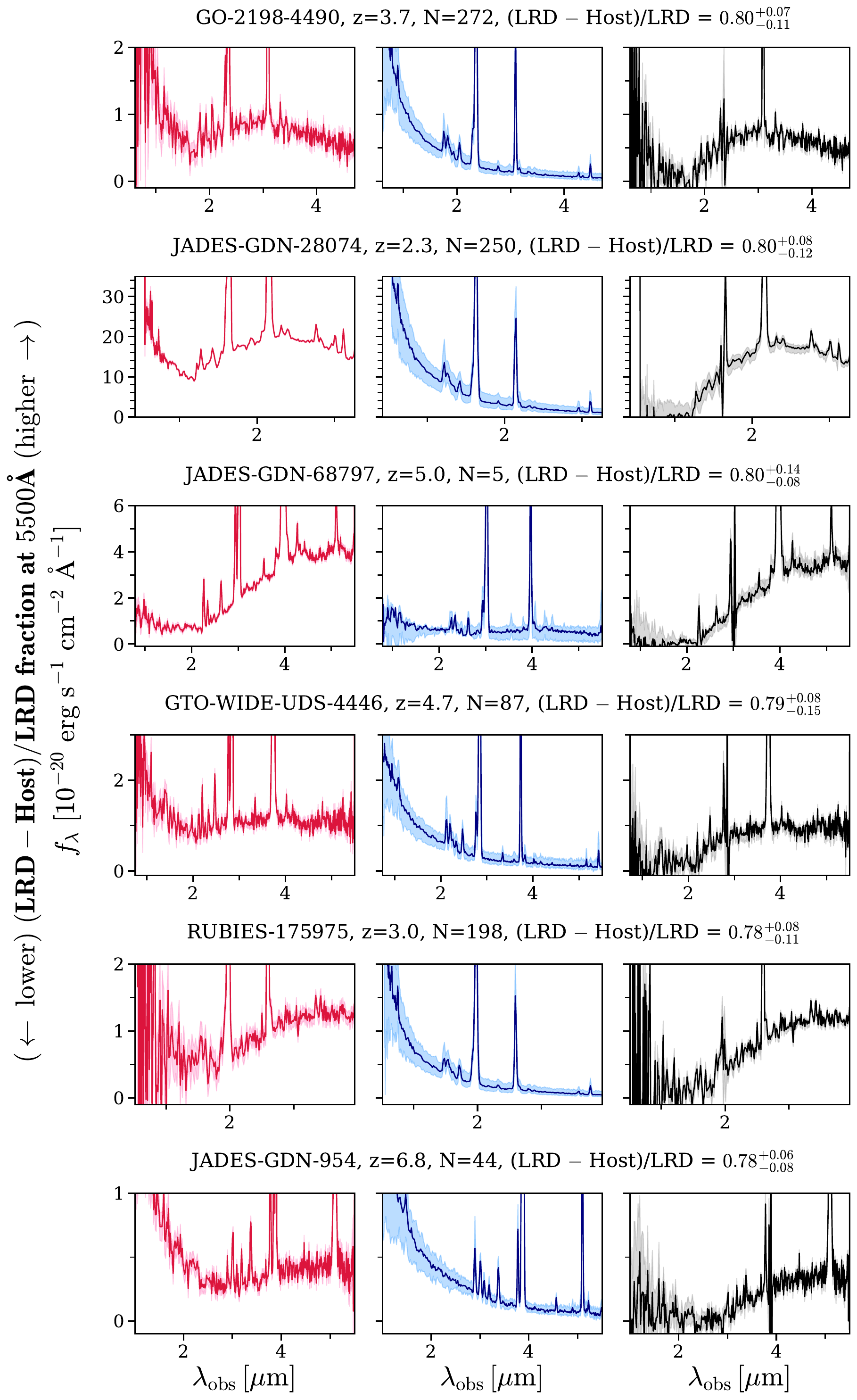}
    \caption{Same as Figure \ref{fig:all_lrd_decomposition}.}
\end{figure*}

\begin{figure*}
    \centering
    \includegraphics[width=0.75\linewidth]{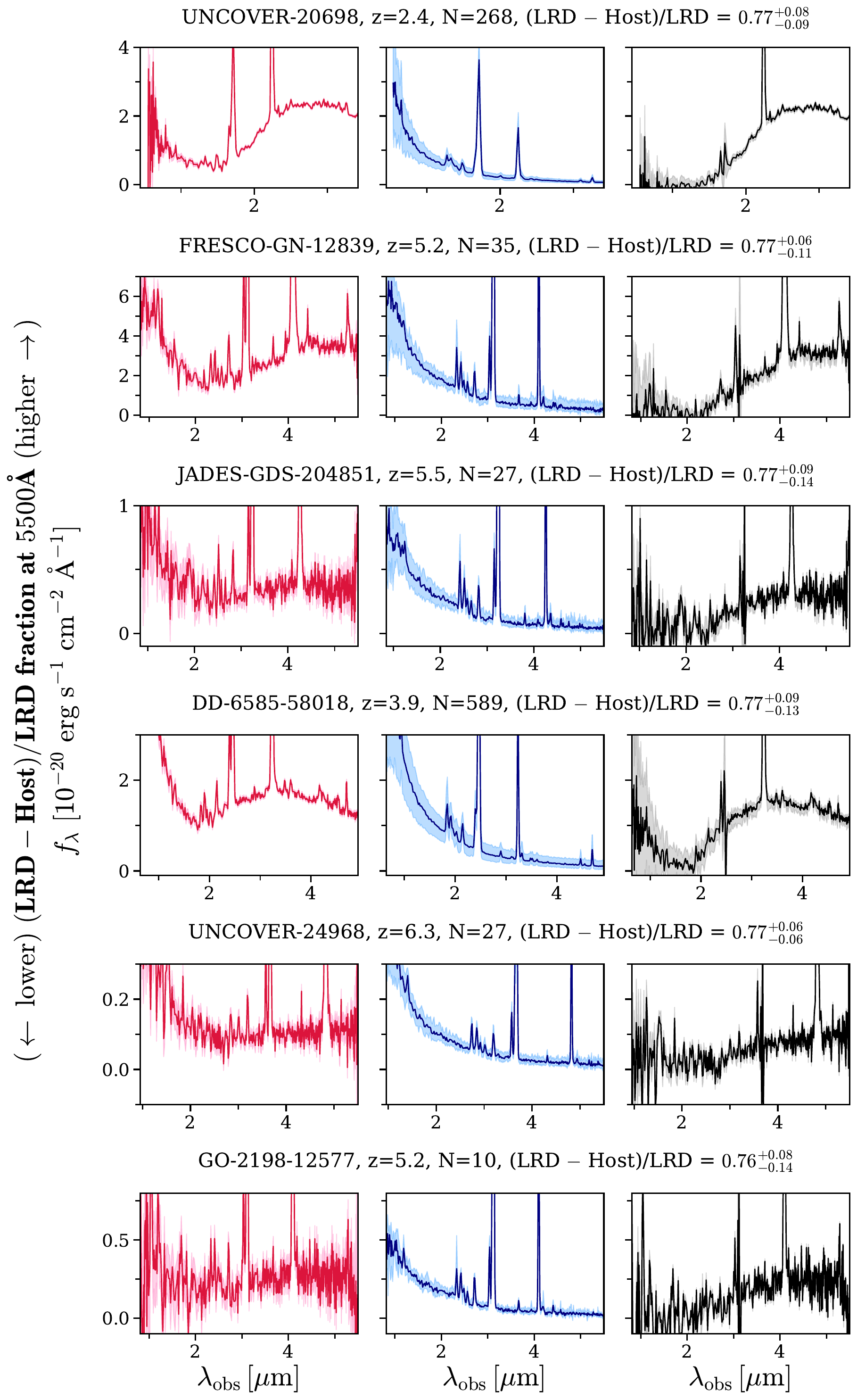}
    \caption{Same as Figure \ref{fig:all_lrd_decomposition}.}
\end{figure*}

\begin{figure*}
    \centering
    \includegraphics[width=0.75\linewidth]{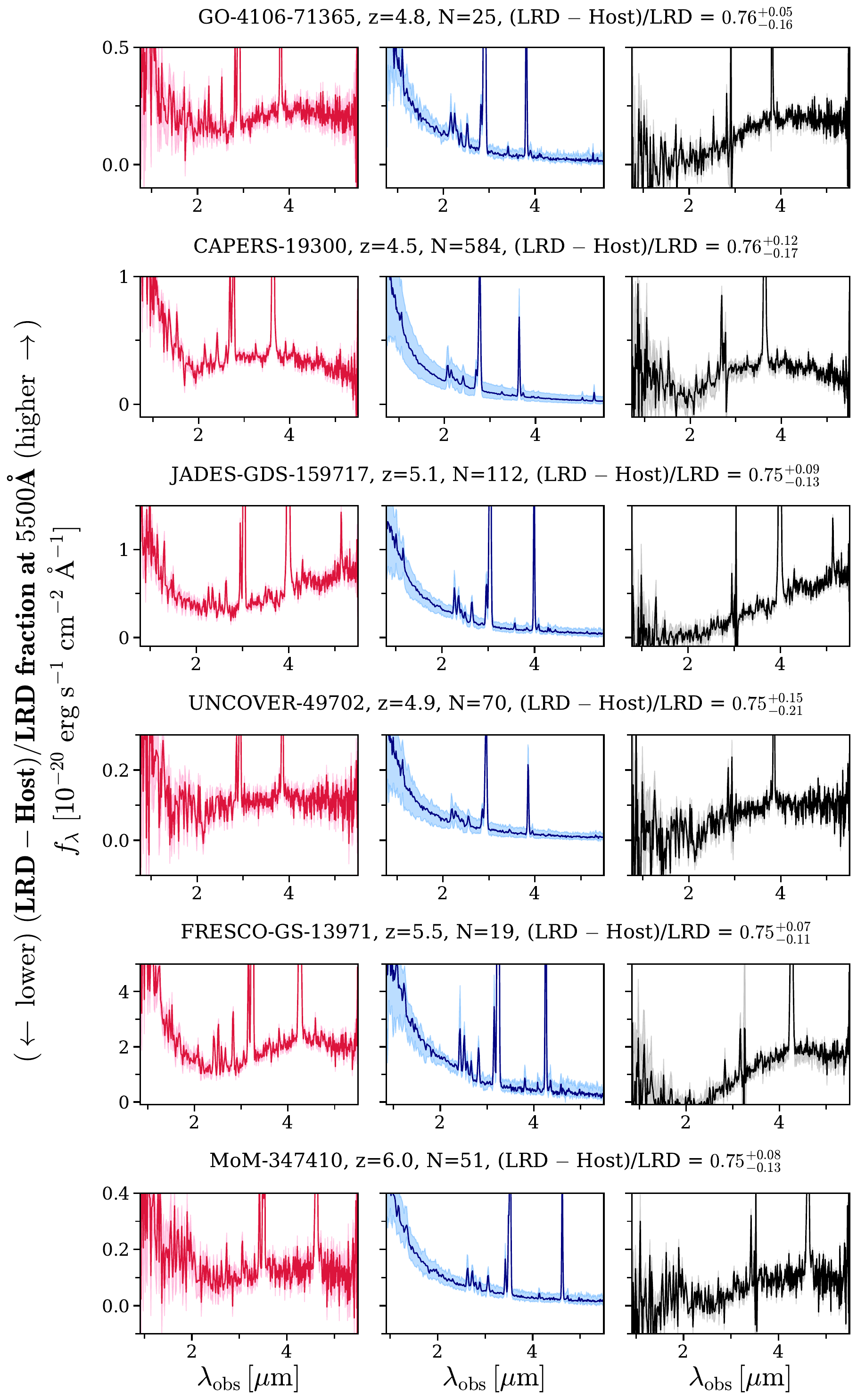}
    \caption{Same as Figure \ref{fig:all_lrd_decomposition}.}
\end{figure*}

\begin{figure*}
    \centering
    \includegraphics[width=0.75\linewidth]{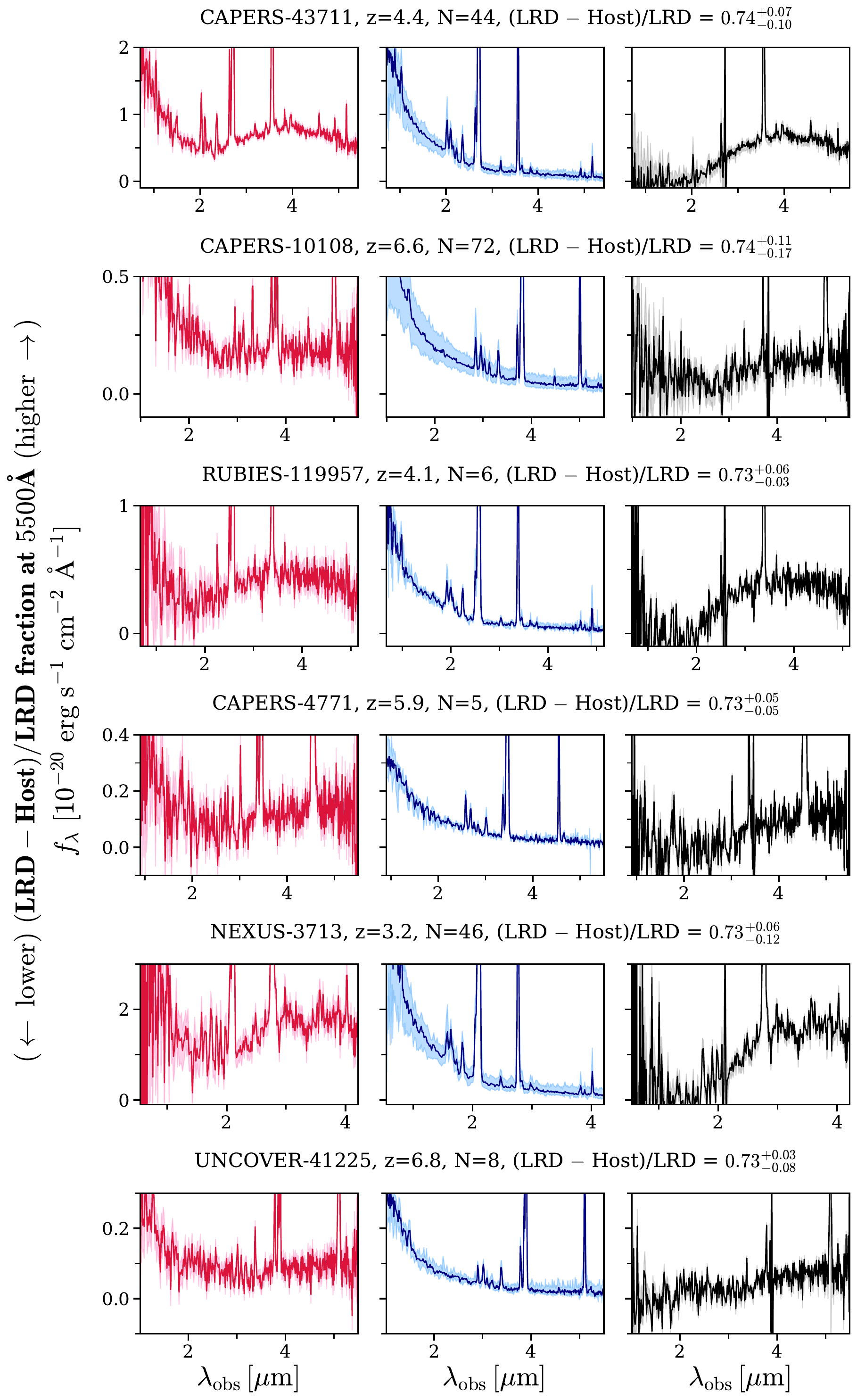}
    \caption{Same as Figure \ref{fig:all_lrd_decomposition}.}
\end{figure*}

\begin{figure*}
    \centering
    \includegraphics[width=0.75\linewidth]{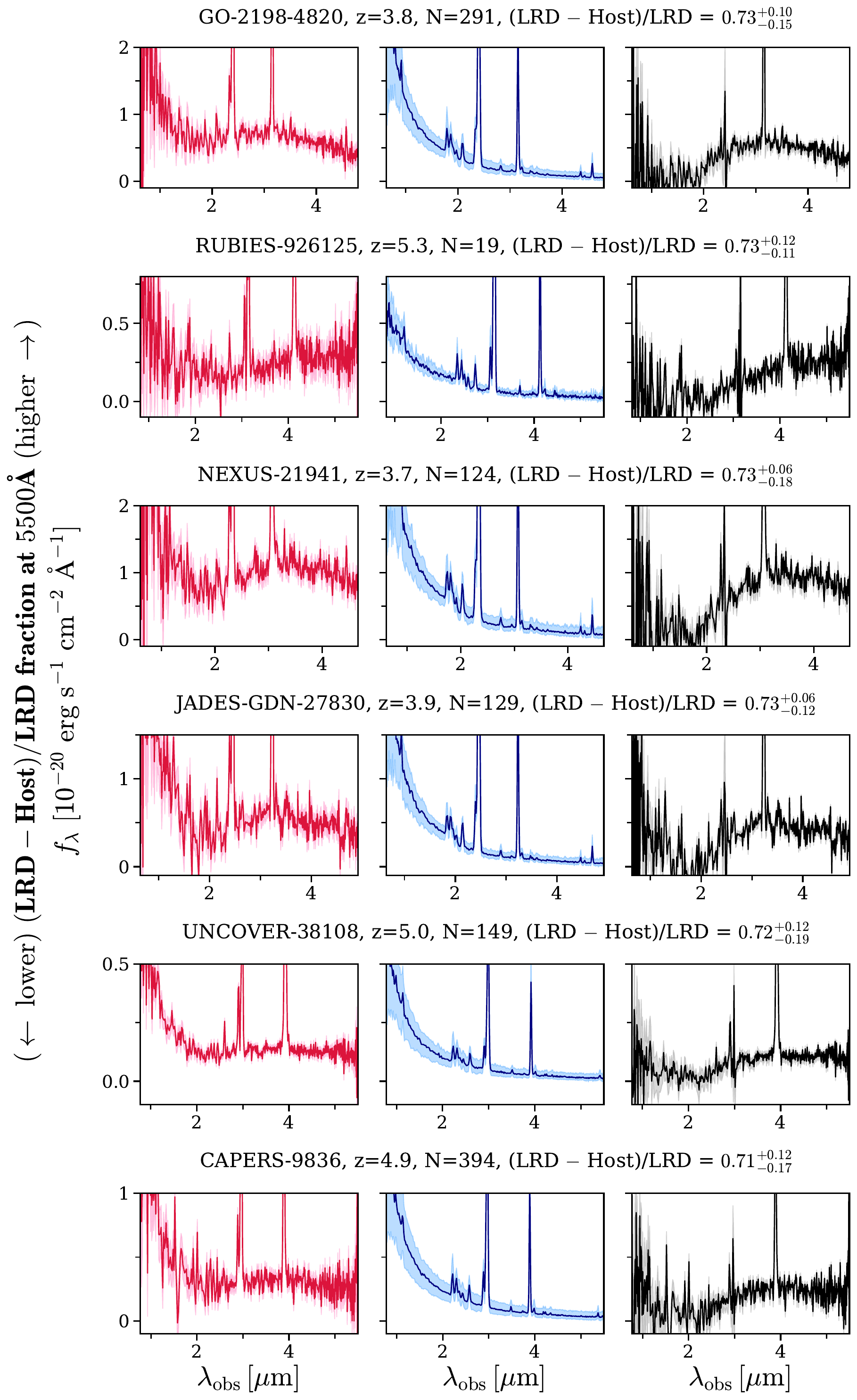}
    \caption{Same as Figure \ref{fig:all_lrd_decomposition}.}
\end{figure*}

\begin{figure*}
    \centering
    \includegraphics[width=0.75\linewidth]{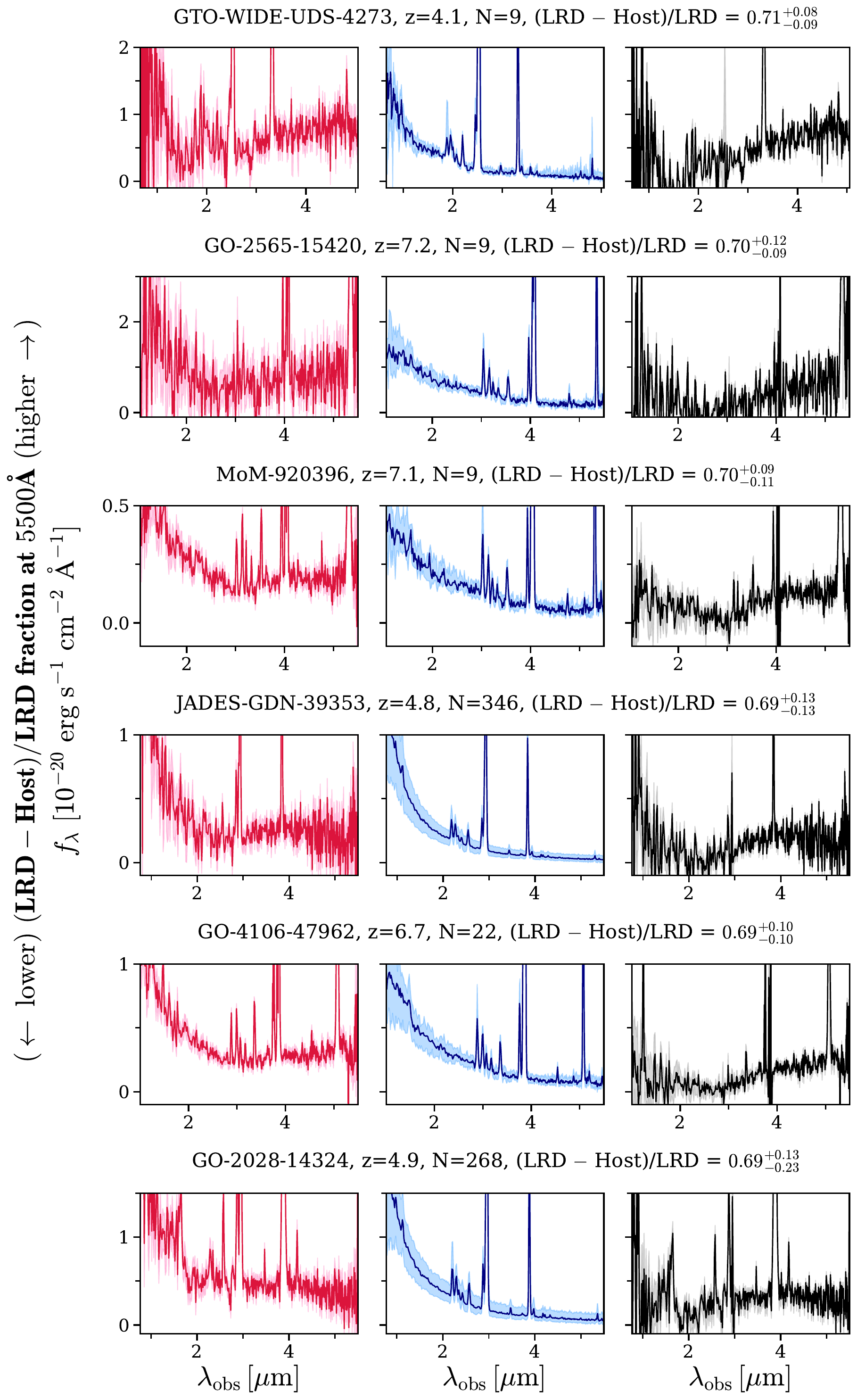}
    \caption{Same as Figure \ref{fig:all_lrd_decomposition}.}
\end{figure*}

\begin{figure*}
    \centering
    \includegraphics[width=0.75\linewidth]{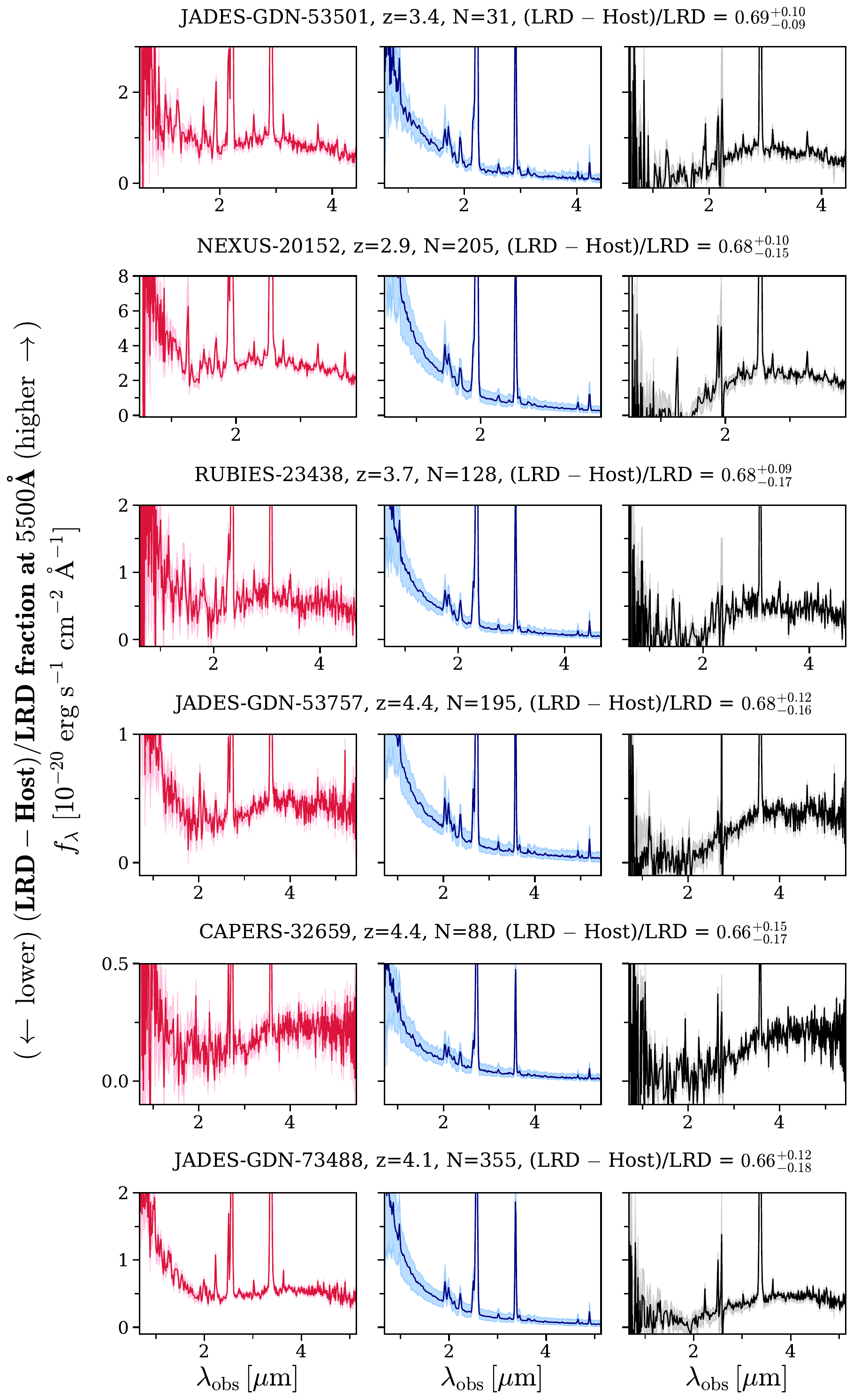}
    \caption{Same as Figure \ref{fig:all_lrd_decomposition}.}
\end{figure*}

\begin{figure*}
    \centering
    \includegraphics[width=0.75\linewidth]{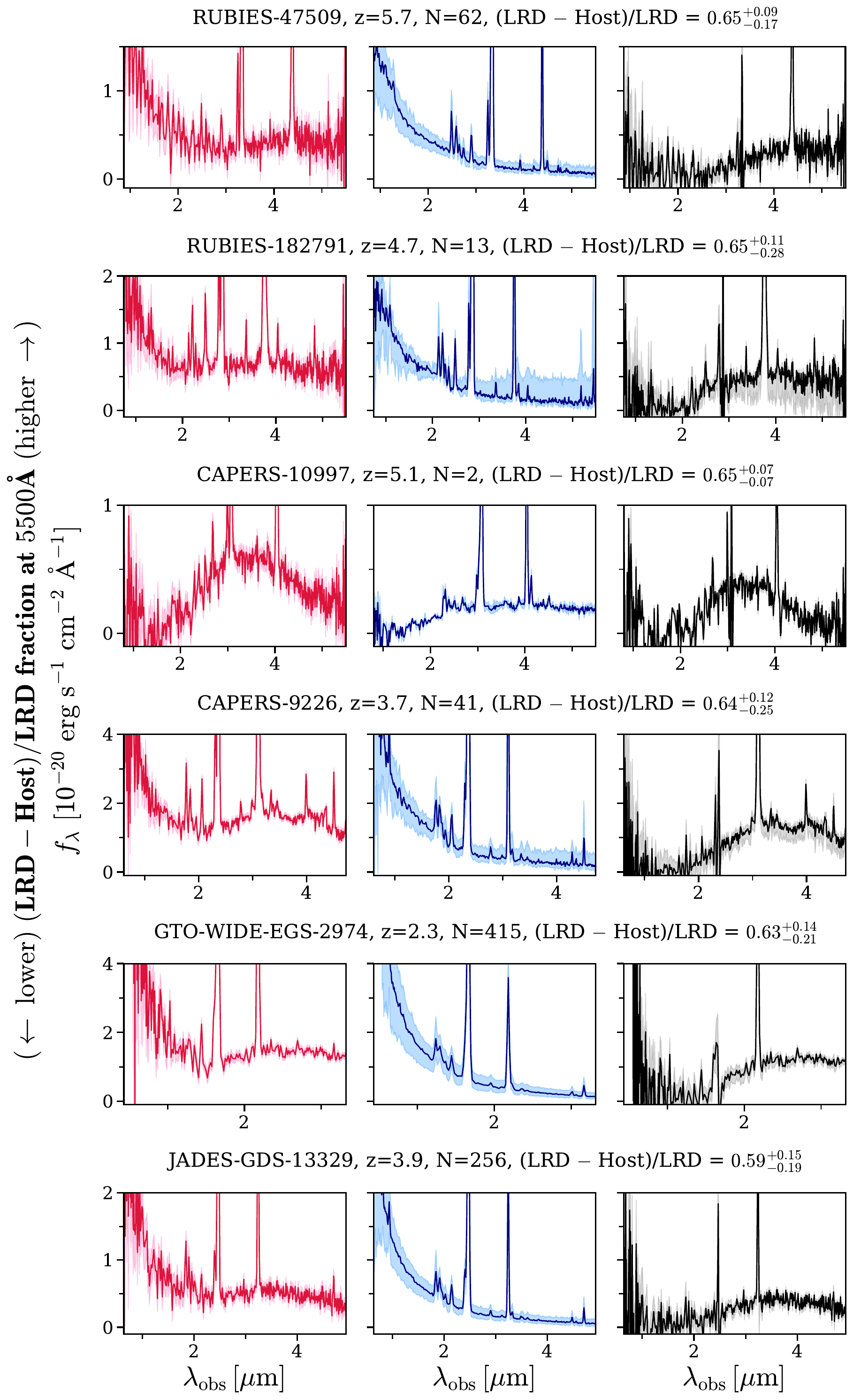}
    \caption{Same as Figure \ref{fig:all_lrd_decomposition}.}
\end{figure*}

\begin{figure*}
    \centering
    \includegraphics[width=0.75\linewidth]{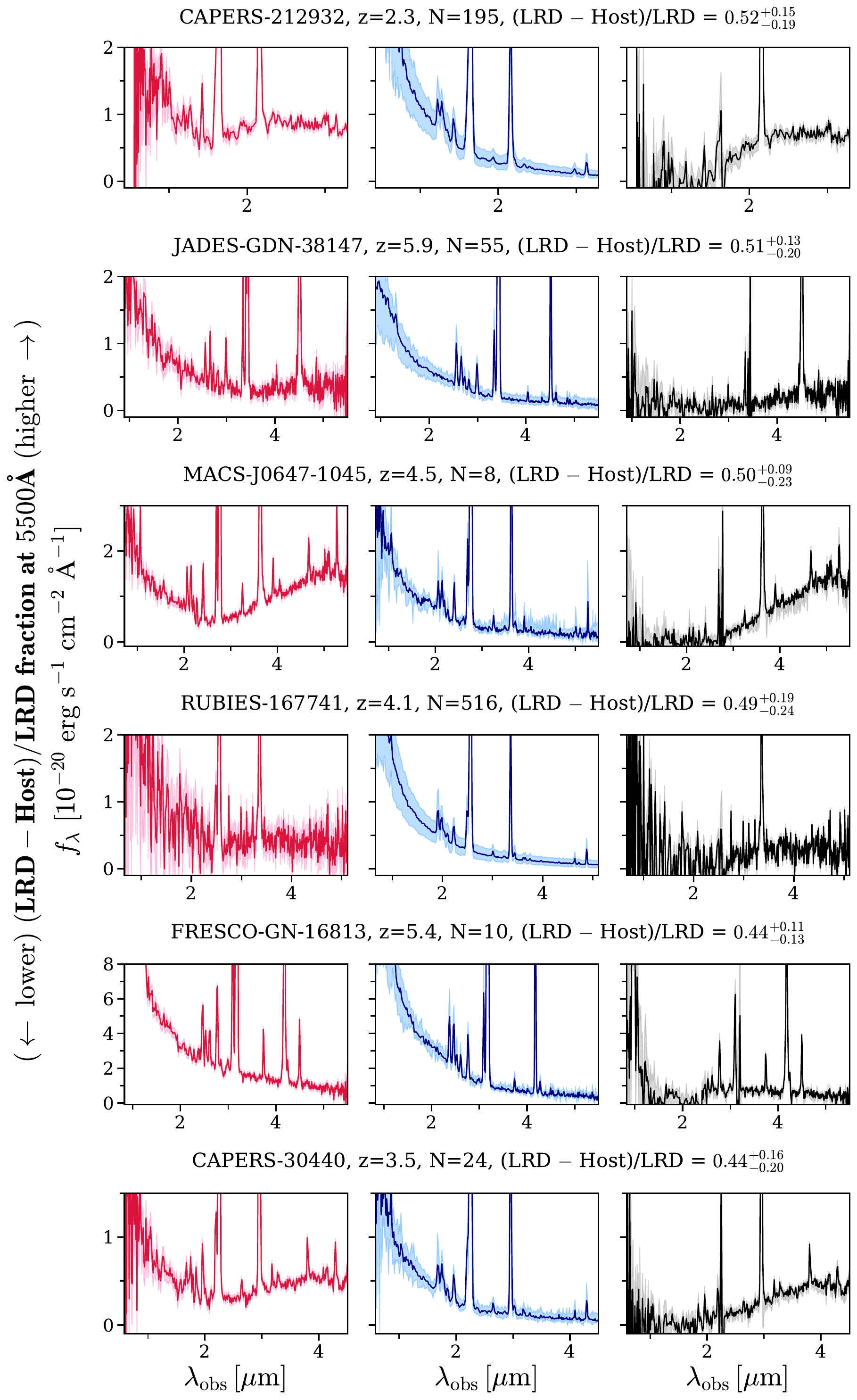}
    \caption{Same as Figure \ref{fig:all_lrd_decomposition}.}
\end{figure*}

\begin{figure*}
    \centering
    \includegraphics[width=0.75\linewidth]{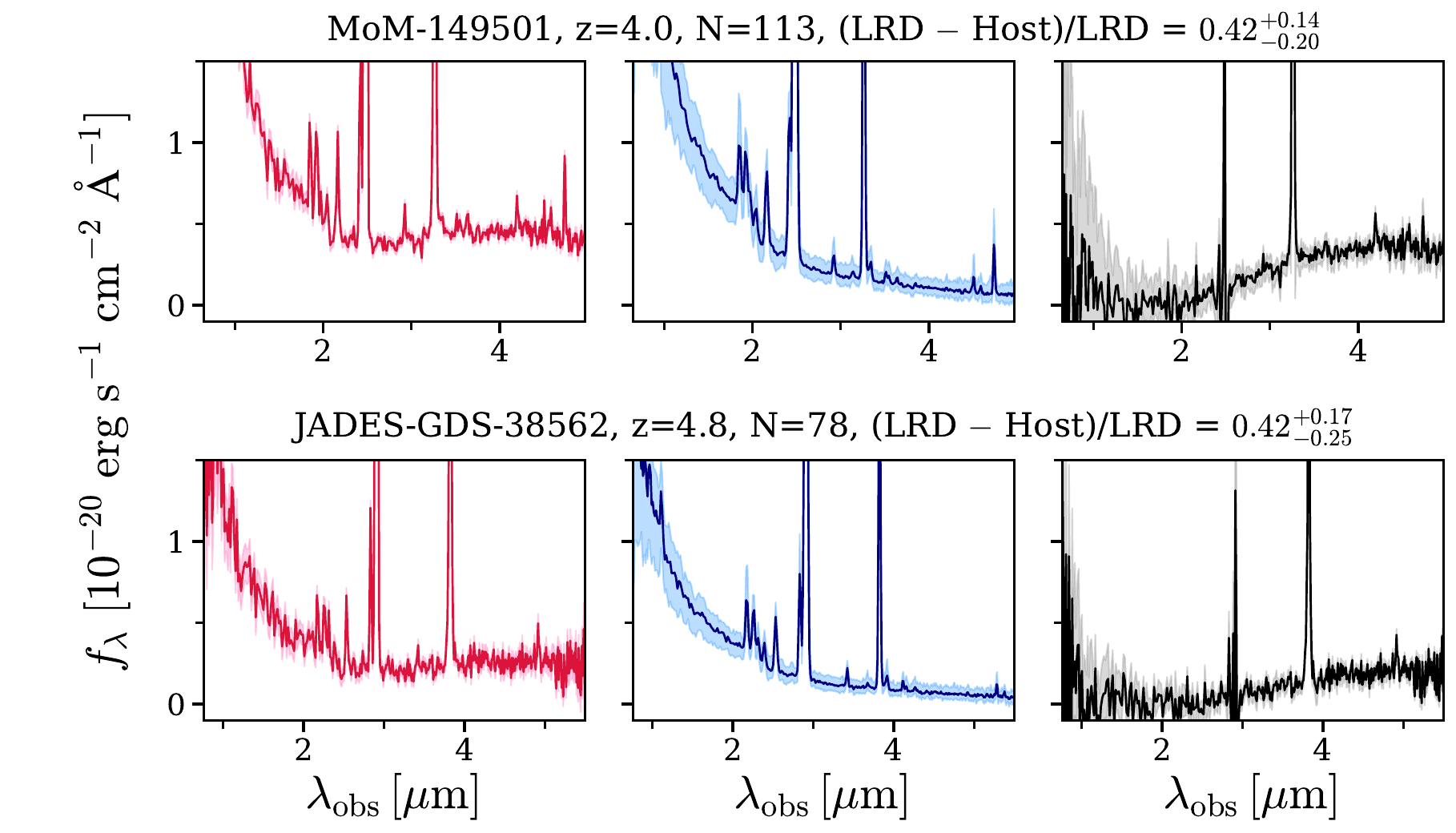}
    \caption{Same as Figure \ref{fig:all_lrd_decomposition}.}
    \label{fig:final_lrd_decomposition}
\end{figure*}

\section{Program Information for LRD Spectra}
In Table \ref{tab:programinfo} we list the fifteen JWST programs that we draw the 98 LRD prism spectra from that are studied in this work.

\begin{deluxetable*}{lr}
\tabletypesize{\footnotesize}
\tablecaption{Program information for the LRD spectra used in this work (98 spectra)}
\tablehead{\colhead{Program} & \colhead{\# LRD Spectra}}
\startdata
\label{tab:programinfo}
\vspace{-0.3cm}\\
CAPERS (GO-6368, PI: M. Dickinson) & $17$ \\
DD-2750 \citep[][]{ArrabalHaro23} & $1$ \\ % ceers-ddt-v4
DD-6585 (PI: D. Coulter) & $2$ \\ % cosmos-transients-v4
FRESCO IFU \citep[][]{Matthee24, Torralba25IFU} & $5$ \\
GO-2028 \citep[][]{Wang24j0910} & $1$ \\ % j0910-wang-v4
GO-2198 \citep[][]{Barrufet25} & $3$ \\ % gds-barrufet-s67-v4
GO-2565 \citep[][]{Nanayakkara25} & $1$ \\ % glazebrook-cos-obs2-v4
GO-4106 (PI: E. Nelson) & $3$ \\ % egs-nelsonx-v4
GTO-WIDE (GTO-1212, GTO-1213, GTO-1215, PI: N. Luetzgendorf) & $4$ \\
JADES (GTO-1180, \citealp{DEugenio25}; GTO-1181, PI: D. Eisenstein; GTO-1286, PI: N. Luetzgendorf) & $15$ \\
MACS-J0647 \citep[GO-1433,][]{Hsiao24} & $1$ \\
MoM (GO-5224, PIs: R.P. Naidu \& P.A. Oesch) & $10$ \\
NEXUS \citep[GO-5105,][]{Shen24nexus} & $8$ \\
RUBIES \citep[GO-4233,][]{degraaff25rubies} & $17$ \\
UNCOVER \citep[GO-2561,][]{Bezanson24} & $10$
\enddata
\end{deluxetable*}

\end{document}